\documentclass[usenatbib,usegraphicx]{mn2e}

\usepackage{times}
\usepackage{url}
\usepackage{color}
\usepackage{multirow}
\usepackage[export]{adjustbox} 

\newcommand{\be}{\begin{equation}}  \newcommand{\ee}{\end{equation}}
  
\newcommand{\ba}{\begin{eqnarray}}\newcommand{\ea}{\end{eqnarray}}

\title[RCSLenS]{RCSLenS: The Red Cluster Sequence Lensing Survey\thanks{Based on observations obtained with MegaPrime/MEGACAM, a joint project of CFHT and CEA/DAPNIA, at the Canada-France-Hawaii Telescope (CFHT) which is operated by the National Research Council (NRC) of Canada, the Institut National des Sciences de l'Univers of the Centre National de la Recherche Scientifique (CNRS) of France, and the University of Hawaii.}}

\author[H.~Hildebrandt et al.]{H.~Hildebrandt,$^{1}$\thanks{Email: hendrik@astro.uni-bonn.de}
A.~Choi,$^2$
C.~Heymans,$^2$
C.~Blake,$^3$
T.~Erben,$^1$
L.~Miller,$^4$
\newauthor
R.~Nakajima,$^1$
L.~van~Waerbeke,$^5$
M.~Viola,$^6$
A.~Buddendiek,$^1$
J.~Harnois-D\'{e}raps,$^{2,5}$
\newauthor
A.~Hojjati,$^5$
B.~Joachimi,$^7$
S.~Joudaki,$^3$
T.~D.~Kitching,$^8$
C.~Wolf,$^9$
S.~Gwyn,$^{10}$
\newauthor
N.~Johnson,$^{11}$
K.~Kuijken,$^6$
Z.~Sheikhbahaee,$^1$
A.~Tudorica$^1$
and
H.~K.~C.~Yee$^{12}$
\\
$^1$Argelander-Institut f\"ur Astronomie, Auf dem H\"ugel 71, 53121 Bonn, Germany\\
$^2$Scottish Universities Physics Alliance, Institute for Astronomy, University of Edinburgh, Royal Observatory, Blackford Hill,\\ Edinburgh EH9 3HJ, UK\\
$^3$Centre for Astrophysics \& Supercomputing, Swinburne University of Technology, PO Box 218, Hawthorn, VIC 3122, Australia\\
$^4$Department of Physics, Oxford University, Keble Road, Oxford OX1 3RH, UK\\
$^5$University of British Columbia, Department of Physics and Astronomy, 6224 Agricultural Road, Vancouver, B.C. V6T 1Z1, Canada\\
$^6$Leiden Observatory, Leiden University, Niels Bohrweg 2, 2333 CA Leiden, The Netherlands\\
$^{7}$Department of Physics and Astronomy, University College London, Gower Street, London WC1E 6BT, UK\\
$^{8}$Mullard Space Science Laboratory, University College London, Holmbury St Mary, Dorking, Surrey RH5 6NT, U.K\\
$^9$Research School of Astronomy and Astrophysics, Australian National University, Canberra ACT 2611 Australia\\
$^{10}$NRC Herzberg Astronomy and Astrophysics, 5071 West Saanich Road, Victoria, BC, V9E 2E7, Canada\\
$^{11}$EPCC, University of Edinburgh, King's Buildings, Edinburgh, UK, EH9 3JZ\\
$^{12}$Department of Astronomy \& Astrophysics, University of Toronto, 50 St. George St., Toronto, Ontario, Canada, M5S3H4\\
}

\date{Released 2014}

\pagerange{\pageref{firstpage}--\pageref{lastpage}} \pubyear{2014}

\begin{document}
\setlength{\voffset}{-12mm}

\label{firstpage}

\maketitle
\begin{abstract}
  We present the Red-sequence Cluster Lensing Survey (RCSLenS), an application of the methods developed for the Canada France Hawaii Telescope Lensing Survey (CFHTLenS) to the $\sim$785\,deg$^2$, multi-band imaging data of the Red-sequence Cluster Survey 2 (RCS2). This project represents the largest public, sub-arcsecond seeing, multi-band survey to date that is suited for weak gravitational lensing measurements. With a careful assessment of systematic errors in shape measurements and photometric redshifts we extend the use of this data set to allow cross-correlation analyses between weak lensing observables and other data sets. We describe the imaging data, the data reduction, masking, multi-colour photometry, photometric redshifts, shape measurements, tests for systematic errors, and a blinding scheme to allow for more objective measurements. In total we analyse 761 pointings with $r$-band coverage, which constitutes our lensing sample. Residual large-scale B-mode systematics prevent the use of this shear catalogue for cosmic shear science. The effective number density of lensing sources over an unmasked area of 571.7\,deg$^2$ and down to a magnitude limit of $r\sim24.5$ is 8.1 galaxies per arcmin$^2$ (weighted: 5.5\,arcmin$^{-2}$) distributed over 14 patches on the sky. Photometric redshifts based on 4-band $griz$ data are available for 513 pointings covering an unmasked area of 383.5\,deg$^2$. We present weak lensing mass reconstructions of some example clusters as well as the full survey representing the largest areas that have been mapped in this way. All our data products are publicly available through CADC at \url{http://www.cadc-ccda.hia-iha.nrc-cnrc.gc.ca/en/community/rcslens/query.html} in a format very similar to the CFHTLenS data release. 
\end{abstract}

\section{Introduction}
\label{sec:introduction}
Observational cosmology has succeeded in establishing a widely accepted standard model based on general relativity and inflation that describes all observations on large scales with surprising accuracy. An integral part of this model is the existence of a so-called dark sector that contains most of the matter in the Universe and is responsible for its accelerating expansion, but is neither observed in the laboratory nor described by the other pillar of modern theoretical physics, the standard model of particle physics. There is hope that increasingly detailed observations will help to better understand this dark sector and yield some guidance for theoreticians working on physics beyond the standard models.

One promising way to study this dark sector relies on the fact that all cosmic mass perturbations, whether they are visible or dark, deflect light rays. This gravitational lensing effect introduces characteristic patterns on the sky that can be extracted from astronomical images. For the vast majority of the sky, these patterns are extremely weak and can only be measured by statistically averaging over large areas on the sky. However, it is also these large areas that - if imaged to sufficient depth - correspond to large cosmological volumes and carry a lot of information about how the Universe has evolved over cosmic time. Hence this field of weak gravitational lensing has developed into one of the major tools in observational cosmology.

Weak gravitational lensing effects can be best measured from extremely sharp images with the highest achievable resolution. Furthermore, one needs to know the distance to the celestial objects being imaged to fix the lensing geometry and in order to study the lensing effect as a function of time. These requirements naturally lead to the design of large area imaging surveys that are observed under the best seeing conditions and obtain data in multiple bands to allow for the estimation of photometric redshifts (photo-$z$). Examples of such ongoing weak-lensing oriented surveys are the Kilo Degree Survey \citep{kuijken/etal:2015,2015A&A...582A..62D}, the Dark Energy Survey \citep[see for example][]{2015arXiv150705603J}, and Hyper Suprime-Cam \citep{2015ApJ...807...22M}.

Here we present the Red-sequence Cluster Lensing Survey (RCSLenS\footnote{\url{http://www.rcslens.org}}), the largest multi-band imaging survey with sub-arcsecond seeing to date, that is ideally suited to measure cross-correlations between cosmological weak gravitational lensing signals and other probes. The methods applied here are derived from the Canada France Hawaii Telescope Lensing Survey \citep[CFHTLenS;][]{2012MNRAS.427..146H,2013MNRAS.433.2545E}, which was observed with the same telescope and camera as RCSLenS under the Canada France Hawaii Telescope Legacy Survey (CFHTLS) program. 

In Sect.~\ref{sec:data} we describe the data set and the data reduction. Sect.~\ref{sec:shape_measurements} deals with the shape measurement technique. In Sect.~\ref{sec:photo} we present the photometry and photometric redshifts. Sect.~\ref{sec:systematics} is dedicated to tests checking for systematic errors.  We present dark matter maps in Sect.~\ref{sec:DM_maps} and an investigation of residual B-modes in Sect.~\ref{sec:Bmode}.  In Sect.~\ref{sec:summary} we summarise our results and give an outlook on the scientific exploitation of this data set. The data release is described in appendix~\ref{sec:release}.

\section{Data set and reduction}
\label{sec:data}

\subsection{The RCS2 data}
\label{sec:RCS2_data}
\begin{figure*}
\centering
\includegraphics[width=11cm,angle=270]{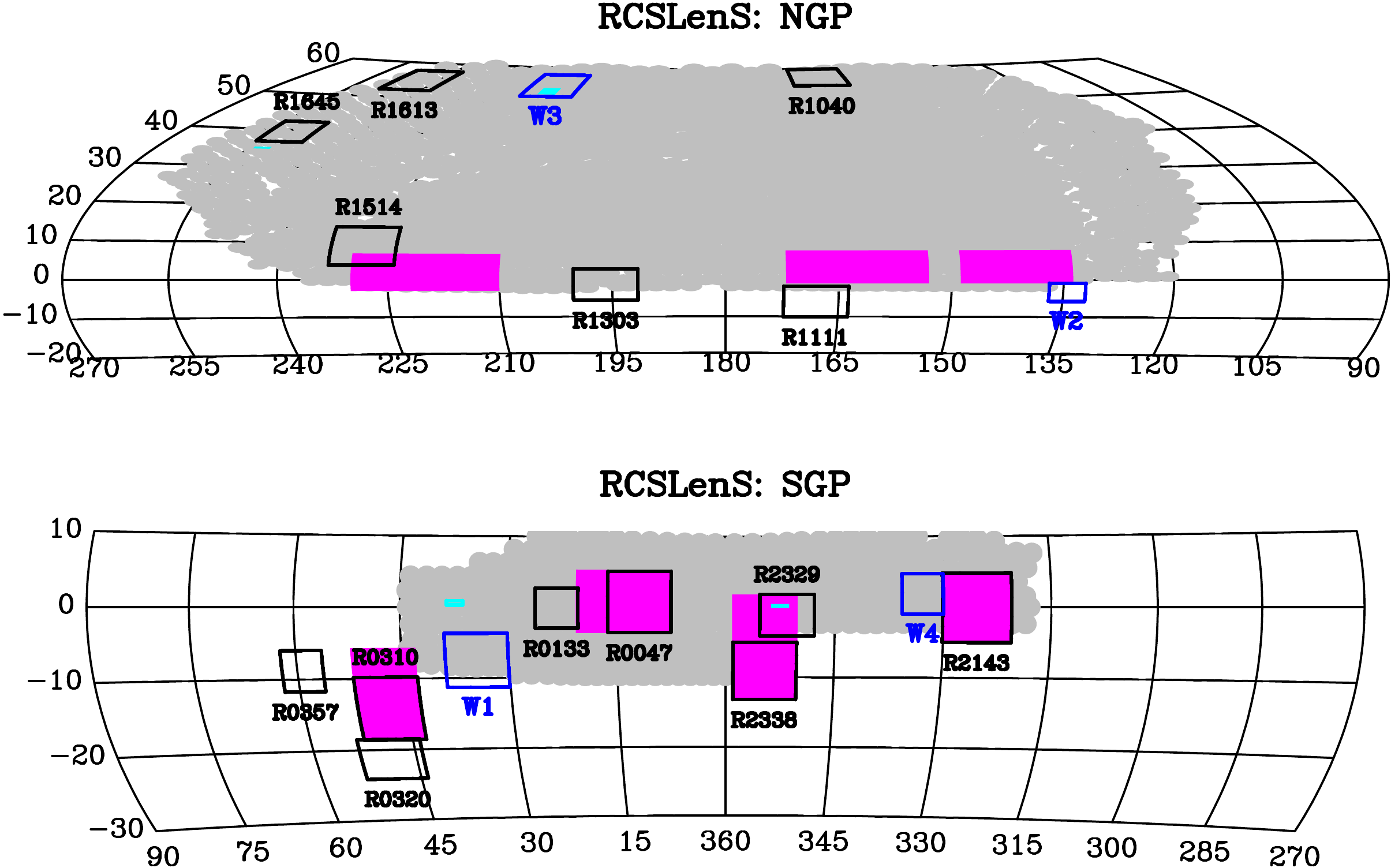}
\caption{Footprint of the RCS2 (black squares), CFHTLS (blue squares), SDSS (grey circles), WiggleZ (pink), and DEEP2 (cyan) in the North Galactic Cap (top) and the South Galactic Cap (bottom).}
\label{fig:footprint}
\end{figure*}

The Red-sequence Cluster Survey 2 \citep[RCS2;][]{2011AJ....141...94G} is a multi-band imaging survey in the $griz$-bands\footnote{For a unique identification of the RCSLenS filter names with the official CFHT filter identifiers see Table~\ref{tab:filter_names}.} over an area of $\sim$785~deg$^2$ to a depth of $\sim 24.3$ mag in the $r$-band (for a point source at $7\sigma$) carried out with the MegaCam imaging camera \citep{2003SPIE.4841...72B} mounted on the Canada France Hawaii Telescope (CFHT). The area is divided into 14 patches, the largest being $10\times10$~deg$^2$ and the smallest $6\times6$~deg$^2$. In Fig.~\ref{fig:footprint} the footprint of the survey is shown. Each square represents a mosaic, which consists of multiple pointings of the $\sim1\,{\rm deg}^2$ camera field-of-view. The different bands share the same pointing strategy, but not all pointings were completed in all bands. Each RCS2 pointing is observed with one single exposure in each band with an exposure time of 4, 8, 8, and 6 minutes in the $g$, $r$, $i$, and $z$-bands, respectively.

RCS2 was mainly designed to optically select a very large sample of galaxy clusters over a wide redshift range and has already been successful in that programme \citep[see, for example,][]{2015A&A...579A..26V,2015arXiv150603817V}. The data have already been reduced and analysed by the RCS2 team, as detailed in \citet{2011AJ....141...94G}. A weak lensing analysis concentrating on galaxy-galaxy-lensing is presented in \cite{2011A&A...534A..14V}. The purpose of the RCSLenS project is to re-analyse the data with a dedicated weak lensing pipeline that is derived from the one used for CFHTLenS \citep{2012MNRAS.427..146H,2012MNRAS.421.2355H,2013MNRAS.433.2545E,2013MNRAS.429.2858M}.

All bands were taken under superb seeing conditions \citep[see Fig.~4 of][]{2011AJ....141...94G}. Essentially all images have a seeing FWHM$<1\farcs0$. The $r$-band data represent the best compromise between seeing and number density of objects because of the longest exposure time (8min), so they are used to estimate the shapes of faint galaxies for weak lensing applications.

\subsection{Overlap with spectroscopic surveys}
\label{sec:overlap_specz}
RCS2 overlaps with several spectroscopic surveys, thereby allowing for additional data calibration and cosmological analyses based on combined probes. These surveys are:

\subsubsection{SDSS}
The Sloan Digital Sky Survey \citep[SDSS;][]{2000AJ....120.1579Y} is a combined photometric and spectroscopic survey covering an area of 15\,000\,deg$^2$, mostly in the Northern sky. A total of $\sim400$\,deg$^2$ overlaps with RCS2. We are mainly interested in the SDSS photometry to calibrate our own RCSLenS photometry (see Sect.~\ref{sec:photo}) as well as in the spectroscopic samples from BOSS (Baryon Oscillation Spectroscopic Survey), a spectroscopic follow-up project in SDSS-III \citep{2011AJ....142...72E}. In particular we are using the sample of highly-biased Luminous Red Galaxies (LRGs; galaxy bias of $\sim1.8$) as lenses in several RCSLenS science papers. A total number of $\sim50\,000$ BOSS LRGs lies in an unmasked area of 184\,deg$^2$ of overlap between RCSLenS and SDSS.

\subsubsection{WiggleZ}
The WiggleZ Dark Energy Survey \citep{2010MNRAS.401.1429D} is a redshift survey of emission line galaxies carried out with the Anglo-Australian-Telescope (AAT). It covers an area of 1\,000\,deg$^2$ in the Southern sky and was actually partly pre-selected with RCS2 data. Out of the $\sim200\,000$ emission line galaxies in the redshift range $0<z<1$ approximately $80\,000$ lie in the overlap area of RCSLenS and WiggleZ (181\,deg$^2$ of unmasked area). This gives us a second lens sample with a smaller galaxy bias \citep[$\sim1$; see e.g.][]{2016MNRAS.456.2806B}.

\subsubsection{DEEP2}
\label{sec:deep2}
RCS2 also overlaps with the DEEP2 galaxy redshift survey \citep{2013ApJS..208....5N} in the 23h field. While covering a much smaller area than the other two spectroscopic surveys this data set is particularly useful to test and characterise the performance of our photo-$z$. DEEP2 is fairly complete down to the magnitude limit that we are interested in for weak lensing studies \citep[see Figs.~31\&~31 in][]{2013ApJS..208....5N} so that it is ideally suited for photo-$z$ calibration. Besides this technical aspect it is also used to study intrinsic alignments, one of the main astrophysical systematic error source for cosmic shear tomography. The precise environmental information from the spectroscopy can be used in concert with our accurate shape measurements from RCSLenS to constrain the dependence of intrinsic galaxy shape on local density, which is so far largely unknown. In total RCSLenS overlaps with 5639 sources from DEEP2 over an area of $\sim1$\,deg$^2$

\subsection{Additional imaging data}
\label{sec:additional_data}
\subsubsection{CFHTLenS}
Due to the similarity of the data sets and the data handling some RCSLenS science projects include data from CFHTLenS. The main difference between the two data sets is that CFHTLenS features an additional $u$-band and that the coadded data are deeper by $\sim1\,{\rm mag}$. The latter also relates to the fact that CFHTLenS images are stacks\footnote{Note that shapes are still measured on individual exposures in CFHTLenS.} created from multiple exposures whereas RCSLenS consists of single exposures in each pointing and band. This has important consequences for the systematic errors in the shape measurements as detailed in Sect.~\ref{sec:shear_syst}.

Shapes of galaxies are measured in the $i$-band of CFHTLenS. Since the $i$-band only covers 70\% of the RCS2 area, the $r$-band is used for shape measurements in RCSLenS. For details on the CFHTLenS data processing, photometry/photo-$z$, and shape measurements we refer the reader to \cite{2013MNRAS.433.2545E}, \cite{2012MNRAS.421.2355H}, and \cite{2013MNRAS.429.2858M}, respectively.

\subsubsection{DEEP2 fields}
Some of the fields from the DEEP2 redshift survey (see Sect.~\ref{sec:deep2}) are not included in the RCS2 footprint as can be seen in Fig.~\ref{fig:footprint}. However, there are very similar MegaCam data available for those fields which are archived. In order to increase the area for our intrinsic alignment studies, we decided to include those fields in our data processing. Some properties of these fields are different (e.g. depth in some bands), and those details will be presented in the science papers using those data. Note that these additional fields are not part of the data release but can be made available upon request.

\subsection{N-body simulations}
\label{sec:simulations}
Most of the cosmological projects with the RCSLenS data require a dedicated suite of N-body simulations to correctly interpret the measurements. These simulations are similar to the CFHTLenS ``clone'' simulations described in \cite{2012MNRAS.427..146H} and \cite{2012MNRAS.426.1262H}. In particular, the simulations are used to estimate covariance matrices for the different data vectors as well as to create mock catalogues for the galaxy samples used in the analyses. This new N-body suite and the ray-tracing simulation products are referred to as the SLICS (Scinet LIght Cone Simulations) and are described in \cite{2015MNRAS.450.2857H}. Here we summarise the main properties of these simulations.

Shear, convergence and density maps are created from light cones that are extracted from the dark matter distribution along the line of sight. The weak lensing quantities $\kappa$ and $\gamma$ are estimated using the Born approximation in the flat sky limit. Oweing to the larger data volume and the larger angular scales probed by RCSLenS (compared to CFHTLenS) we increased the box size of the SLICS as well as the number of independent realisations. Instead of individual patches of $\sim$ 12deg$^2$ in the CFHTLenS clone these new simulations cover 60deg$^2$ per patch, and instead of 184 patches we now achieve a total number of 1000 patches. The redshift resolution was decreased from 26 to 18 slices in the range $0 < z < 3$ to limit the amount of storage space. However, we increased the spatial resolution by a factor of $\sim$ 2 to be able to estimate shear and $\kappa$ maps with a resolution of $\sim$ 0.'1.

Shear maps are finally used to create mock galaxy catalogues following the method described in \cite{2012MNRAS.427..146H} and \cite{2016MNRAS.456.2806B}.

\subsection{Data reduction with THELI}
\label{sec:data_reduction}
The data reduction is carried out with the THELI pipeline \citep{2005AN....326..432E,2013ApJS..209...21S} and closely resembles the handling of the CFHTLenS data set \citep{2009A&A...493.1197E,2013MNRAS.433.2545E}. We perform the following steps:
\begin{enumerate}
\item The pre-reduced individual exposures are downloaded from the Canadian Astronomy Data Centre (CADC). Those were processed with the ELIXIR pipeline \citep{2004PASP..116..449M} and are already corrected for overscan and bias, have been flat-fielded, fringes have been removed from the redder bands, and a correction for the scattered light\footnote{A revised version of the scattered-light correction is now available within ELIXIR, but the RCS2 data have not been processed with this new version yet.} has been applied. Furthermore, crucial data like photometric zeropoints\footnote{Note that the ELIXIR photometric zeropoints are based on stellar photometry with {\sc SExtractor} MAG\_AUTO apertures. These instrumental magnitudes are then matched to reference catalogues. Extracting stellar photometry with different apertures from our RCSLenS stacks will lead to offsets with respect to the same reference catalogue.}, extinction coefficients, colour terms, and gain values are provided for each exposure.
\item All 36 chips of each exposure are checked to identify chips with no data (due to e.g. read-out problems) or an excessive amount of saturated pixels (e.g. due to a bright star). Those chips are flagged and do not enter into subsequent reductions.
\item Fields with bad or insufficient data are excluded from further processing. Out of the 785 fields only 765 were observed in the $r$-band. Since this is the band that we use for our shape measurements (see Sect.~\ref{sec:shape_measurements}) we exclude the 20 fields without $r$-band data. Out of the remaining 765 fields we reject another four fields for different reasons (shallow data, background gradients, etc.).
This leaves us with 761 fields to process.
\item The sky background is subtracted.
\item Weight images are created that encapsulate information about unusable areas (cosmic rays, hot/cold pixels, satellite tracks, reflections, etc.). Pixels affected by such defects are assigned a weight of zero. All other pixels are assigned a weight corresponding to their estimated inverse sky background variance. We note that we need to identify all image defects on the single frame images as we typically only have one observation per pointing and per filter. The algorithms to identify various defects and the peculiarities of our weight images are discussed in detail in \cite{2009A&A...493.1197E,2013MNRAS.433.2545E}.
\item From the single frames we extract catalogues for our later astrometric calibration with {\sc SExtractor} \citep{1996A&AS..117..393B}. Sources are detected with the criterion of consisting of at least 5 consecutive pixels with a value that is 5-$\sigma$ above the local sky background variation. We also reject objects having non-zero {\sc SExtractor} flags or pixels that are flagged within their area.
\item The astrometric and \emph{relative} photometric calibration is performed with the {\sc SCAMP} software \citep{ber06}. We calibrate simultaneously all individual images of one RCSLenS patch. In this way we can make optimal use of overlapping sources between individual pointings which helps to constrain astrometric distortions and relative photometric offsets between pointings. However, as RCSLenS only obtained one single image per pointing and per filter with a small overlap ($\sim0\farcm5-1'$) the gain compared to a pointing-wise calibration is only minor for this survey. This becomes especially problematic for the photometric accuracy which was revised significantly with external data (see below). The details of this step and the following image co-addition with the {\sc SWARP} software \citep{2003SWarp} are identical to the CFHTLenS processing which is described in \citet{2013MNRAS.433.2545E}. Due to RCS2 being a single-exposure survey, ``coaddition'' with SWarp means only re-sampling to a new pixel grid.\footnote{Some pointings and filters were actually observed with more than one exposure. The reasons for such a repeat visit can be numerous. Whenever possible we coadd these multiple exposures so that about 10\% of the images we produce are based on more than one exposure.}
\item The absolute photometric calibration is initially based on the photometric zeropoints provided by CADC, properly weighted to account for the results of the relative photometric calibration of the previous step. We later re-calibrate the photometry using information from SDSS and stellar locus regression (see Sect.~\ref{sec:abs_photo_cal}).
\end{enumerate}

\subsection{Initial star selection}
\label{sec:SG_sep}
In order to model the point spread function (PSF), we require an initial list of candidate stars for input into \emph{lens}fit, the shape measurement code used for CFHTLenS (further described in Sect.~\ref{sec:lensfit}).  We create the source samples, per pointing, in
the following way: 
\begin{enumerate}
\item We run SExtractor on individual exposure chips with a high detection threshold (DETECTION\_MINAREA / DETECTION\_THRESH is set to 5 / 5) and we only consider clean, unflagged detections henceforth. 
\item Candidate stellar sources are identified on the stellar locus in the size-magnitude plane. 
\item We perform a standard PSF analysis with the Kaiser, Squires and Broadhurst algorithm \citep[KSB;][]{1995ApJ...449..460K}. This involves estimating weighted second-order brightness moments for all candidate
stars and to perform, on the chip level, a two-dimensional second order polynomial fit to the PSF anisotropy. The fit is done iteratively with outliers removed to obtain a clean sample of bright, unsaturated stars suitable for PSF analysis.\footnote{Note that the fit is only performed to remove outliers and come up with a pure star sample. The fit is not use in any further step of the analysis.}
\item All objects surviving the previous step are transferred to the candidate list of \emph{lens}fit stars. 
\item In case a pointing consists of more than one individual exposure we report all sources that are in at least one of the lists from step (iv). In that case we catch candidate stars that are not reported in an exposure due to defects (e.g. bad pixels). We note that it is not our intention to provide a \emph{complete} list of stellar sources per pointing but we want to collect a pure sample of bright, unsaturated and randomly distributed star candidates. Our prescription leads to at least 60 candidate stars (significantly more in most cases) per chip in the RCSLenS area.
\end{enumerate}

\subsection{Masking}
\label{sec:masking}
Image defects that are not registered in the weight maps mentioned above are detected with automatic masking algorithms that have been applied in CFHTLenS.  These algorithms are mainly used to mask bright stars and their reflection halos. We refer to external catalogues of stars, GSC-1 \citep[complete from $r\simeq10$ to $\simeq16$,][]{lasker/etal:1996} and UCAC4 \citep[complete from $r\simeq10$ to $\simeq16$,][]{zacharias/etal:2012} and mask the stars with a polygon template that fits the shape of the diffraction spikes in the MegaCam images (down to $r=17.5$), scaled to the magnitude of the star. For the brightest stars ($r<10.35$ or $r<11.2$, see Table~\ref{tab:mask_bits}) we also mask a circular region of $7\farcm5$ diameter that is affected by a reflection halo. Additionally we detect regions in the images that show a severe under-density of objects \citep{2007A&A...470..821D}. This typically happens at the chip edges, but the fraction of such areas is small.

Due to RCSLenS being a sparse single-exposure survey a lot of image defects that typically do not occur at the same sky position in dithered exposures (cosmic ray hits, hot/cold pixels, etc.) cannot be rejected in RCSLenS. Thus, we need to include information about these defects in the masks. This information is based on the flag images produced by THELI.

All of the steps mentioned above are run automatically. These automatic masks are then checked visually and modified accordingly. Manual masks for missed asteroid/satellite streaks, as well as missed bright stars and their halos (due to incompleteness of the stellar catalogue or variable stars), are added.  Finally, the manual masking is inspected by a single person for uniformity.

Adjacent pointings of RCSLenS overlap by a small amount to allow for better astrometric and photometric cross-calibration across the survey area. In order to have a unique assignment of objects to pointings, we introduce cuts in right ascension and declination to separate pointings\footnote{Note that these cuts at constant RA and Dec correspond to curved lines in pixel space.}. These cuts are based on the object catalogues and typically lie in the middle of the overlap region of two pointings. Pixels in the images and objects in the catalogues that lie outside these cuts for a given pointing are also masked.

The masks are provided as FITS files which use bit coding to preserve information about the reason behind masking a particular pixel. The bit coding is summarised in Table~\ref{tab:mask_bits}. After masking pixels with a mask value $>1$ but allowing for bits 32, 256, 1024, and 2048 the total area of the RCSLenS data set is 571.8\,deg$^2$. This masking scheme is useful in the case that only the shape information from the $r$-band and no photo-$z$ data are used, since bits 32, 256, 1024, and 2048 correspond to $giyz$ data.

\subsection{Catalogue creation}
\label{sec:catalogues}
Sources are detected on the $r$-band data using SExtractor. We require sources to have five consecutive pixels that are at least 1.5$\sigma$ above the local background. This source catalogue is then used as the input catalogue for the shape measurements and the multi-colour photometry which are described in more detail in Sect.~\ref{sec:shape_measurements} and Sect.~\ref{sec:photo}, respectively.

We apply the cuts in sky coordinates (Sect.~\ref{sec:masking}) to enable the catalogues from adjacent fields to fit seamlessly together. Using the masks (Sect.~\ref{sec:masking}) together with these sky coordinate limits, we construct random catalogues of object positions that can be used to estimate angular correlation functions from the data. These random catalogues contain between $5\times10^5$ and $10^6$ objects per pointing to minimise shot noise in correlation function measurements.

\subsection{Data sanity checks}
\label{sec:sanity}
After processing each field, we run detailed sanity checks to control the quality of the data set. A comprehensive one-page summary  of all tests is compiled for each field (see Fig.~\ref{fig:CDE0047B0_sanity} for an example) and checked visually for outliers. The individual tests comprise:
\begin{itemize}
\item Sky distribution of the objects in the $r$-band detected catalogue.
\item Sky distribution of galaxies with non-zero \emph{lens}fit weight (see Sect.~\ref{sec:lensfit}) and PSF stars.
\item Whisker plot of the stellar ellipticity as a function of position in the pointing.
\item Map of the galactic extinction \citep[taken from][]{1998ApJ...500..525S}.
\item Colour-colour diagrams of observed stars with predicted stellar loci calculated from the spectral energy distribution library of \citet{1998PASP..110..863P}.
\item Redshift distributions for bright and faint objects. Shown are the histograms of the photo-$z$ point estimates (Z\_B) as well as the stacked $P(z)$ curves.
\item Magnitude number counts in all available bands.
\item Angular auto-correlation function of galaxies with $22<r<23$ and \emph{lens}fit weight$>0$.
\end{itemize}
Several glitches in the data handling can be discovered by inspecting these check plots by eye (e.g., zeropoint magnitude errors in the colour-colour diagram, masking errors in the auto-correlation function, etc.) and those are repaired accordingly. It should be noted that this scheme ensures the sanity of the data on the pointing level but cannot check for homogeneity on large scales. For the survey design of RCSLenS, with individual pointings being observed independently of each other, it makes sense to regard each pointing as a photometrically independent unit and treat it as such. However, for measurements on the largest scales we refer to other methods to check for systematic errors, some of which are covered in Sect.~\ref{sec:systematics}, while others are detailed in the respective scientific papers \citep[e.g.][]{2015arXiv151203626C}.

\section{Shape measurements}
\label{sec:shape_measurements}

\subsection{The \emph{lens}fit code}
\label{sec:lensfit}
{\it Lens}fit \citep{2007MNRAS.382..315M,2013MNRAS.429.2858M,2008MNRAS.390..149K} is a forward model-fitting shape measurement code specifically designed for cosmological weak gravitational lensing applications that require accurate correction of the effects of the PSF.

The PSF is measured from the star catalogue described in Sect.~\ref{sec:SG_sep}. The pixelised images of these stars are used directly after centroiding, i.e. the individual pixels of the PSF images across the field of view are fit by a second order polynomial. The resulting pixelised PSF model at a galaxy's position is convolved with analytical models describing the brightness profiles of galaxies. The latter are chosen to be composites of a de Vaucouleurs bulge and an exponential disk with fixed relative scale-lengths but variable bulge-to-disk ratios, variable absolute scale-lengths, and variable ellipticities.

This model library is then fit to the data leaving also the centroid of the galaxy as a free parameter \citep[constrained by a prior as explained in][]{2013MNRAS.429.2858M}. This then yields a joint likelihood function for these four parameters ($e_1$, $e_2$, bulge-to-disk ratio, scale-length). Priors for the ellipticity, scale-length, and bulge-to-disk ratio are taken from external data sets \citep[for details see][]{2013MNRAS.429.2858M}. Marginalising over all other parameters yields mean likelihood estimates of the galaxy ellipticities $e_1$ and $e_2$ and an associated inverse variance weight. These quantities represent the main observables in all weak lensing shear applications.

\subsection{Differences to CFHTLenS}
\label{sec:differences_to_CFHTLenS}
The main differences between RCSLenS and CFHTLenS in terms of shape measurements - apart from obvious differences like the seeing distribution - are that RCSLenS is observed with single exposures whereas CFHTLenS features several dithered exposures per pointing, and that the $r$-band is used for shape measurements in RCSLenS as opposed to the $i$-band in CFHTLenS.

\subsubsection{Effect of single exposures}
Measuring the PSF and galaxy shapes on a single exposure instead of multiple exposures that are dithered with respect to each other means that one loses resolution. Dithering inevitably causes relative shifts by fractions of a pixel which lead to a better sampling of the image. We do not have this advantage with RCSLenS, so we expect some stronger systematic effects, e.g. larger $c$ calibrations (see below), due to this. This is analysed in detail in Sect.~\ref{sec:noise_bias}.

\subsubsection{Transition from $i$- to $r$-band}
Substituting the $i$-band with the $r$-band requires one to adjust the size prior mentioned above. While we assume that the distributions of the ellipticity and bulge-to-disk ratio are the same for both bands we adapt the prior in the scale-length to account for the fact that galaxies at a given $r$-band magnitude show different sizes than galaxies at the same $i$-band magnitude. The procedure to construct this new $r$-band prior is described in \cite{2013MNRAS.429.2858M} for the CFHTLenS $i$-band prior. Further details can be found in \cite{kuijken/etal:2015} who use the same new prior for an analysis of the Kilo Degree Survey. We investigate the importance of the size prior by running \emph{lens}fit twice, once with the new $r$-band prior and once with the old, CFHTLenS $i$-band prior. Ellipticities change by $\la1\%$ on average.

\subsubsection{Blinding}
In the era of precision cosmology systematic effects become increasingly important. Corrections for a number of systematic effects (in the shape measurements but also in the photometry and photo-$z$) are discussed in this paper. In order to avoid manipulation of some part of the data analysis pipeline based on premature inspection of the measured signals, we implement a blinding scheme. This scheme helps to largely avoid confirmation bias and works in the following way.

The raw shear measurements from the \emph{lens}fit code are included in the catalogues with three additional shear columns that slightly perturb the raw shear column. This is done by an external blind setter through random number selection. The amplitude of the perturbation is chosen such that cosmological parameters from a non-tomographic cosmic shear measurement would vary around the best fit Planck values within a 10$\sigma$ confidence interval (with the confidence interval taken from Planck). It is not revealed to the team which of the four columns (labelled A, B, C, and D) is the correct one. All science analyses are carried out on all four shear columns, also called ``blindings''. Plots and numerical results in science papers that depend on the shear catalogue are shown four times for each of the four different blindings but without the labels A, B, C, D. Only when a science paper is ready for submission, i.e. after having been fully reviewed by all co-authors, will the lead author unblind themself and the perturbed plots and numbers be removed. The other co-authors are not told which of the four blindings is the correct one.

\subsection{Calibration shear measurement biases}
\label{sec:noise_bias}
We employ a two-stage scheme to calibrate our shape catalogue, modelling the calibration corrections to shear measurement in terms of a multiplicative term $m$ and an additive term $c$ such that

\be
g_i^{\rm obs} = (1+m) g_i^{\rm true} + c_i \,,
\ee
with $g_i$ being the $i$th component of the reduced shear estimated from a weighted average of the ellipticities measured by \emph{lens}fit.

\subsubsection{Multiplicative bias}
First, image simulations are used to identify possible multiplicative and additive biases in our shape measurement technique \citep[see][]{2006MNRAS.368.1323H}. Here we rely on the simulations that were created for CFHTLenS, detailed in \cite{2013MNRAS.429.2858M}. While these simulations do not exactly match the RCSLenS catalogues in terms of the distributions of magnitude, size and ellipticity, the dependence of the multiplicative bias on size and signal-to-noise ratio (SNR) is not expected to change significantly for RCSLenS. It is important that the image simulations cover the whole parameter space (in SNR, size, and possibly PSF quantities) spanned by the observations. Once this is ensured the actual multiplicative bias can be estimated from either re-sampling the simulations\footnote{See the mock catalogue re-sampling analysis in \cite{kuijken/etal:2015}, which is performed for a different survey, but should give an idea of the size of the effect.} or from using a parametric fit to the observables. Since for RCSLenS we use the same shape measurement technique and the same camera as for CFHTLenS we use their SNR and size-dependent calibration correction given by
\be
m(\nu_{\rm SN}, r_{\rm d}) = \frac{\beta}{\log_{10} \nu_{\rm SN}} \exp(-\alpha \, r_{\rm d} \, \nu_{\rm SN})
\label{eq:mcalmod}
\ee
where $\nu_{\rm SN}$ is the SNR, $r_d$ is the size measured in arcseconds, $\alpha=0.306$ and $\beta=-0.37$.    This corresponds to an average calibration correction to the RCSLenS ellipticities of 5\%. Note that this simulation-based estimate of the multiplicative bias mostly depends on the shape measurement technique (which did not change compared to CFHTLenS) whereas the empirical estimate of the additive bias described in the next section is much more sensitive to the actual properties of the data (depth, number of exposures, seeing, PSF ellipticity).

\subsubsection{Additive bias}
\begin{figure}
\includegraphics[width=\hsize]{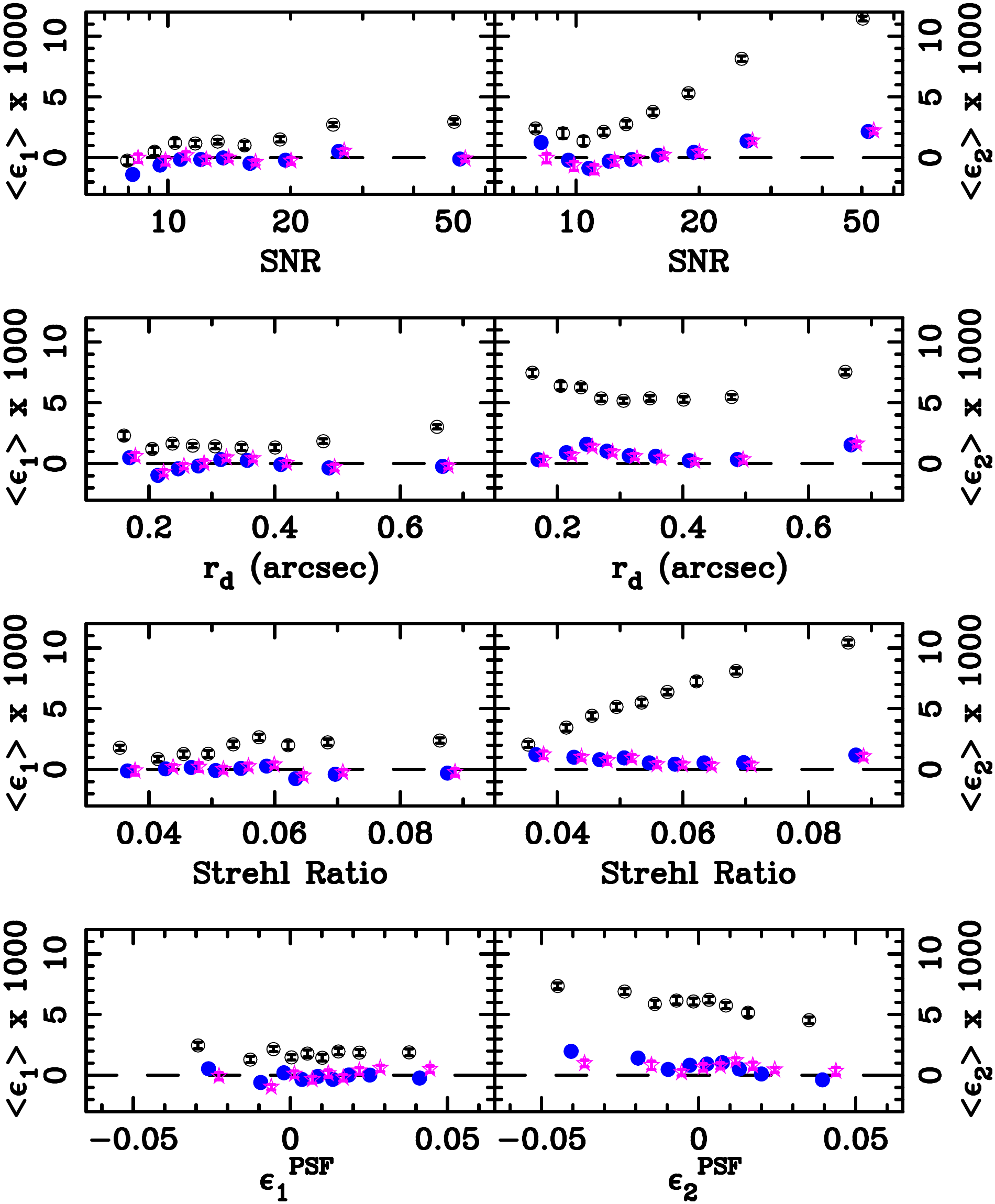}
\caption{Empirical calibration of the additive shear measurement bias for the $\epsilon_1$ (left) and $\epsilon_2$ (right) ellipticity components for one of the blindings. Shown are the residual mean $\epsilon_{1/2}$ in bins of SNR, size, PSF Strehl-ratio, and PSF ellipticity before correction (black), after a detector bias correction (blue) and after a detector and noise bias correction (pink).}
\label{fig:ecorr}
\end{figure}

Whilst the image simulations do not reveal any additive bias significantly different from zero, we find in both the RCSLenS and the CFHTLenS data a small but significant additive term at the level of a few times $10^{-3}$. Fig.~\ref{fig:ecorr} shows the weighted mean ellipticity for each component, $\langle \epsilon_1 \rangle$ (left panels, black symbols) and $\langle \epsilon_2 \rangle$ (right panels, black symbols), as a function of galaxy SNR, size, PSF pseudo Strehl ratio\footnote{The pseudo Strehl ratio in \emph{lens}fit is defined as the fraction of light in the PSF model that falls into the central pixel, and is therefore a measure of the level of under-sampling in the observed pixellated PSF, with better seeing data having larger Strehl ratios.}, and PSF ellipticity. 

This shows that the residual ellipticity bias is dependent on SNR and size of the galaxy, as seen in the analysis of CFHTLenS. However, there are significant differences. First of all, for CFHTLenS the residual bias in the $\epsilon_1$ component was consistent with zero, which is not the case for RCSLenS. Secondly, the trends in $\epsilon_2$ are quite different for the RCSLenS data set. While large, high-SNR galaxies did not show any significant bias in CFHTLenS \citep[see Fig.~3 of ][]{2012MNRAS.427..146H}, the situation for RCSLenS is opposite. The bias in the $\epsilon_2$ component in RCSLenS is largest for these galaxies.  Furthermore we see a strong dependence of the bias in $\epsilon_2$ on the Strehl ratio of the PSF, which was not observed in CFHTLenS. Due to the sparsity of RCSLenS in comparison to CFHTLenS (see Sect.~\ref{sec:differences_to_CFHTLenS}) we expect RCSLenS to be more susceptible to under-sampling errors in our modelling of the galaxy and PSF and our hypothesis is that this is the root cause of this additive calibration bias.

The lowest panel of Fig.~\ref{fig:ecorr}  shows that in addition to the trends with SNR, Strehl and galaxy size, we also find a weak dependence on the mean galaxy ellipticity as a function of PSF ellipticity.  This is an expected side-effect of `noise-bias' \citep{2014MNRAS.439.1909V}, but it was not previously detected in the analysis of CFHTLenS.

In order to model the complex behaviour of the residual ellipticity bias we employ a two-stage process. The first stage we call a ``detector bias correction'' since our hypothesis is that this signal originates from a yet unknown effect present in the MegaCam CCDs\footnote{Note that no such effect is seen in the KiDS data \citep{kuijken/etal:2015} which are based on a different camera (OmegaCam@VST) but reduced with a very similar pipeline.}. Here we split the data into $7\times 7 \times 7$ bins in SNR, Strehl and galaxy size, with the bins selected so that there are roughly equal numbers of galaxies in each bin.  An empirical correction is then determined for each bin, given by the weighted average ellipticity in that bin.    We found the structure in this correction too complex to model with functional form in SNR, Strehl and size and so we simply apply a single correction to every galaxy within a bin.  The residual mean ellipticities for the full sample are greatly reduced as shown in Fig.~\ref{fig:ecorr} (blue symbols) where the binning is chosen on purpose to differ from the grid upon which the correction was initially derived.

Once this detector-level bias is corrected for we run a first pass of the star-galaxy cross-correlation test for residual systematics (see. Sect.~\ref{sec:shear_syst}). All fields that pass this test are then used to refine the detector bias correction using the same approach as above. We attribute the remaining bias to noise-bias \citep{2014MNRAS.439.1909V}.   In order to correct for this effect we need to be cautious.  Any error in the determination of a correction that is dependent on a spatially-correlated quantity such as a PSF could lead to a strong systematic error in the final shear catalogue.  We split the data into $7\times 7 \times 7$ bins in SNR, Strehl and now PSF ellipticity, with the bins again selected so that there are roughly equal numbers of galaxies in each bin.   After the first detector-level correction has been applied as described above, we find that the residual additive term is strongest at low-SNR, as expected for noise-bias, and can be well fit by
\be
c_i^{\rm noise \, bias} = \frac{{\rm A}_i \epsilon^*}{{\rm S}^{2.3} \log_{10}(\nu_{\rm SN})} \exp^{-{\rm B}_i \nu_{\rm SN}}
\ee
where $S$ is the Strehl ratio, $\nu_{\rm SN}$ is the SNR, $\epsilon_i^*$ is the PSF ellipticity component $i=1,2$.  The free parameters are fit to the data, and we find ${\rm A}_1=-0.057$, ${\rm A}_2=-0.007$, ${\rm B}_1=0.662$, ${\rm B}_2=0.416$.  We apply this small correction to the catalogue, the impact of which is shown in Fig.~\ref{fig:ecorr} by the pink symbols.

The star-galaxy cross-correlation test is then re-run applying both the detector-level and noise-bias correction to see if any fields change their status (pass/fail). The results of this final test are presented in Sect.~\ref{sec:shear_syst}.  This final result for the additive correction in both ellipticity components is added to the catalogues according to the size, SNR, Strehl-ratio and PSF of an object and should be subtracted from the measured ellipticities in all measurements. We keep both terms, the detector-level correction and the noise-bias correction, separate in the catalogues so that further tests can distinguish between the two. In Table~\ref{tab:mean_e} we report the mean ellipticities in the different stages of the correction process averaged over the whole survey (pass fields) taking the \emph{lens}fit weight into account.

\begin{table}
\caption{Mean ellipticities before correction (2nd column), after detector-level correction (3rd column), and after detector-level plus noise-bias correction (4th column) for both ellipticity components.}
\label{tab:mean_e}
\begin{tabular}{lrrr}
\hline
\hline
                   & no correction & detector correction & detector \& noise-\\
                   &                       &                                & bias correction\\
\hline
$\left<\epsilon_1\right>$ &  $0.0018\pm 0.0001$ & $0.0001\pm 0.0001$ & $0.0000\pm 0.0001$\\
$\left<\epsilon_2\right>$ &  $0.0060\pm 0.0001$ & $0.0008\pm 0.0001$ & $0.0007\pm 0.0001$\\
\end{tabular}
\end{table}

Note that the correction presented here is only valid for the full source sample. If cuts are applied (e.g. in SNR or sky position) the correction has to be determined again in principle or additive biases can become larger than what is presented in Table~\ref{tab:mean_e}.

\subsection{Number density}
\label{sec:number_density}
The weighted number density of objects with shape measurements in our RCSLenS catalogues using the definition by \cite{2012MNRAS.427..146H} is 5.5 galaxies per square arcminute with an ellipticity dispersion per component of $\sigma_{\epsilon}=0.251$. This is calculated over an area of 571.1\,deg$^2$ where $r$-band data is available. Over the 383.5\,deg$^2$ of area where photo-$z$ (see Sect.~\ref{sec:photo-z}) are available, i.e. the area that has full $griz$-band coverage, the number density is very similar with 4.9 galaxies per square arcminute. If one further restricts the photo-$z$ range to $z_{\rm phot}>0.4$ (see Sect.~\ref{sec:photo-z}) the number density drops to 2.9 galaxies per square arcminute. The objects with the most reliable photo-$z$ in the range $0.4<z_{\rm phot}<1.1$ correspond to a number density of 2.2 galaxies per square arcminute, i.e. $\sim 45$\% of the full photo-$z$ sample.

Another definition of the weak lensing source density was proposed by \cite{2013MNRAS.434.2121C} which gives slightly lower numbers. For our full sample this yields 4.9 galaxies per square arcminute. An overview of the different number densities also in comparison to CFHTLenS, KiDS, and the DES SV data \citep{2015arXiv150705603J} can be found in Table~\ref{tab:number_density}. For further discussion of the different definitions we refer the reader to \cite{kuijken/etal:2015}.

\begin{table}
\caption{Number densities of weak lensing source galaxies drawn from RCSLenS, CFHTLenS, KiDS, and DES. The third column shows the raw number density, i.e. including all objects that a shape was measured for, the fourth column shows the definition of Heymans et al. (2012, H12), and the fifth column shows the definition by Chang et al. (2013, C13). The two numbers per column for DES correspond to the two different shape measurement algorithms NGMIX and IM3SHAPE, with the former yielding higher number densities than the latter.} 
\label{tab:number_density}
\begin{tabular}{lrrrr}
\hline
\hline
Sample & area  & raw &  H12 & C13\\
       & [deg$^2$] & \multicolumn{3}{c}{[arcmin$^{-2}$]}\\
\hline
RCSLenS, full                    & 571.7 & 8.1 & 5.5 & 4.9 \\
RCSLenS, $griz$                  & 383.5 & 7.2 & 4.9 & 4.3 \\
CFHTLenS                         & 125.7 & 17.8 & 15.1 & 14.0 \\
KiDS DR2                         &  68.5 & 8.8 & 6.0 & 4.5 \\
DES SV data                      & 139.0 & 4.2 / 6.9 & 4.1 / 6.8 & 3.7 / 5.7 \\
\end{tabular}
\end{table}

\section{Photometry and photometric redshifts}
\label{sec:photo}
The methods to extract the multi-colour photometry from the calibrated images, to estimate photometric redshifts, and to calculate absolute magnitudes and stellar masses are very similar to those used in CFHTLenS. Details can be found in \citet[photometry and photo-$z$]{2012MNRAS.421.2355H} and \citet[absolute magnitudes and stellar masses]{2014MNRAS.437.2111V}.

\subsection{Multi-colour photometry}
\label{sec:multi_colour}
Measuring accurate colours for galaxies requires a careful correction for the varying PSF between bands. Here we follow an approach that first convolves the images with a position-dependent kernel that converts the local non-Gaussian PSF with varying PSF size into a Gaussian PSF with constant size (i.e. constant across the image and between images of the same pointing in the different bands). The convolution kernels are estimated from shapelet-based PSF measurements of stars in the field \citep{2008A&A...482.1053K}.

These images in the different bands of one pointing that now have the same Gaussian PSF are then used with SExtractor in dual-image mode. The unconvolved $r$-band image is used for detection whereas the Gaussianised images in all bands are used for measuring the fluxes. This ensures that the same physical apertures are used in all bands. Since the apertures are defined on an unconvolved image but fluxes are measured on convolved images, the fluxes might be underestimated. Thus, this procedure provides accurate colours, but not total magnitudes. We also measure fluxes with SExtractor on the unconvolved $r$-band image to have one band where we can estimate reliable total magnitudes. Under the assumption that there are no colour gradients the total magnitudes in the other bands can be calculated from the $r$-band magnitude and the colours.

\subsection{Absolute photometric calibration}
\label{sec:abs_photo_cal}
The absolute photometric calibration is first based on the nightly zeropoints provided by CFHT through CADC (see Sect.~\ref{sec:data_reduction}). We improve on this initial calibration by using data from the SDSS where available. In regions that do not overlap with SDSS we employ a method similar to the stellar locus regression \citep{2009AJ....138..110H}.

Out of the 761 tiles used in our data processing, 604 overlap with the SDSS. We compare the magnitudes of stars in the RCSLenS catalogues to data from the eighth data release of SDSS \citep[DR8][]{2011ApJS..193...29A}, which represents the full imaging data of SDSS. For RCSLenS we calculate the magnitudes from the total $r$-band magnitudes (SExtractor MAG\_AUTO) and the colours based on isophotal magnitudes. This is necessary because we detect sources and define the isophotal apertures on the unconvolved $r$-band image, but then extract the photometry from images that are convolved with a PSF-homogenising kernel. For SDSS we use the PSF magnitudes provided in the public catalogue.

\begin{figure}
\includegraphics[width=0.48\hsize]{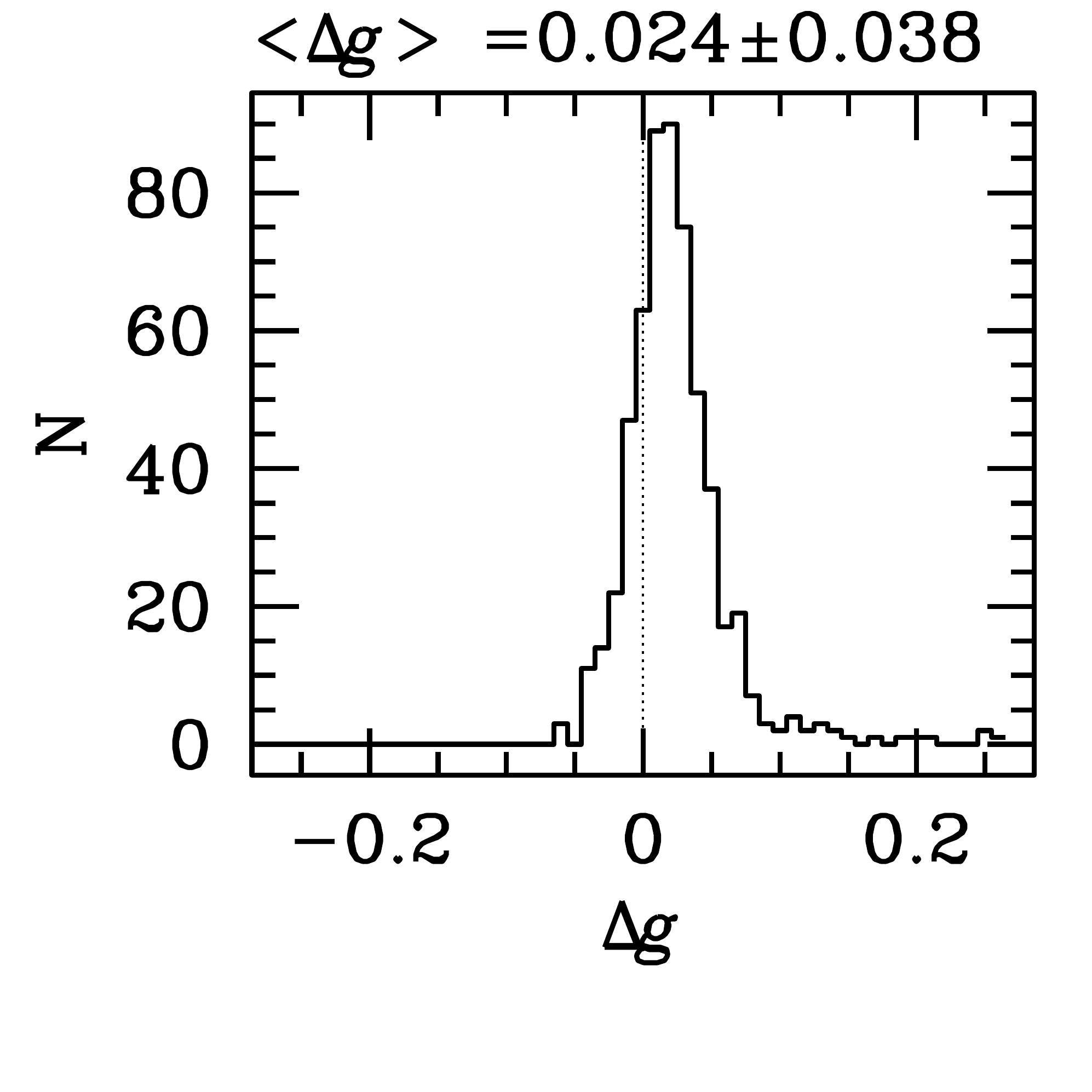}
\includegraphics[width=0.48\hsize]{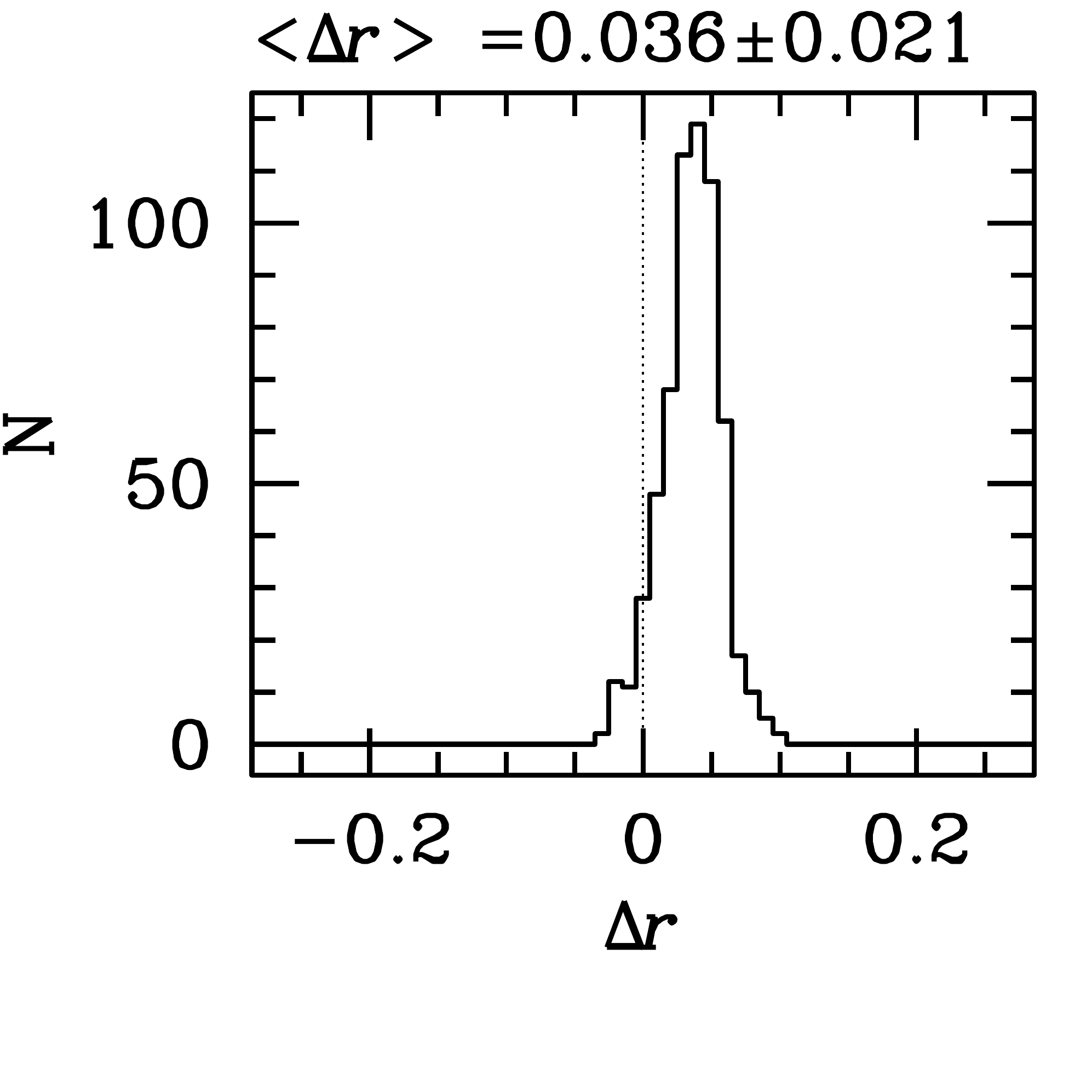}\\
\includegraphics[width=0.48\hsize]{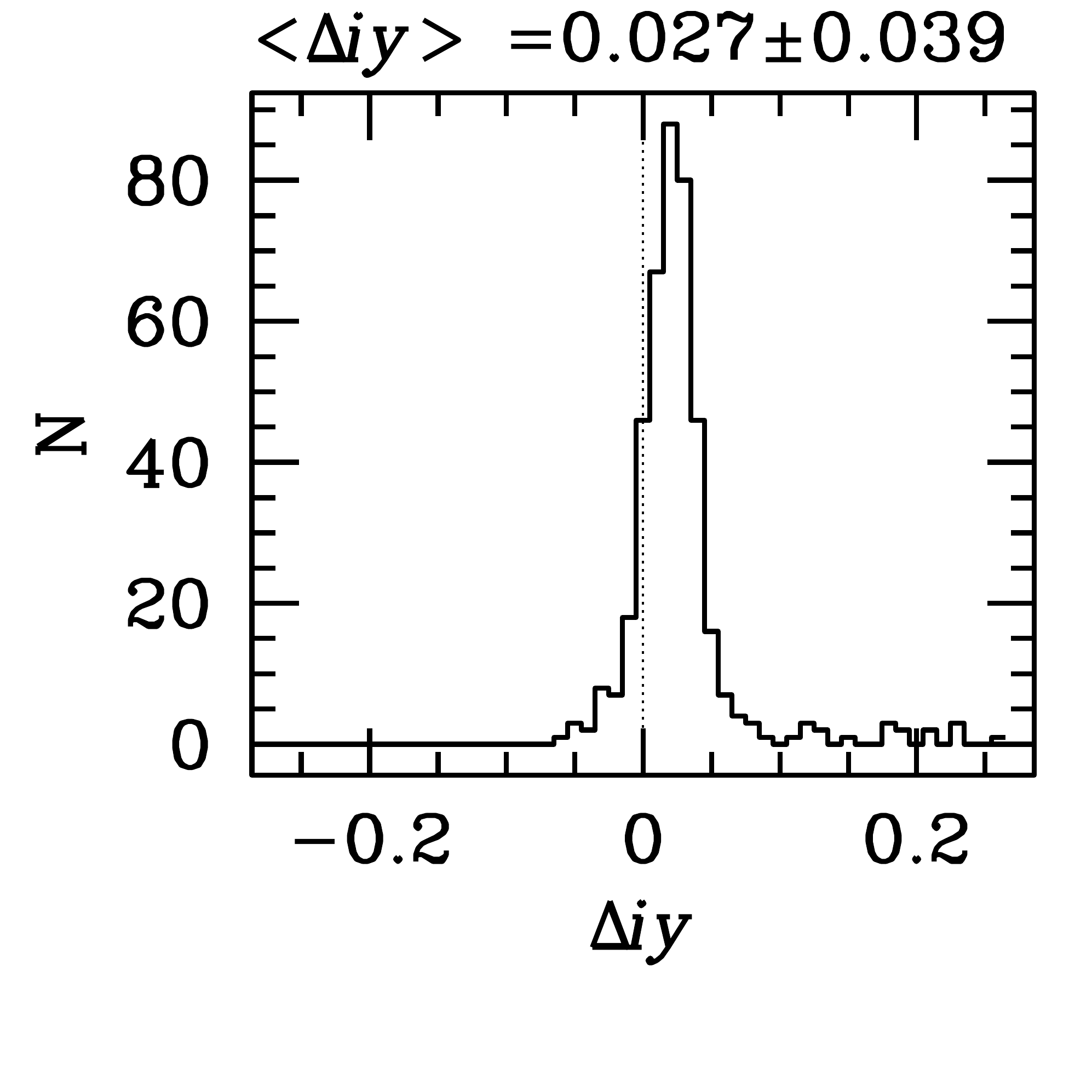}
\includegraphics[width=0.48\hsize]{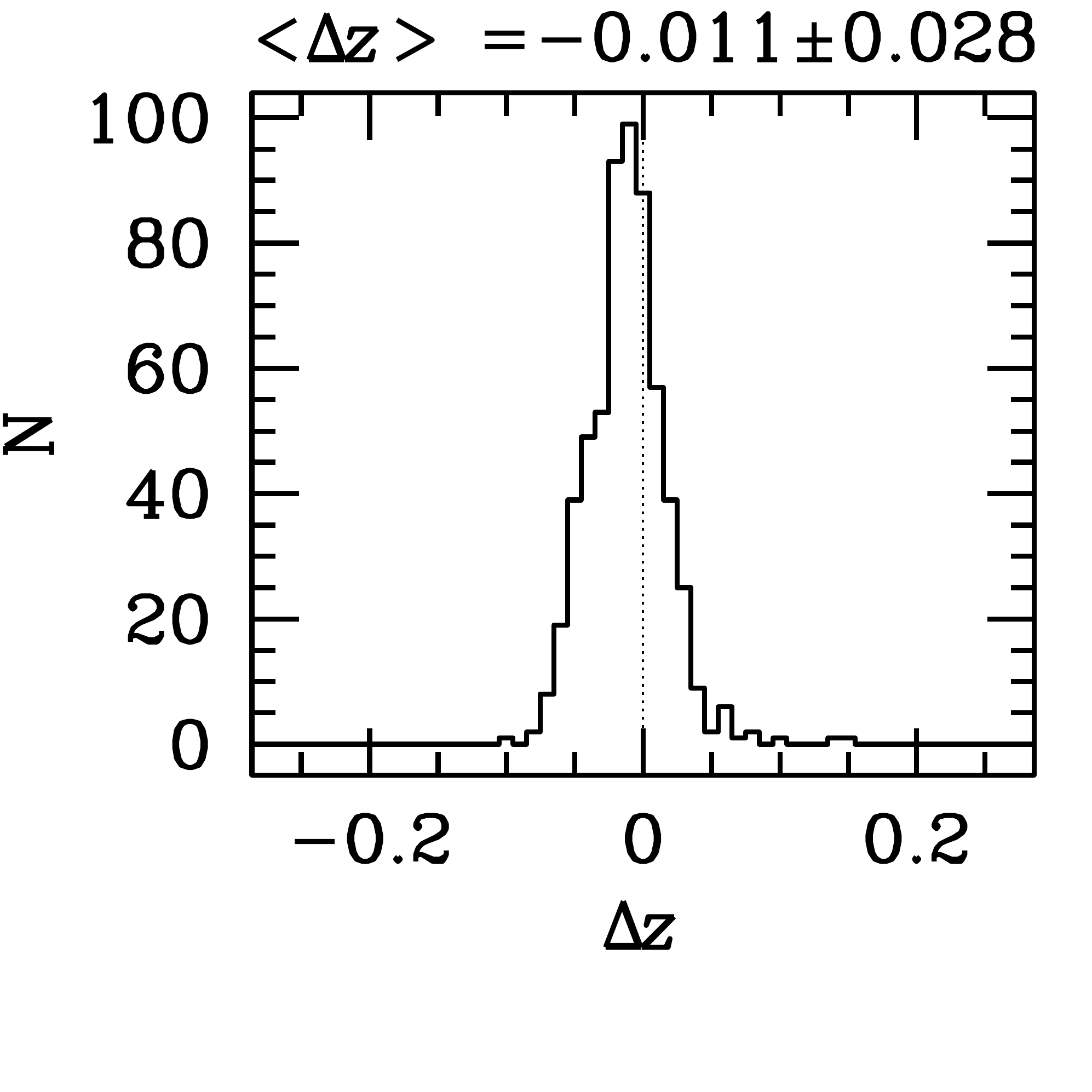}\\
\caption{Distributions of field-wise magnitude differences with respect to SDSS of stars in the four RCSLenS filters ($iy$ indicates that two different $i$-band filters are used in RCSLenS; both are quite similar in terms of throughput to the SDSS $i$-band filter). The mean and standard deviation are reported on top of the panels.}
\label{fig:SDSS_photo_recal}
\end{figure}

The magnitude offsets between SDSS and RCSLenS are averaged for each pointing and this average offset is applied as a photometric zeropoint correction. Figure~\ref{fig:SDSS_photo_recal} shows the distribution of these offsets over all 604 fields that overlap with SDSS for the four filters. Note that not all of these fields have full four-band coverage.

We find moderate mean offsets for the four filters in the range $1-4\%$. However, there is a large variation over the survey area with an RMS scatter of $2-4\%$ depending on the band and maximum offsets of up to $\sim0.3$ mag for some extreme outlier fields. This finding motivated the re-calibration of our zeropoints.

Since the SDSS does not overlap with the full survey area of RCS2 we have to rely on some other technique to re-calibrate the remainder of the fields. For this purpose we use a method that is based on stellar locus regression and introduces some new concepts. This advanced technique incorporates not only a pattern matching between the observed and theoretical stellar loci but also a point-to-point assignment of stars in the data and in an external reference catalogue \citep[we use the stellar library by][]{1998PASP..110..863P}. This additional point matching makes the process more robust, and it helps with multi-band data because assignments from one colour-colour diagram are taken into account in other colour-colour diagrams to improve the matching. Details of this algorithm that is based on the softassign procrustes matching method \citep{Rangarajan97thesoftassign} will be presented in Sheikhbahaee (2016 in prep.).

This advanced stellar locus regression calibrates the colours in the fields without SDSS overlap. However, it does not calibrate the absolute flux scale. Since photo-$z$ are most sensitive to the former this is also the most important calibration for our purposes. Fields that are outliers on the absolute flux scale can still be identified by looking at the magnitude number counts (Sect.~\ref{sec:sanity}) but this is only precise at the $\sim0.1$ mag level.

We test the performance of the advanced stellar locus regression by also running this method on the fields that overlap with SDSS. We compare the \emph{colour} calibrations from both methods and find that they agree on the 2-3\% level on average with a small scatter of $\la2\%$ from field to field. Note that this agreement in \emph{colour} is considerably better than the error that one would make without calibration given the \emph{magnitude} offsets and standard deviations shown in Fig.~\ref{fig:SDSS_photo_recal} which would have to be \emph{combined} from two bands in quadrature to estimate the error on a colour.

\subsection{Photometric redshifts}
\label{sec:photo-z}
Photometric redshifts are estimated with the BPZ algorithm \citep{2000ApJ...536..571B}. We employ the template set by \cite{2004_Capak_PhDT} that represents a re-calibrated version of the original BPZ template set that is based on the empirical templates by \cite{1980ApJS...43..393C} and two starburst templates from \cite{1996ApJ...467...38K}. The Bayesian prior is identical to the one used in CFHTLenS \citep{2012MNRAS.421.2355H}, which is an ad-hoc modification of the original BPZ prior from the Hubble Deep Field to alleviate a bias in the photo-$z$ at low redshift.

We compare the photo-$z$ estimated with BPZ against different spectroscopic redshifts using the catalogues described in Sect.~\ref{sec:overlap_specz}. First we concentrate on the peak of the posterior redshift probability of a galaxy as an estimate of its photo-$z$. It is clear that this yields an incomplete picture of the quality of the full posterior probability functions, $P(z)$, that we provide for each galaxy.  More tests checking the robustness of these are described below.

\begin{figure}
\centering
\includegraphics[clip=true, trim= 2.5cm 0cm 3cm 0cm, width=0.6\hsize]{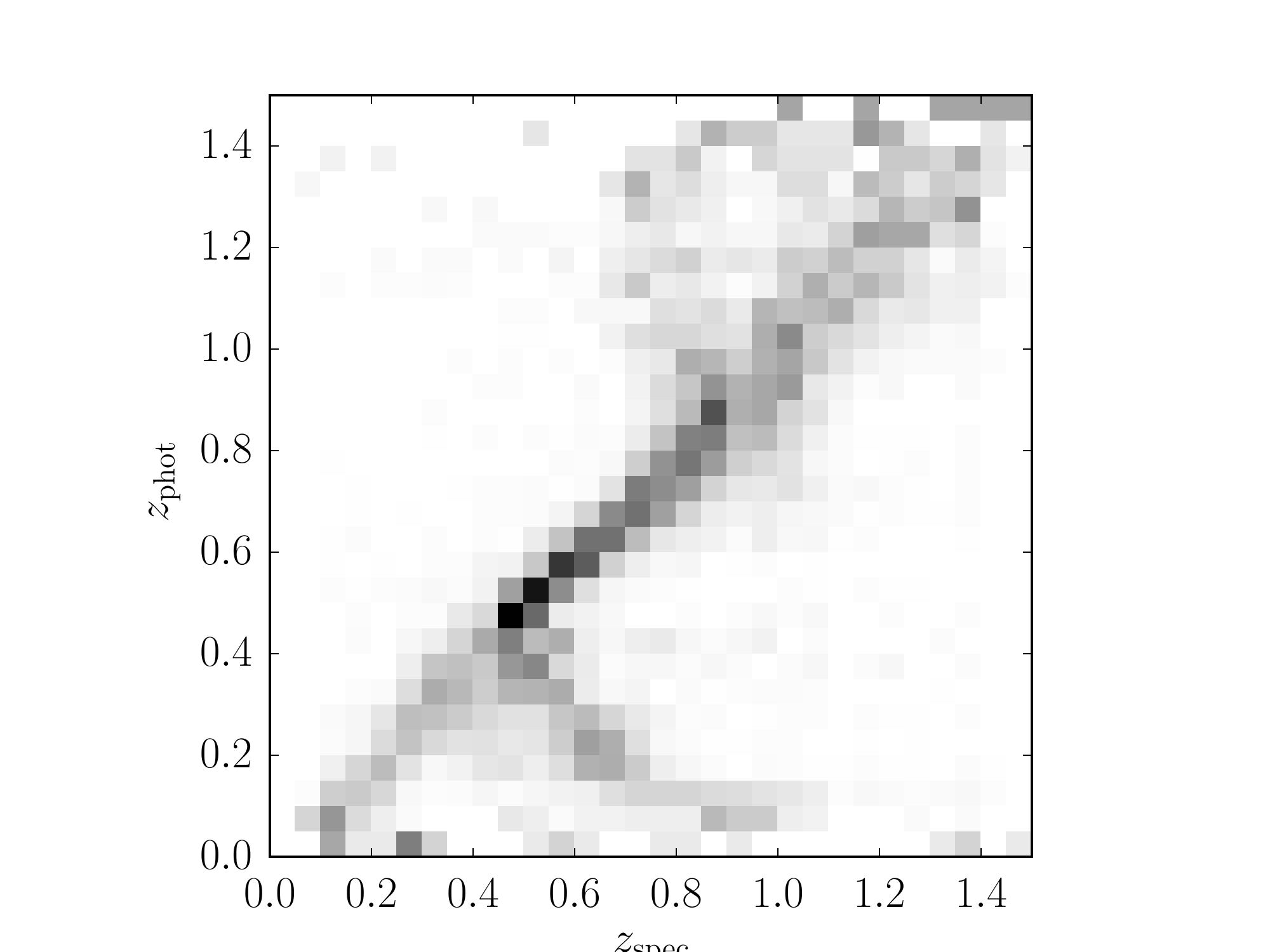}\\
\includegraphics[clip=true, trim= 2.5cm 0cm 3cm 0cm, width=0.6\hsize]{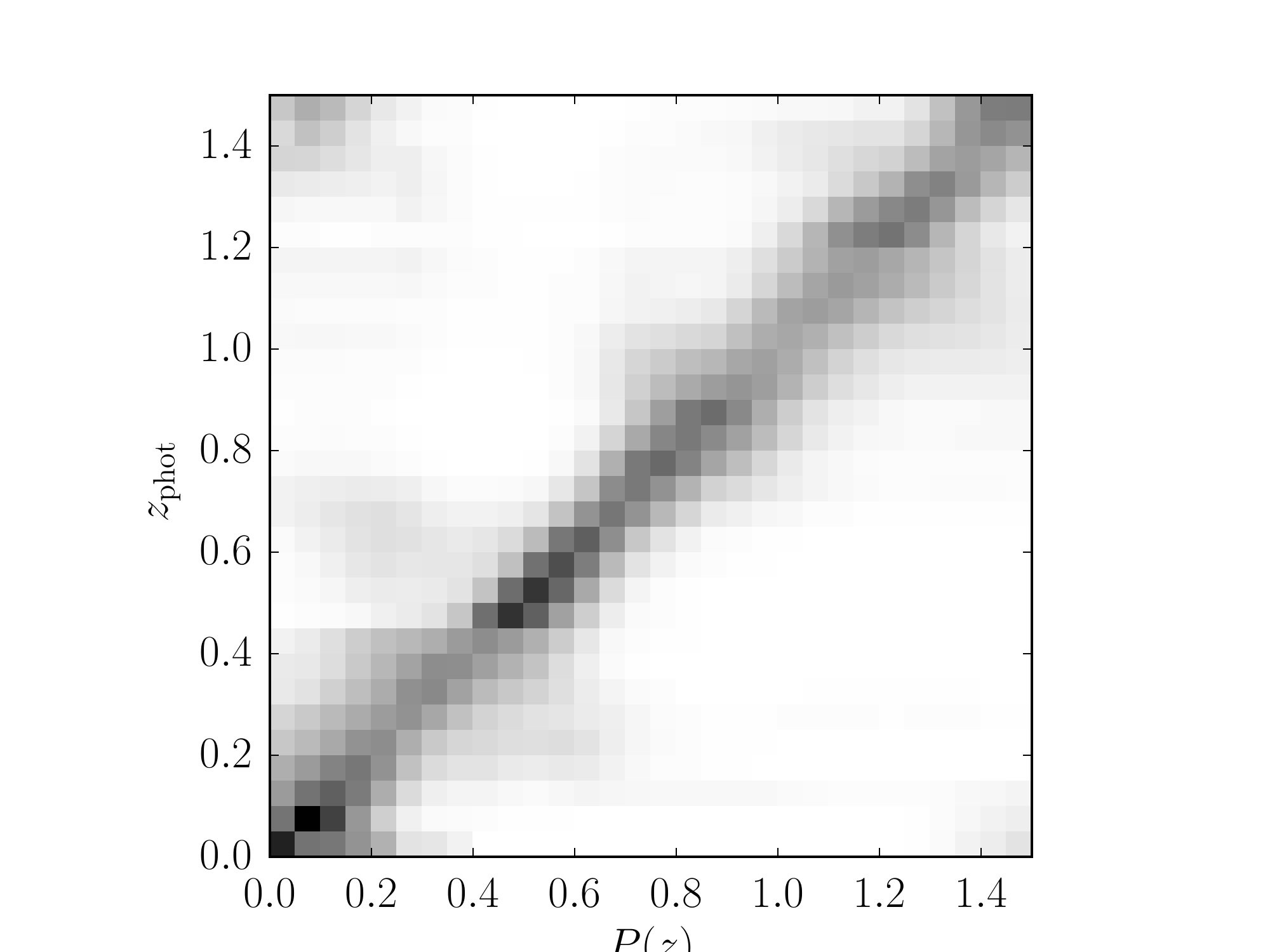}
\caption{Photometric redshifts vs. spectroscopic redshifts (top) and the stacked $P(z)$ in narrow photo-$z$ bins (bottom). Only objects with secure spec-$z$ and reliable detections in the RCSLenS imaging are plotted. Each row in both panels is normalised individually to unity.}
\label{fig:zz}
\end{figure}

In Fig.~\ref{fig:zz} a direct comparison between all available secure\footnote{This corresponds to objects where the spec-$z$ survey teams indicate a probability of at least 95\% that the redshift is correct.} spec-$z$ and our photo-$z$ are shown. It should be noted that the deep spectroscopic catalogue from the DEEP2 survey contains objects that have low SNR in most of our photometric bands. Hence, it can not be expected that their photo-$z$ are very precise. We filter out objects that are not detected in one of the four bands, but this spec-$z$ sample is still quite deep compared to the imaging data. For $z_{\rm phot}\ga0.4$ there is a tight correlation between photo-$z$ and spec-$z$. However, it is clearly visible that the missing $u$-band affects the precision of the photo-$z$ at lower redshift \citep[compared to e.g. CFHTLenS which has a $u$-band; see][]{2012MNRAS.421.2355H}. There is little information in the $griz$-band filter set at these redshifts which leads to uncertain photo-$z$ estimates.

\begin{figure*}
\includegraphics[width=0.4\hsize]{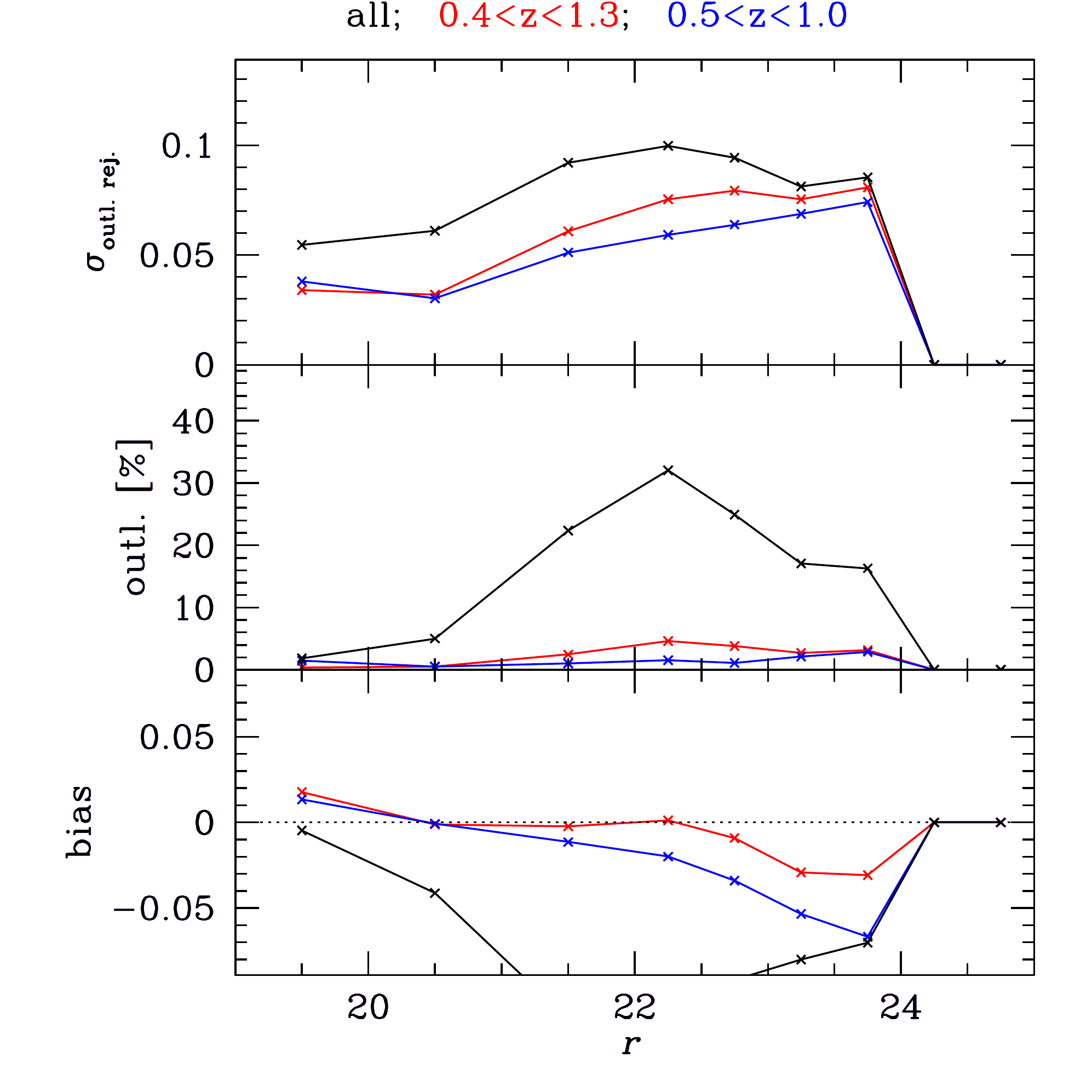}
\includegraphics[width=0.4\hsize]{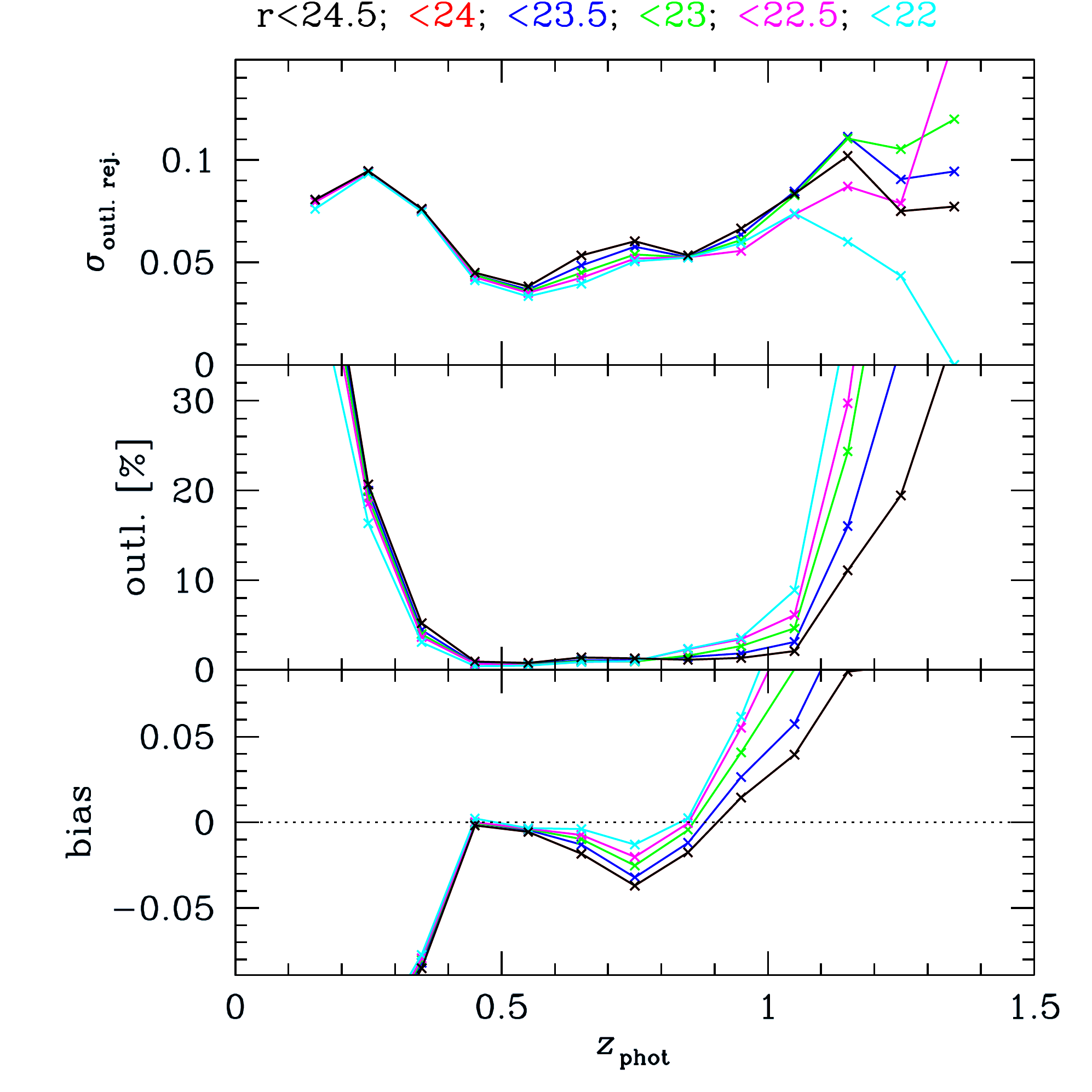}\\
\includegraphics[width=0.4\hsize]{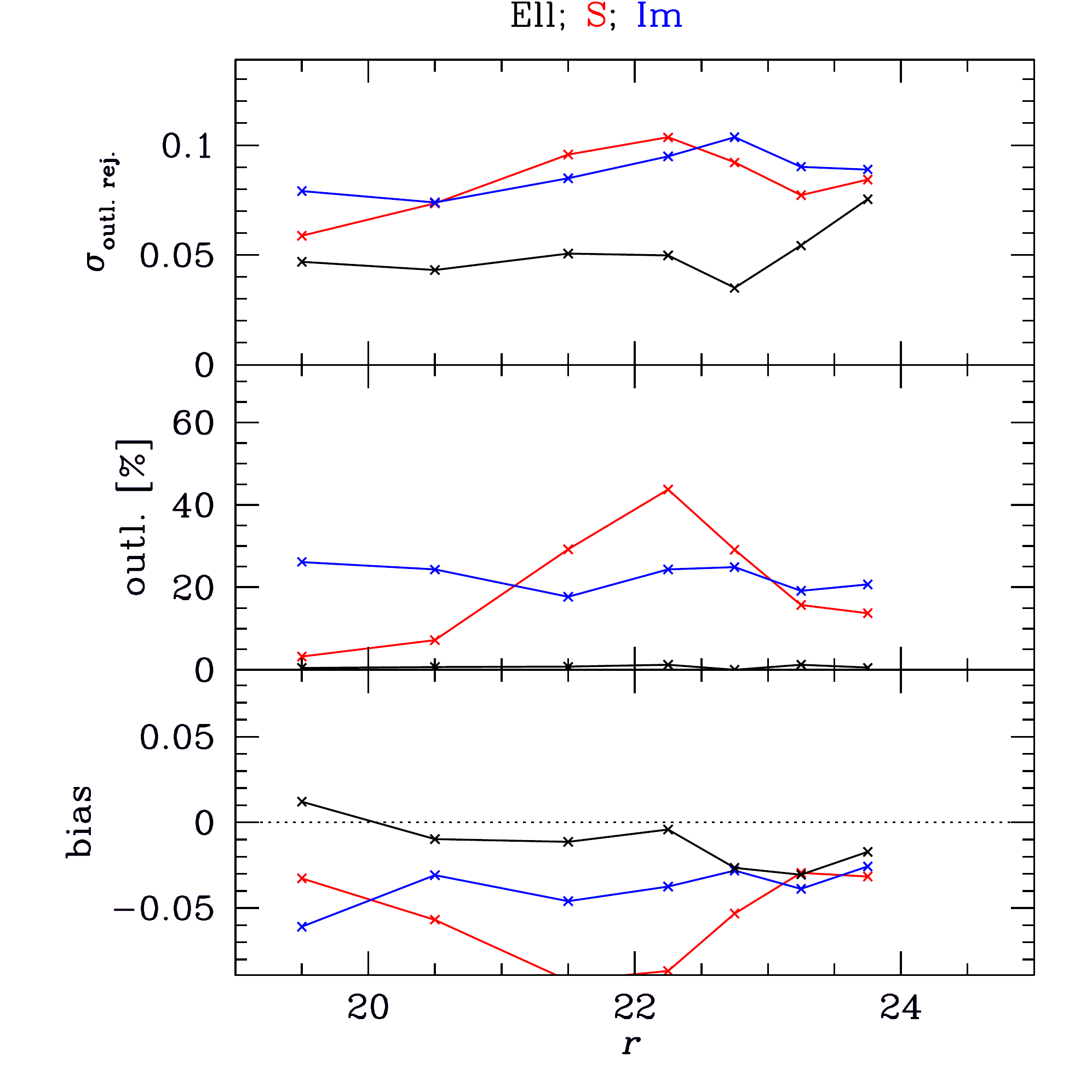}
\includegraphics[width=0.4\hsize]{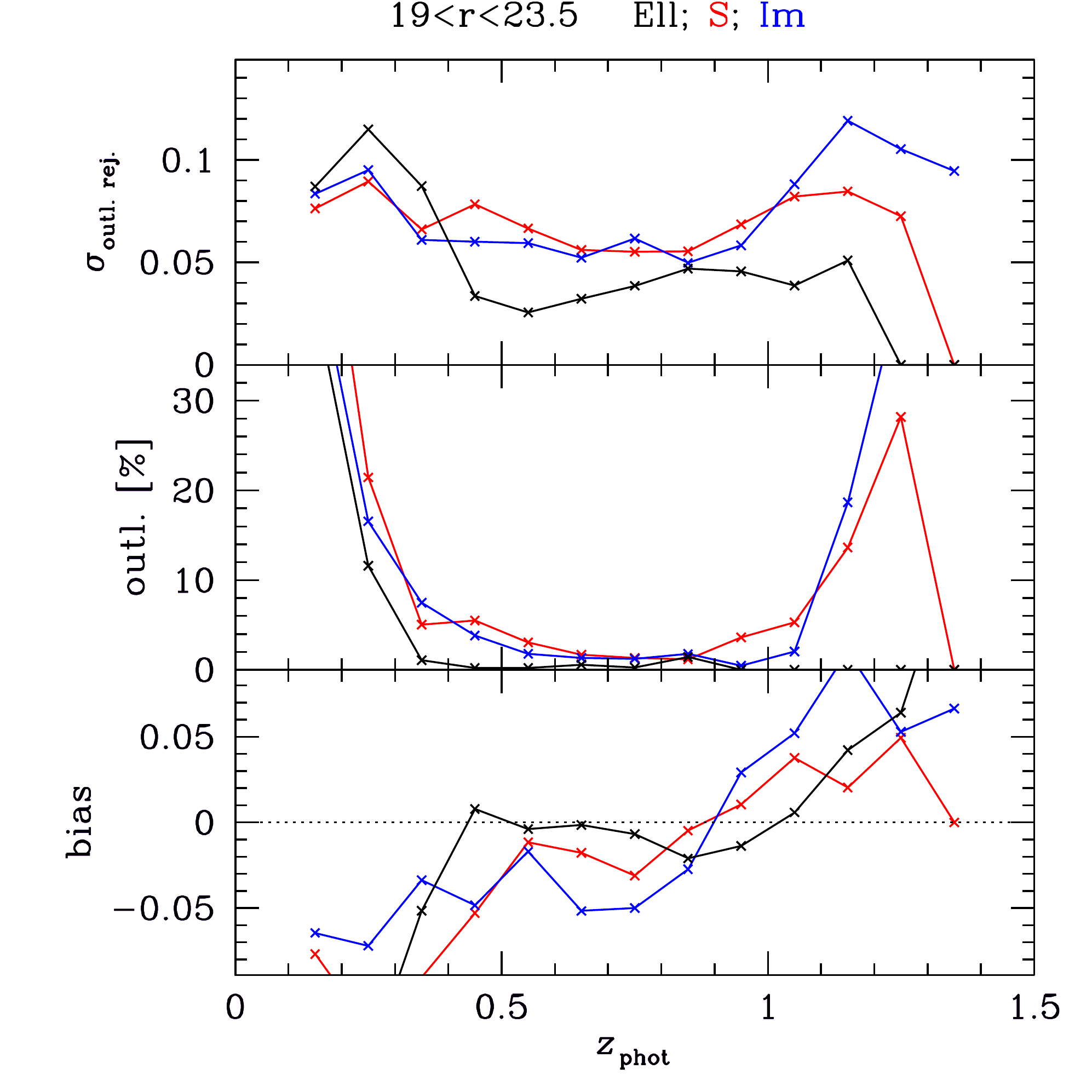}\\
\includegraphics[width=0.4\hsize]{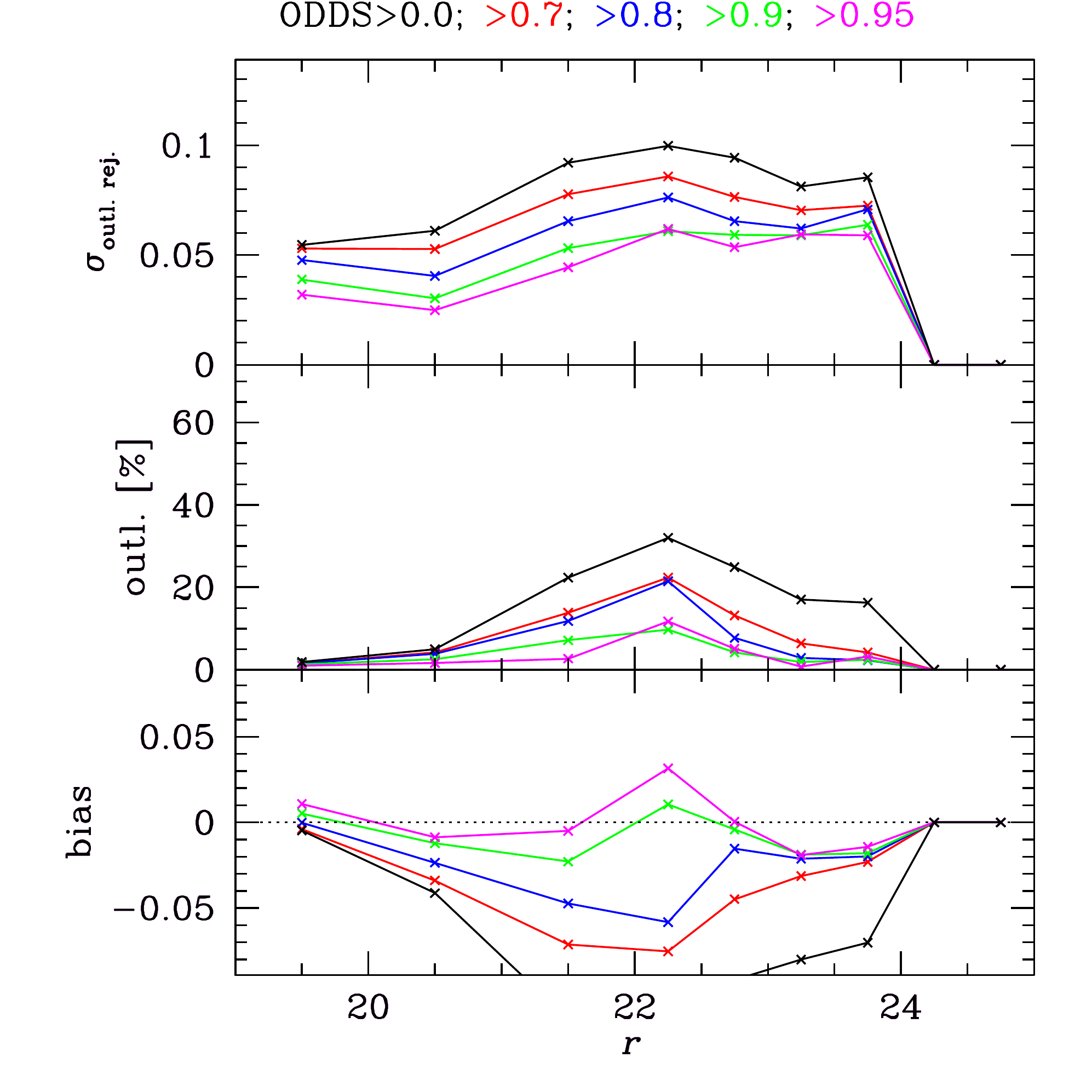}
\includegraphics[width=0.4\hsize]{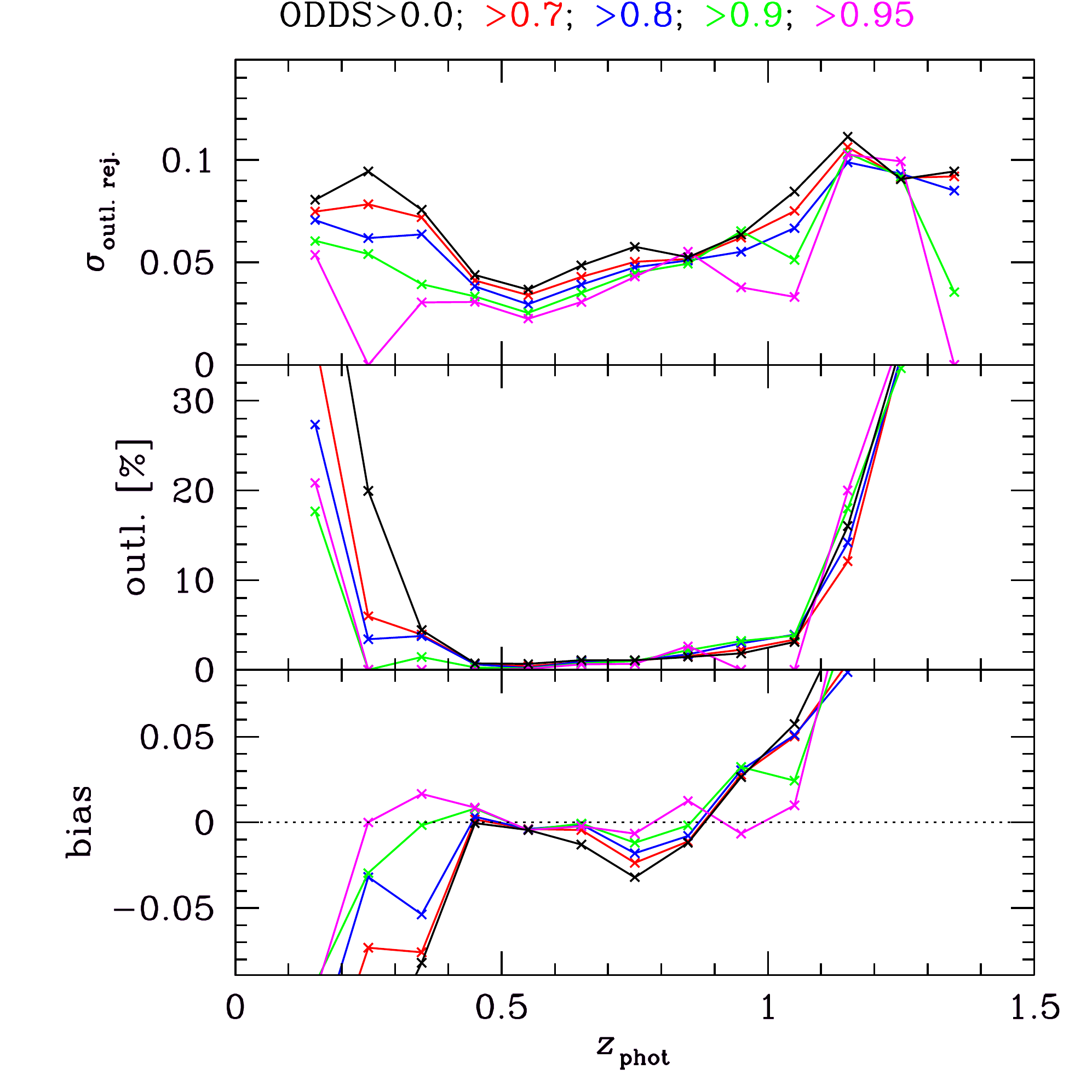}
\caption{Photo-$z$ statistics as a function of magnitude (left) and redshift (right). Shown are the scatter (i.e. the standard deviation of $\Delta_z$), outlier rate and bias (i.e. the mean of $\Delta_z$) for different redshift ranges (top left), different magnitude cuts (top right), different SED types (middle), and different cuts on ODDS (bottom).}
\label{fig:z_stats}
\end{figure*}

A detailed summary of the statistics of the photo-$z$ error distribution is shown in Fig.~\ref{fig:z_stats}. There we plot the mean and the standard deviation of the quantity $\Delta z=(z_{\rm spec}-z_{\rm phot})/(1+z_{\rm spec})$. The latter quantity is reported after outliers (objects with $\Delta z>0.25$)\footnote{In RCSLenS we adopt this more relaxed criterion for outliers \citep[compared to the $\Delta z>0.15$ cut used in][and many other studies]{2012MNRAS.421.2355H} to take into account the larger intrinsic scatter that is caused by fewer filters (4 instead of 5) and noisier photometry.} have been rejected. We also report the outlier rate. Statistics are shown as a function of magnitude and photo-$z$ for cuts on different quantities like the photo-$z$ quality indicator ODDS \citep{2000ApJ...536..571B} or the SED type. The top-left panel of Fig.~\ref{fig:z_stats} confirms that there is a redshift range $0.4<z_{\rm phot}<1.1$ where the outlier rate is very well controlled ($\la2\%$), the scatter is low ($4-8\%$), and the bias stays below 5\%. Interestingly, the middle-left panel shows that the photo-$z$ for galaxies best-fit by an elliptical template can be trusted down to lower redshifts than the ones best-fit by a spiral or starburst template. The opposite is true at the high-redshift edge where the outlier rate of ellipticals starts to rise steeply for $z_{\rm phot}>0.9$ whereas later types are well behaved up to $z_{\rm phot}\sim1.1$. Similar to the findings of \cite{2012MNRAS.421.2355H} the ODDS parameter has very little influence on the overall performance of the photo-$z$ (lower panels of Fig.~\ref{fig:z_stats}) in the region where the photo-$z$ are most informative.

\subsection{Absolute magnitudes and stellar masses}
\label{sec:abs_mag}
Physical parameters of galaxies are estimated with the LePhare code \citep{2006A&A...457..841I} in a very similar way as described in \cite{2014MNRAS.437.2111V}. The redshift of a galaxy is fixed to the most probable Bayesian redshift estimate determined by BPZ. Then LePhare is run with an extensive template library from \cite{2003MNRAS.344.1000B} to find the best-fitting template at the given redshift. For this template the absolute magnitudes in different filters and the stellar mass are calculated.

\section{Tests for systematic errors}
\label{sec:systematics}
In this section we concentrate on tests that are relevant for the weak lensing science that the RCSLenS data will be used for.

\subsection{Photometry}
\label{sec:photometry_systematics}
\begin{figure}
\includegraphics[width=0.48\hsize]{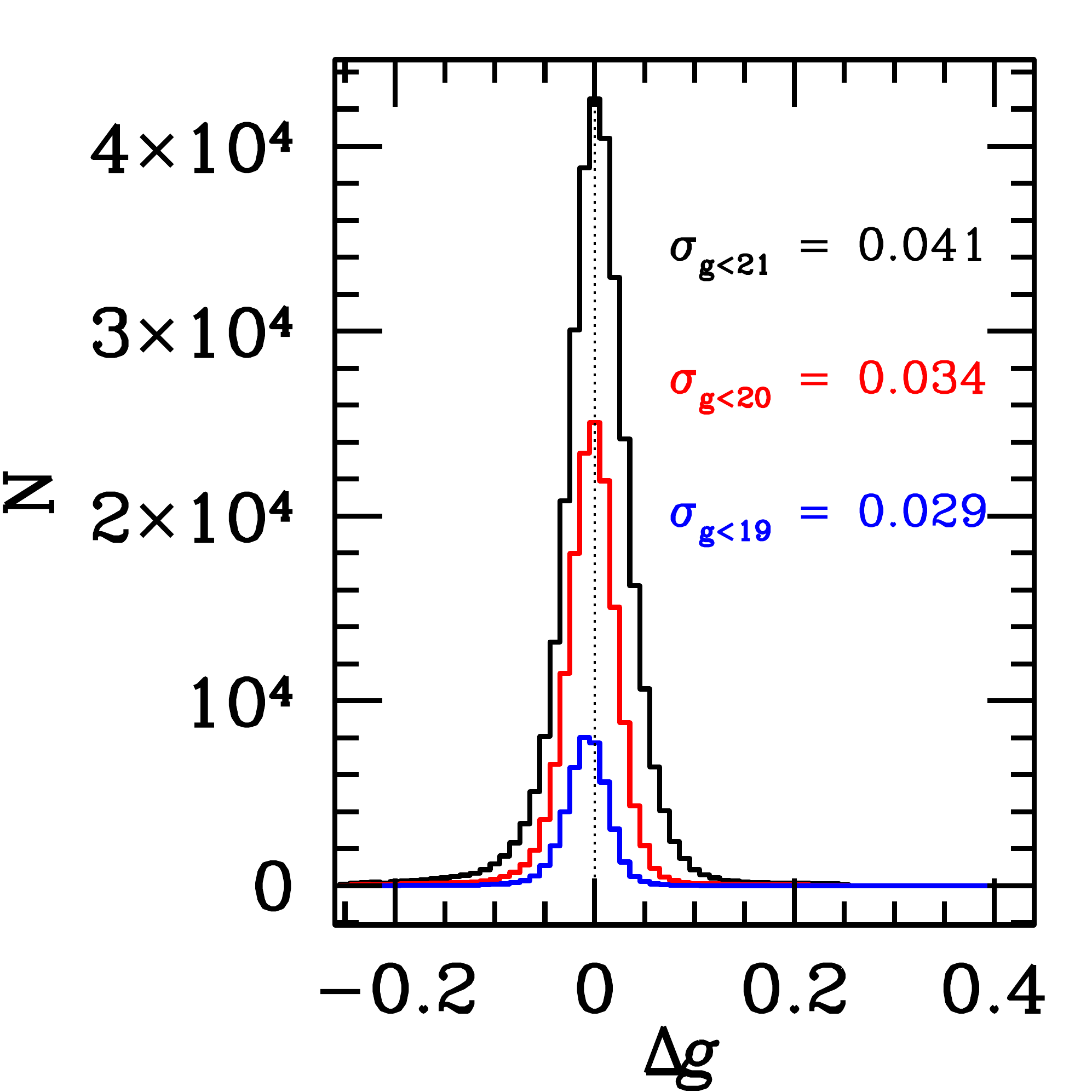}
\includegraphics[width=0.48\hsize]{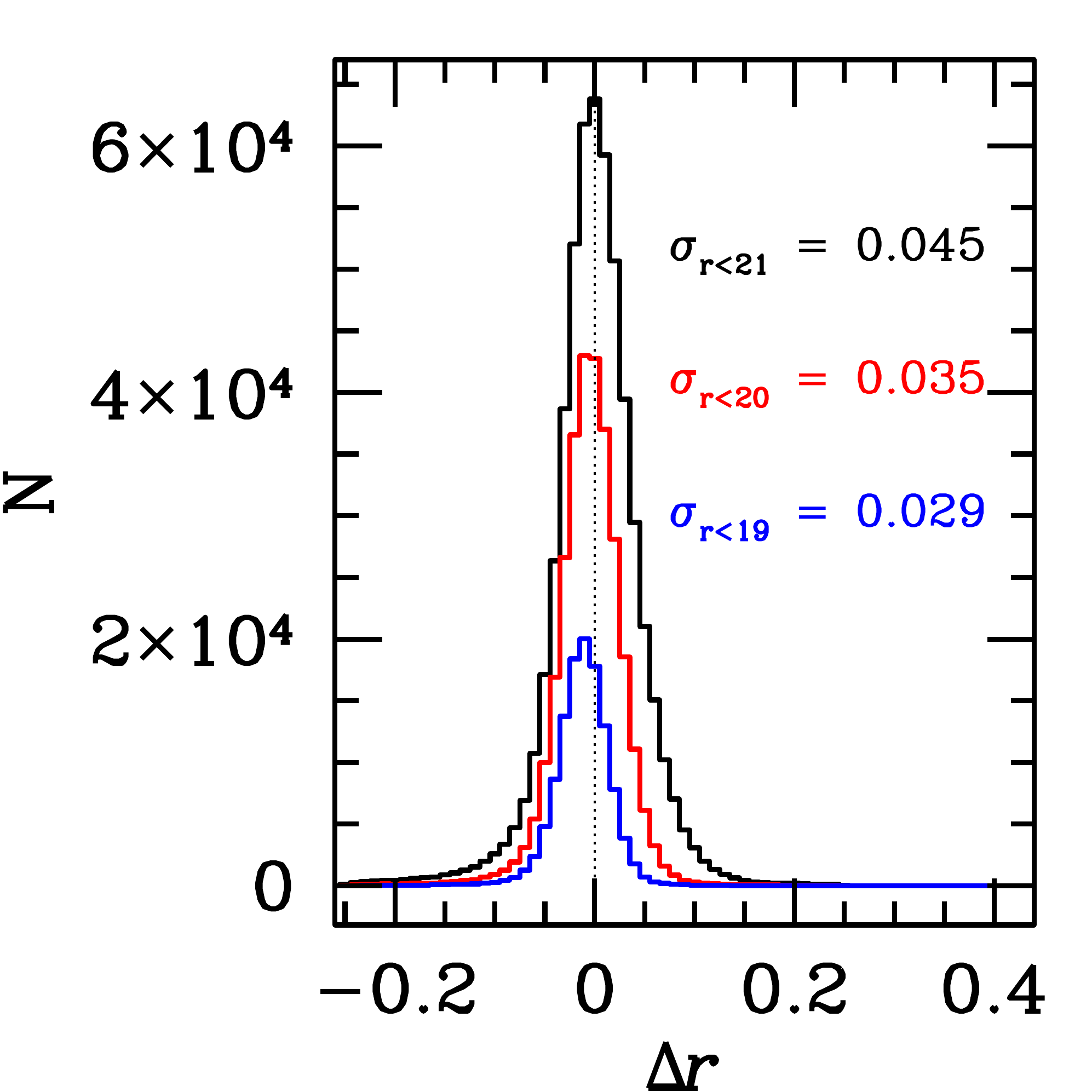}\\
\includegraphics[width=0.48\hsize]{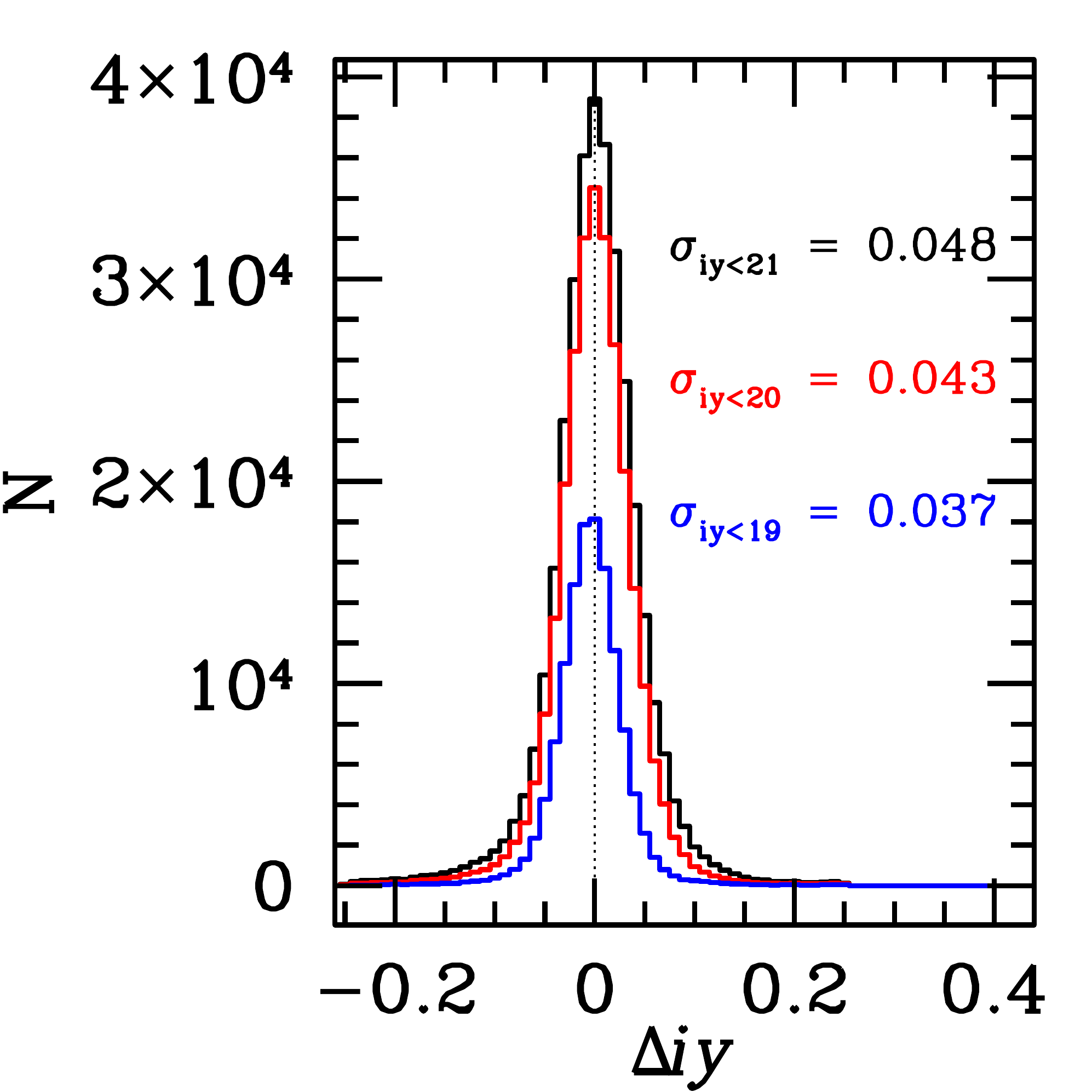}
\includegraphics[width=0.48\hsize]{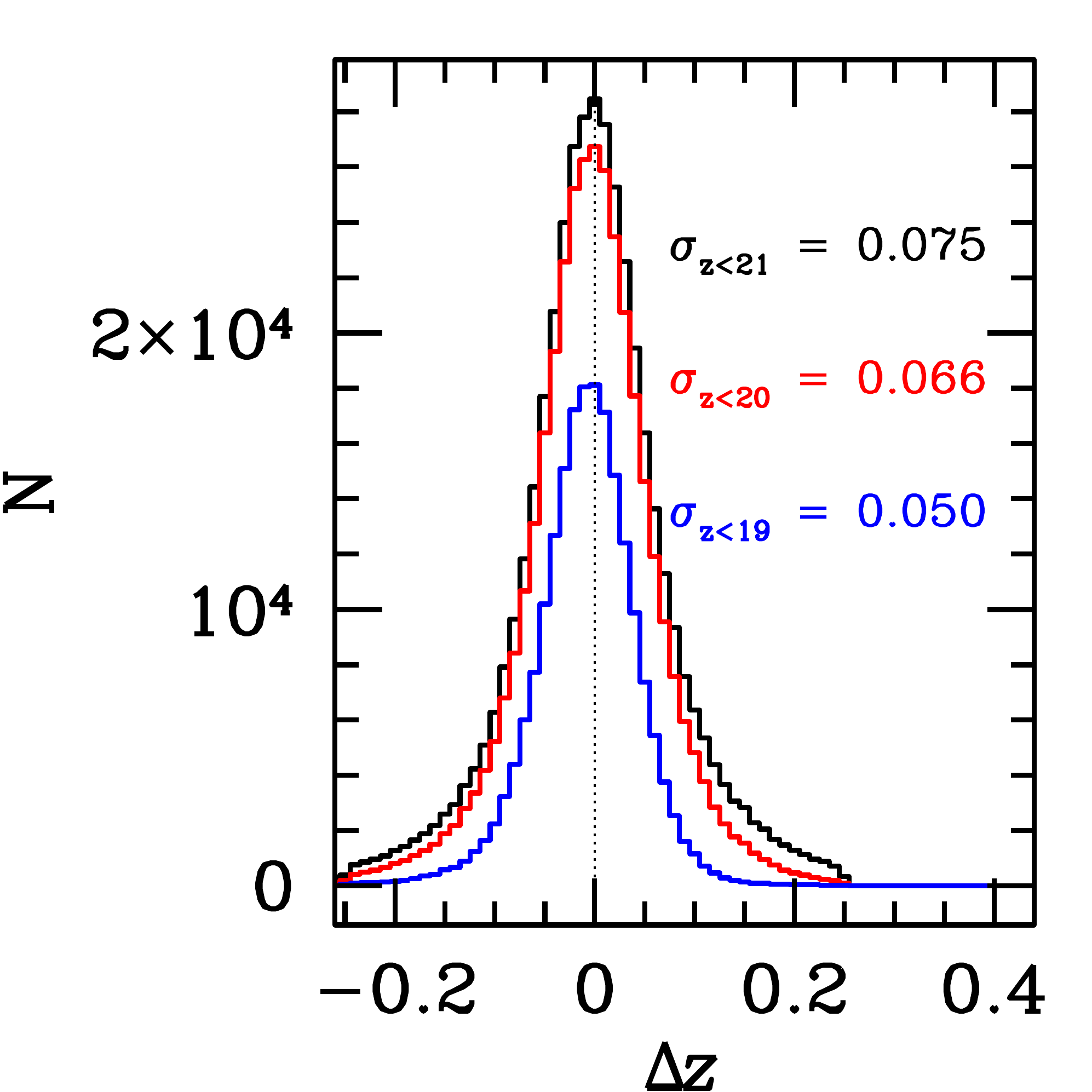}\\
\caption{Distributions of magnitude differences of stars with respect to SDSS in the four RCSLens filters after field-wise photometric re-calibration. The scatter for different magnitude limits is shown, with the brightest sample giving a good estimate of the photometric zeropoint variations with respect to SDSS over the full survey area.}
\label{fig:SDSS_comp}
\end{figure}

In Fig.~\ref{fig:SDSS_comp} we show the distribution of the magnitude differences of bright stars between SDSS and RCSLenS for the four filters after the photometric re-calibration described in Sect.\ref{sec:abs_photo_cal}. We choose different magnitude cuts to check the influence of noise in the SDSS measurements. The mean offset vanishes by construction for the mag$<21$ samples since the mean magnitude difference per field for this sample is already incorporated into the zeropoints (see Sect.~\ref{sec:abs_photo_cal}). We find that the RMS scatter across the survey area is $3-5\%$ in the different bands for the brightest sample (mag$<19$). Besides a small shot noise contribution these numbers represent a combination of the internal photometric consistency of both surveys, SDSS and RCSLenS, at sub-degree scales. One of the main sources of error in RCSLenS in this case is probably the ELIXIR illumination correction, which leads to a similar scatter in the magnitudes over the field-of-view of the MegaCam instrument.\footnote{Note that there is a new, improved ELIXIR illumination for more recent MegaCam data that reduces this scatter to $\sim1\%$ in the $g$- and $r$-bands. However, this pre-reduction is not available for the RCS2 data.}

\begin{figure}
\includegraphics[clip=true, trim=1cm 0cm 2cm 1cm, width=0.48\hsize]{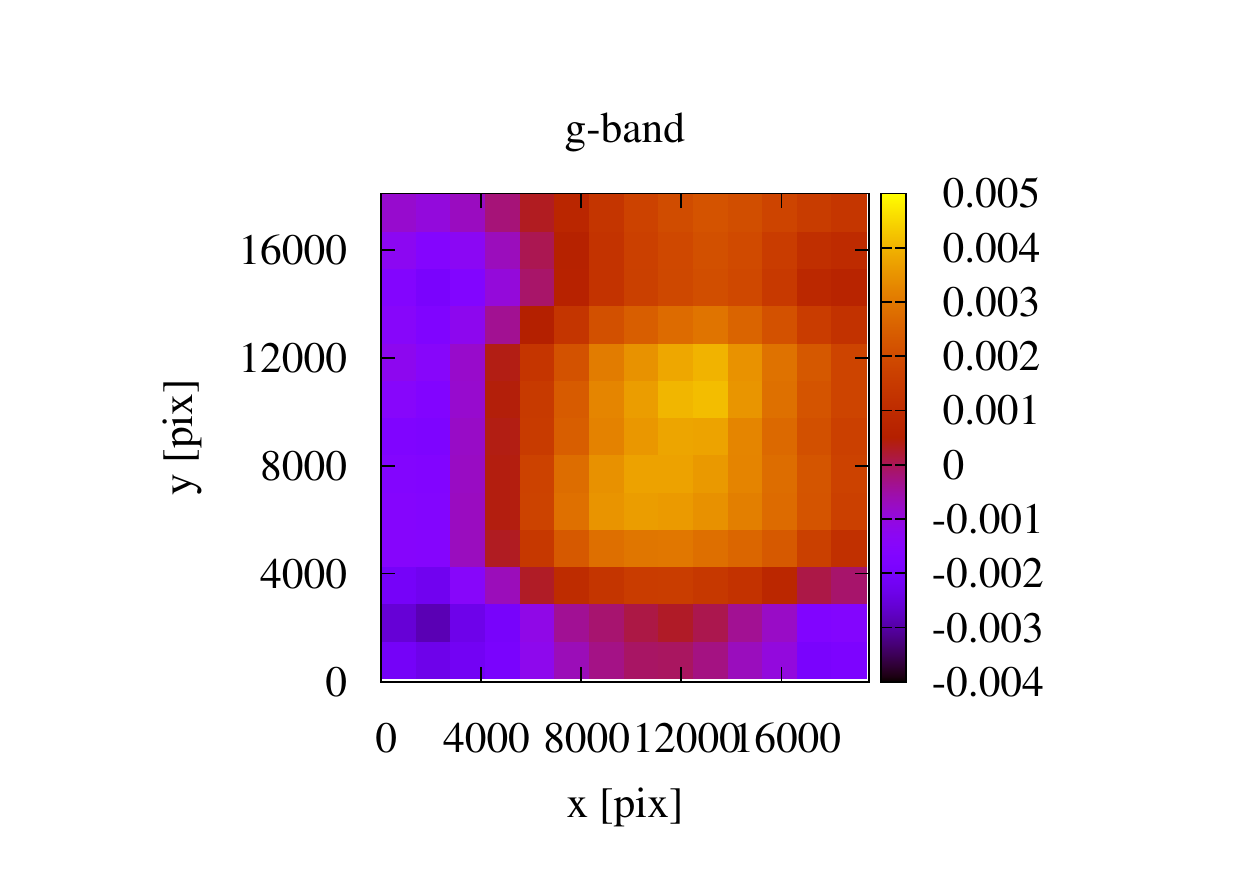}
\includegraphics[clip=true, trim=1cm 0cm 2cm 1cm,width=0.48\hsize]{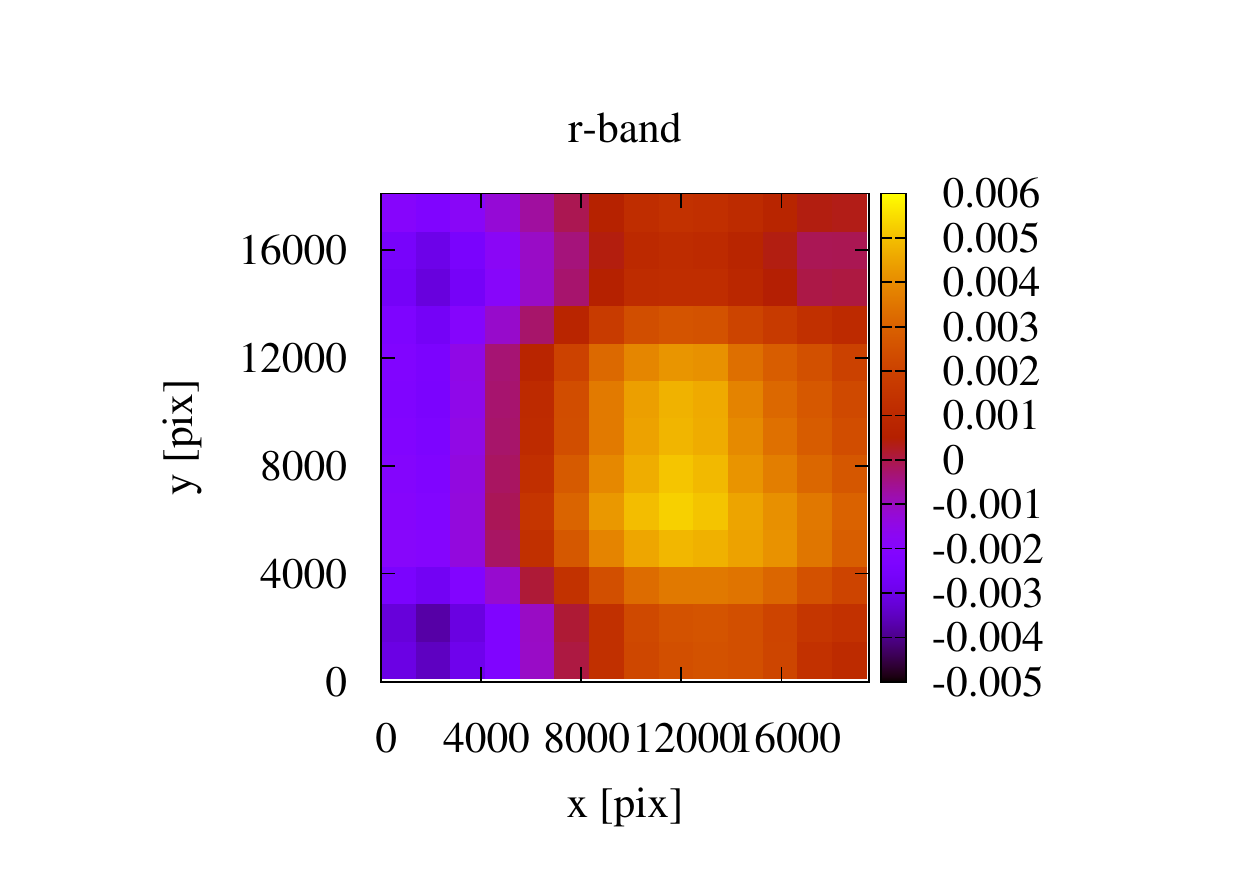}\\
\includegraphics[clip=true, trim=1cm 0cm 2cm 1cm,width=0.48\hsize]{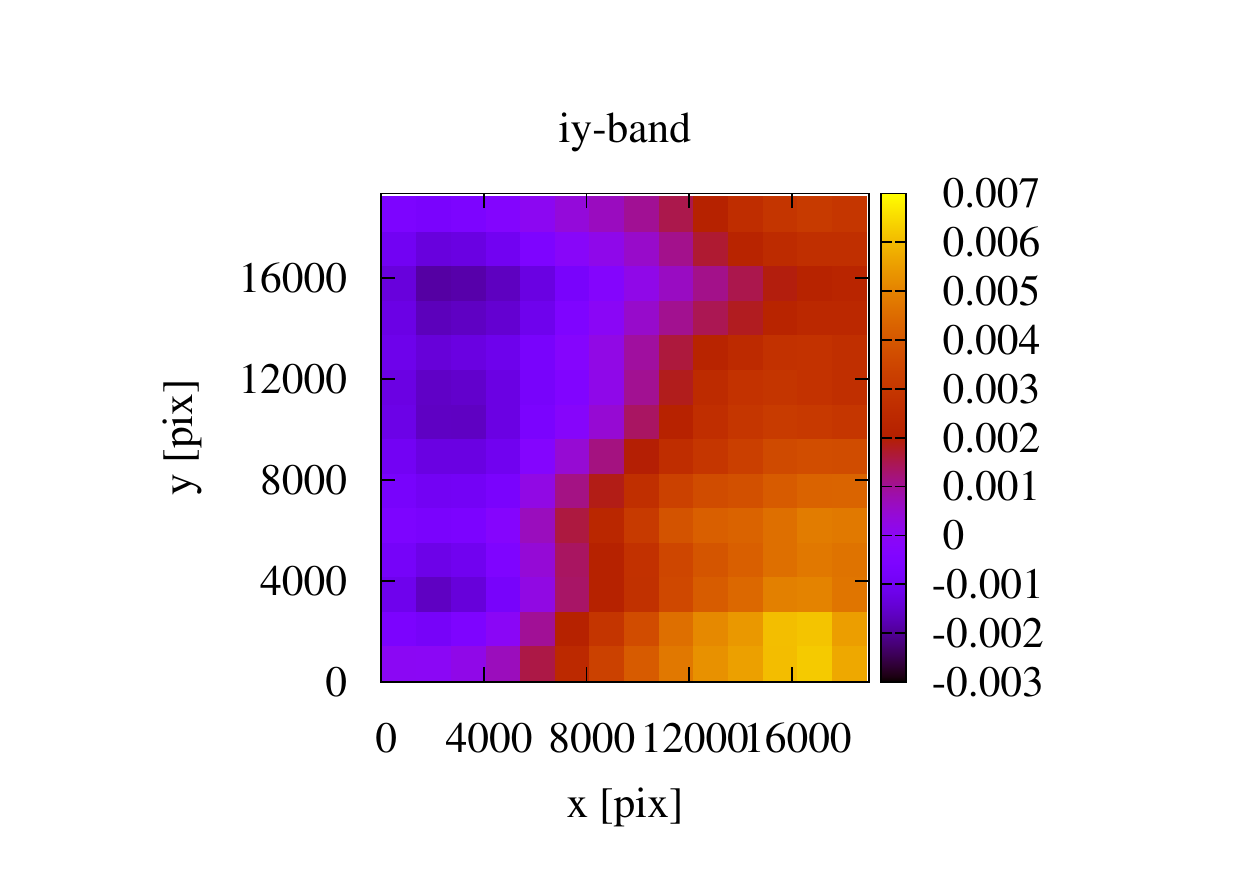}
\includegraphics[clip=true, trim=1cm 0cm 2cm 1cm,width=0.48\hsize]{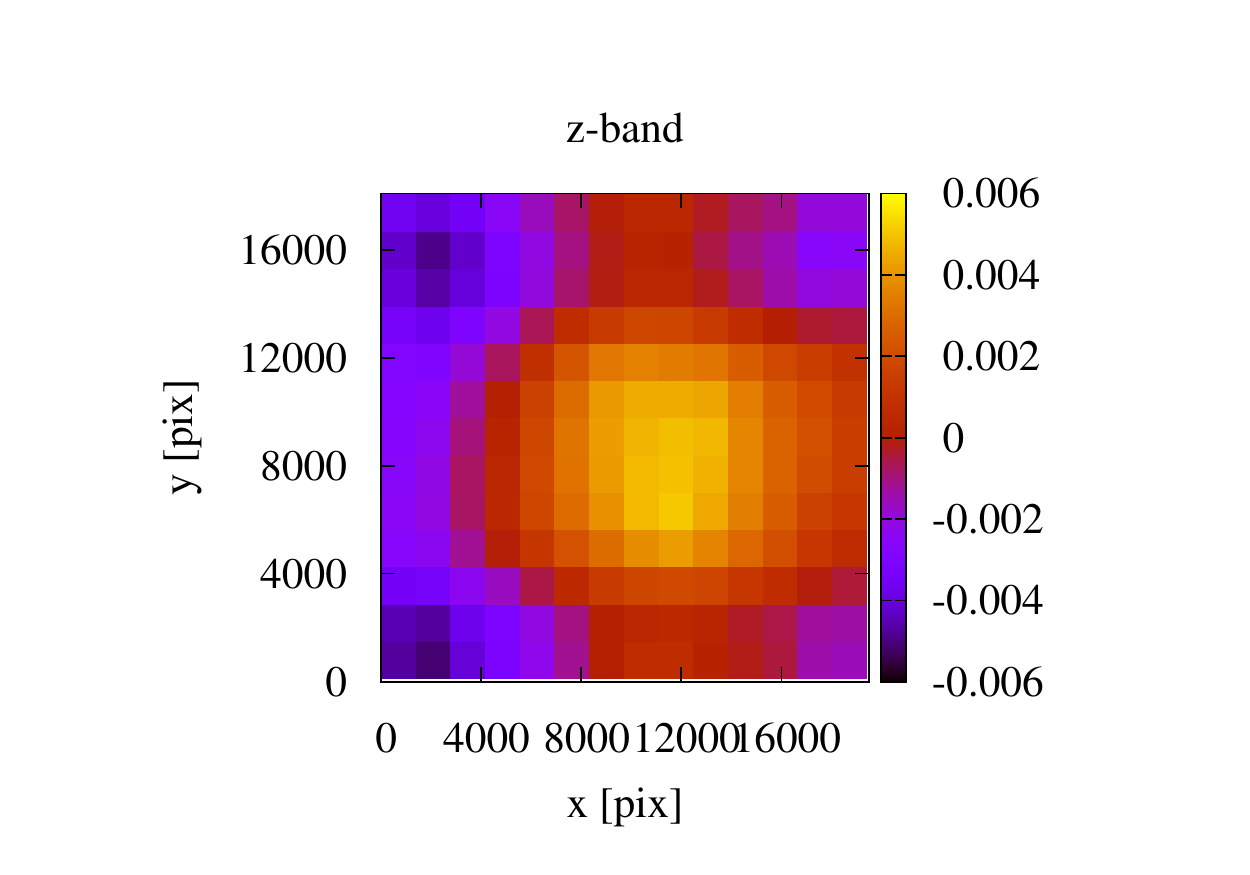}\\
\caption{Distributions of magnitude differences of stars with respect to SDSS in pixel coordinates in the four RCSLens filters after photometric re-calibration. This reveals a pattern in the offset w.r.t. SDSS as a function of position in the CCD mosaic, which is likely caused by an imperfect scattered-light/illumination correction.}
\label{fig:illumination}
\end{figure}

We investigate the residuals in the stellar photometry between SDSS and RCSLenS as a function of pixel position in all RCSLenS fields. Stacking data from all available fields, two-dimensional maps of the residual magnitude difference after re-calibration are created and shown in Fig.~\ref{fig:illumination}. There are clearly some residual structures visible in all bands and the pattern looks very similar for the $grz$-bands while there is some difference for the $i$/$y$-band. One could use such maps to apply a further correction to the zeropoints as a function of position but we neglect this here due to the low level of this effect.

\begin{figure}
\centering
\includegraphics[width=\hsize]{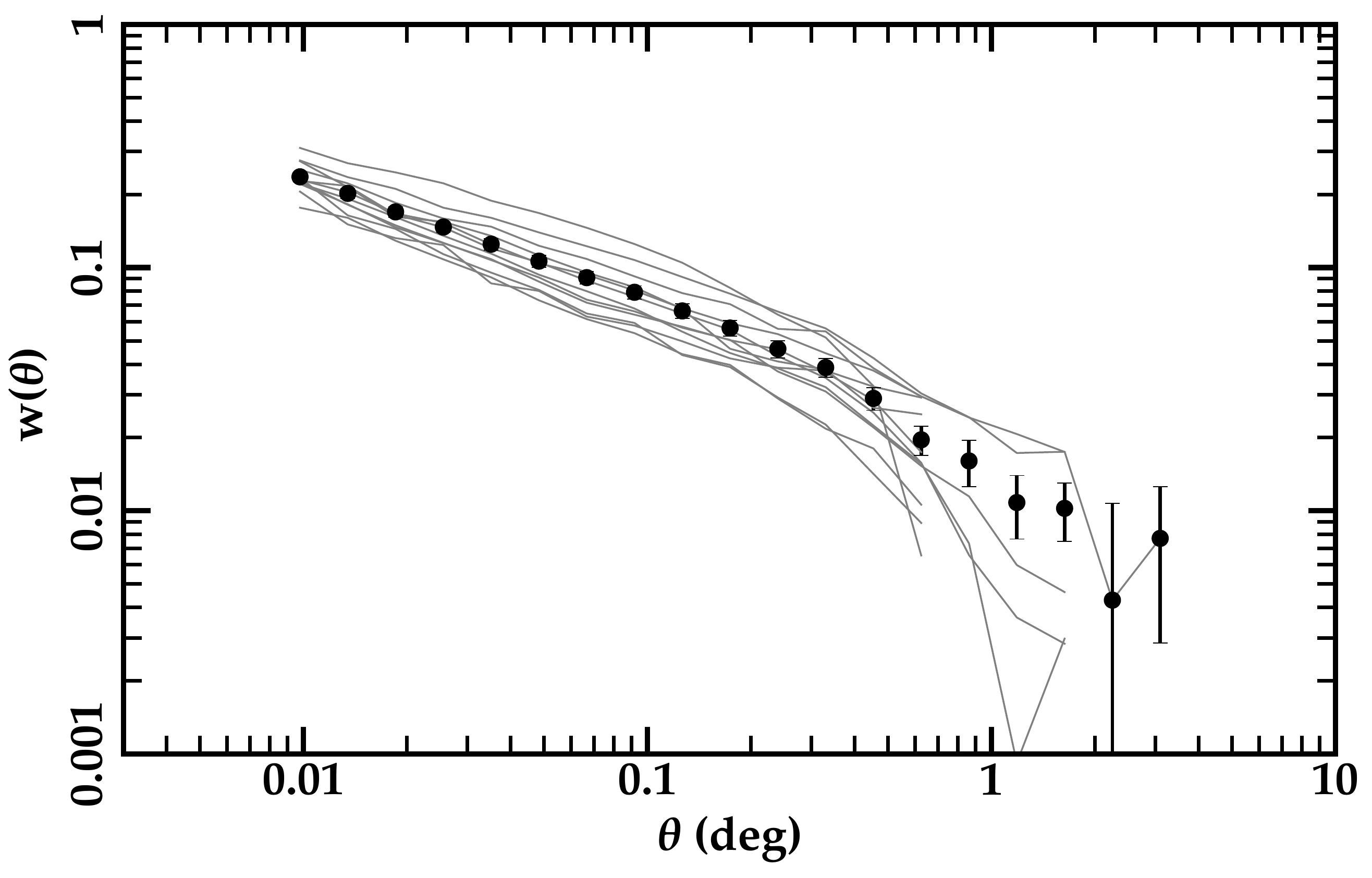}
\caption{Angular auto-correlation function of galaxies with $r<21$. The data points show the survey mean whereas the thin lines represent the 14 different RCSLenS patches.}
\label{fig:auto_corr}
\end{figure}

The angular auto-correlation function of galaxies reacts sensitively to variations in the galaxy photometry \citep[see e.g.][]{2015MNRAS.454.3121M}. As a quick test of the homogeneity of the galaxy photometry we estimate the angular correlation function for bright galaxies with $r<21$. The result for the full survey as well as for the 14 individual patches can be found in Fig.~\ref{fig:auto_corr} showing the expected power-law shape and no peculiarities on scales that correspond to the size of the individual pointings (i.e. $\sim 1$deg).

\subsection{Photo-$z$}
In cosmological weak lensing analyses it is nowadays common to take into account the full redshift probability distribution of each galaxy that is provided by a photo-$z$ code. In the bottom panel of Fig.~\ref{fig:zz} we show the stacked $P(z)$ in narrow photo-$z$ bins. The objects shown are the same ones as in the left-hand panel, i.e. the objects that have a secure spectroscopic redshift. Ideally these two plots should be identical within shot-noise. However, the priors for BPZ are based on magnitude limited samples whereas the sample presented here suffers from severe spectroscopic selection effects. Hence, one can not expect a perfect match. To further assess the $P(z)$, which are used in all lensing studies, we use a modified version of the cross-correlation technique \citep{2008ApJ...684...88N}. Results are presented in \cite{2015arXiv151203626C} revealing significant biases in the $P(z)$ that need to be accounted for or marginalised over. In particular, if the RCSLenS source catalogue is split into four bins in photo-$z$ over the range $0.3<z_{\rm phot}<0.9$ the cross-correlation analysis suggests that the mean of the stacked $P(z)$ in a bin can be biased in the range $-0.095\le \Delta z \le 0.236$. Cross-correlations against BOSS and WiggleZ yield somewhat inconsistent results with the latter yielding considerably smaller $\Delta z$ for all bins than the former.

It should however be noted that the stacked $P(z)$ (right-hand panel of Fig.~\ref{fig:zz}) contain many of the structures that are visible in a direct comparison of photo-$z$ and spectroscopic redshifts (left-hand panel of Fig.~\ref{fig:zz}). This is even partly true outside the redshift range where the photo-$z$ point estimates perform well (see Sect.~\ref{sec:photo-z}). For example, the greatly increased scatter in the photo-$z$ point estimates for $0.2\la z_{\rm phot}\la 0.4$ is replicated in the $P(z)$.

Different measurements are affected differently by these obvious systematic errors in the photometric redshifts. These will be dealt with on a case-by-case basis in the scientific papers.

\subsection{Shapes}
\label{sec:shear_syst}
The main tool to check the measured shapes for systematic errors due to imperfect PSF removal is the cross-correlation function of the corrected ellipticities of galaxies and the ellipticities of stars \citep[see][for a detailed description]{2012MNRAS.427..146H}. The PSF is measured from stars. After PSF deconvolution the galaxy shapes should be unaffected by the shape of the PSF, and the cross-correlation function should be zero. This measurement is carried out on individual pointings so that parts of the survey with a large non-zero cross-correlation function (larger than that which could be caused by cosmic shear) can be excluded from the scientific analyses.

The finite size of each pointing ($\sim$1\,deg$^2$) can give rise to an additional non-zero correlation signal that does not originate from systematic errors in the PSF model but from chance alignments. In principle, the intrinsic ellipticities, the measurement noise and the cosmological shear field can align with the PSF, even if the PSF correction is perfect. Thus, rejecting fields solely based on their non-zero cross-correlation signal would lead to an overly pessimistic rejection scheme and a biased sampling of the true cosmological shear field. In order to avoid this we need to estimate the expected signal from these chance alignments \citep{2012MNRAS.427..146H}. This is done on a large set of mock catalogues that closely resemble the RCSLenS data. These mock catalogues are created from the simulations that are discussed in Sect.~\ref{sec:additional_data}. 

\begin{figure}
\includegraphics[width=\hsize]{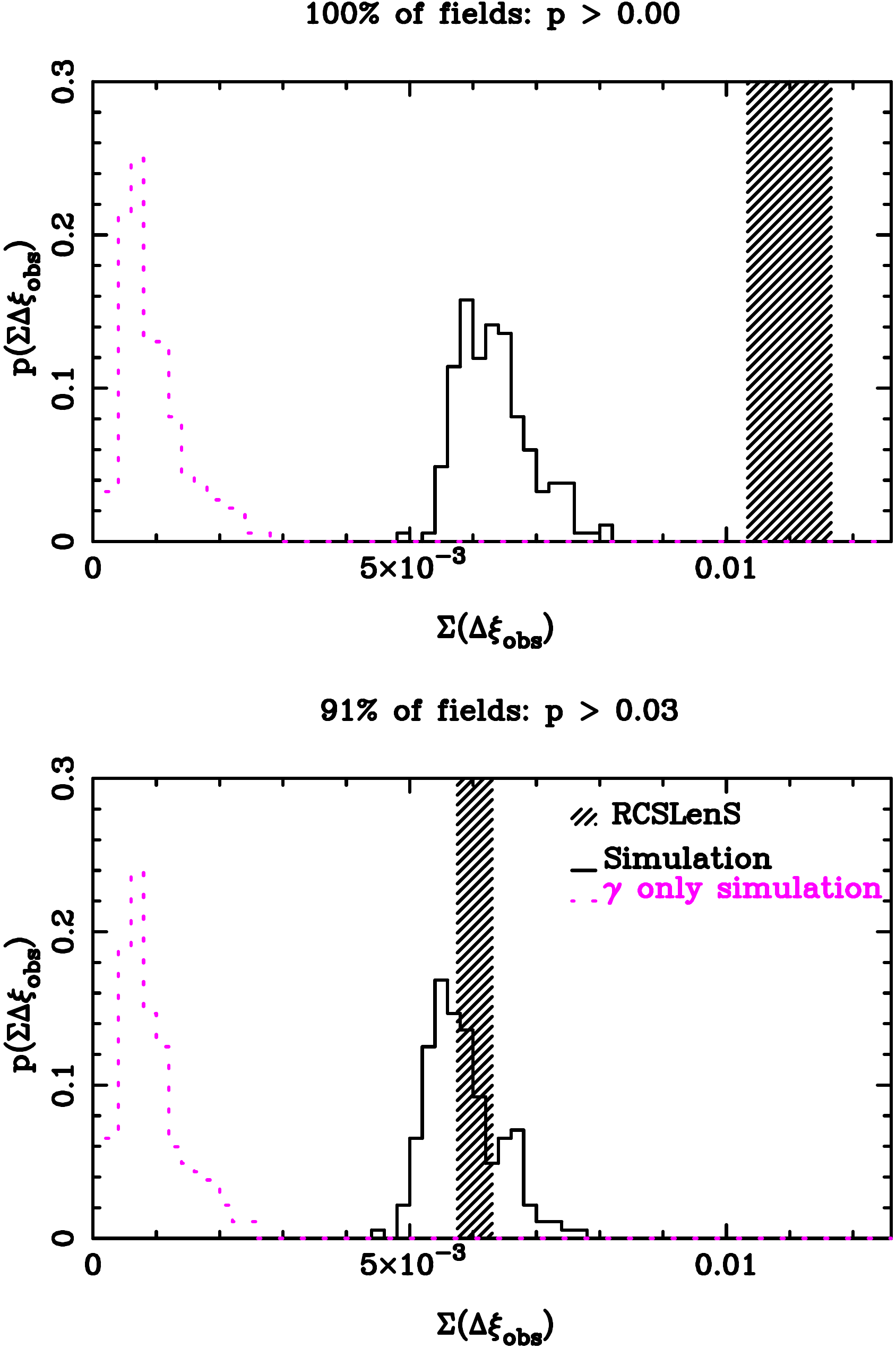}
\caption{The x-axis shows the cumulative (i.e. summed over all fields) amplitude of the cross-correlation function of corrected galaxy ellipticities and stellar ellipticities, $\Sigma(\Delta\xi_{\rm obs})$. Shown is the amplitude at zero lag in the data (shaded region) and the probability distribution ($p(\Sigma\Delta\xi_{\rm obs})$, y-axis) of this quantity over a large set of simulated mock catalogues (solid line). The contribution to this cross-correlation amplitude from chance alignments of the cosmic shear field and the PSF is shown by the dashed pink line. The top panel shows the situation for all RCSLenS pointings whereas in the lower panel the 9\% of pointings with the largest cross-correlation signal have been excluded.}
\label{fig:syst_test}
\end{figure}

Each individual simulation yields an estimate of the residual cross-correlation and the whole set yields a distribution of the amplitude of this cross-correlation function. This amplitude is parametrised by the quantity $\Delta \xi_{\rm obs}$ as defined in \cite{2012MNRAS.427..146H}. By design, the PSF correction is perfect in the simulations. The distribution of the amplitude of the sum of the star-galaxy cross-correlation function at zero lag from the mocks is shown in Fig.~\ref{fig:syst_test} by the solid line. The pink, dashed line shows the contribution from cosmic shear alone. The amplitude measured on the RCSLenS data is shown by the shaded region. It is the sum of $\Delta \xi_{\rm obs}$ over a sample of fields, hence the label $\Sigma(\Delta \xi_{\rm obs})$. If all fields are included the data show an amplitude that is roughly a factor of two higher than the mean amplitude in the simulations. 

Looking at individual pointings in the data it is clear that some particular pointings show very strong signals with a skewed distribution with a long tail towards large amplitudes. Consecutively rejecting the pointings with the largest amplitudes quickly lowers the average signal. A rejection of just 9\% of the pointings leads to an amplitude that is consistent with the expectations from the simulations. We define this set of pointings as our ``pass fields'' and those rejected as ``fail fields''.

It is important to note that we do not simply cut the wings of a Gaussian distribution when rejecting the fields with the highest amplitude. The distribution of amplitudes in the data is highly non-Gaussian and the rejection scheme targets such non-Gaussian outliers.

It is also necessary to stress that the cosmic shear contribution to the star-galaxy cross-correlation function due to chance alignments between the cosmic shear field and the PSF is much smaller (by a factor of $\ga5$) than the other terms, i.e. the contributions due to chance alignments between the intrinsic ellipticities and the measurement noise with the PSF. Hence rejecting pointings in this way does not preferentially reject area with a larger overall cosmic shear signal. This is important to avoid biasing cosmological measurements when only using the pass fields.

\begin{figure}
\centering
\includegraphics[width=0.7\hsize]{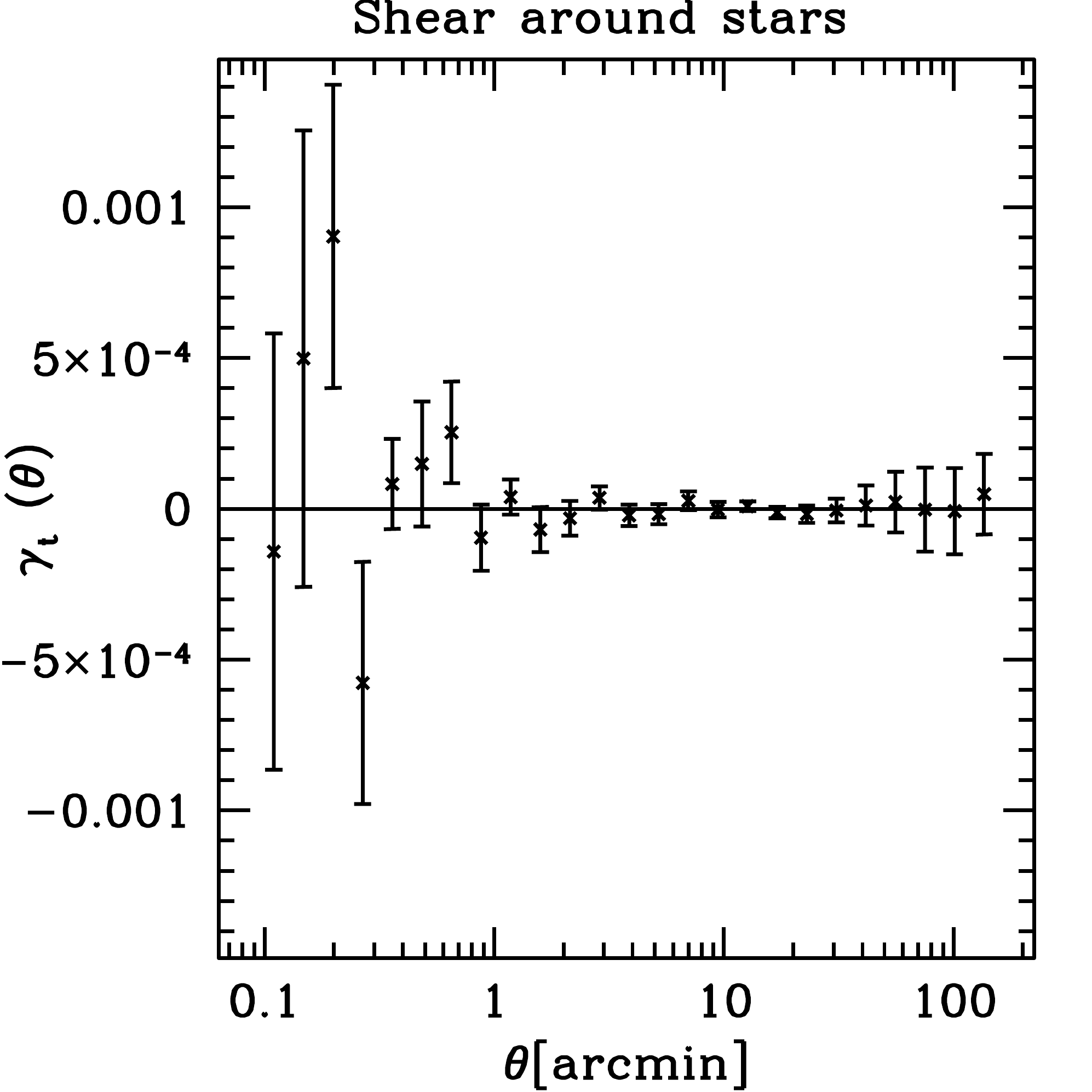}
\caption{Tangential shear of background galaxies around stars as a function of angular separation. A pure star sample is selected by the SG\_FLAG criterion (see Sect.~\ref{sec:SG_sep}) whereas galaxies are selected by \emph{lens}fit (weight$>0$), and a magnitude selection of $19<r<24$ is applied for both samples.}
\label{fig:shear_stars}
\end{figure}

As an additional test we also cross-correlate the star positions with the galaxy ellipticities. The tangential alignment of galaxies with respect to the positions of stars is shown in Fig.~\ref{fig:shear_stars}. The signal is consistent with zero on all relevant angular scales, after the signal around random points has been subtracted. This procedure is common in galaxy-galaxy-lensing measurements to account for a spatially varying additive shear correction without the need to correct for a $c$-term (Sect.~\ref{sec:noise_bias}).

\section{Dark matter maps}
\label{sec:DM_maps}
\begin{figure*}
\includegraphics[valign=t,scale=0.33]{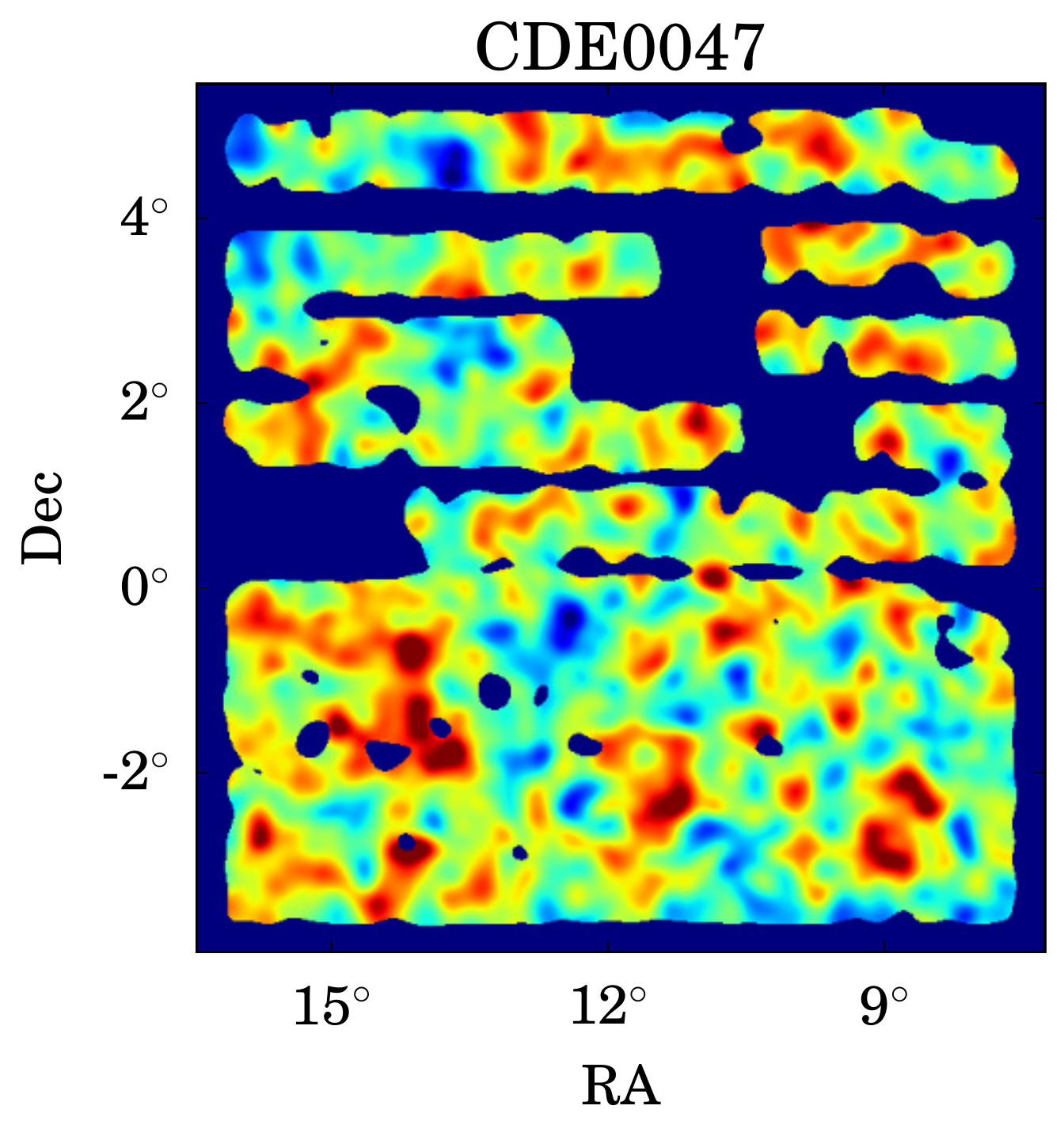}
\includegraphics[valign=t,scale=0.33]{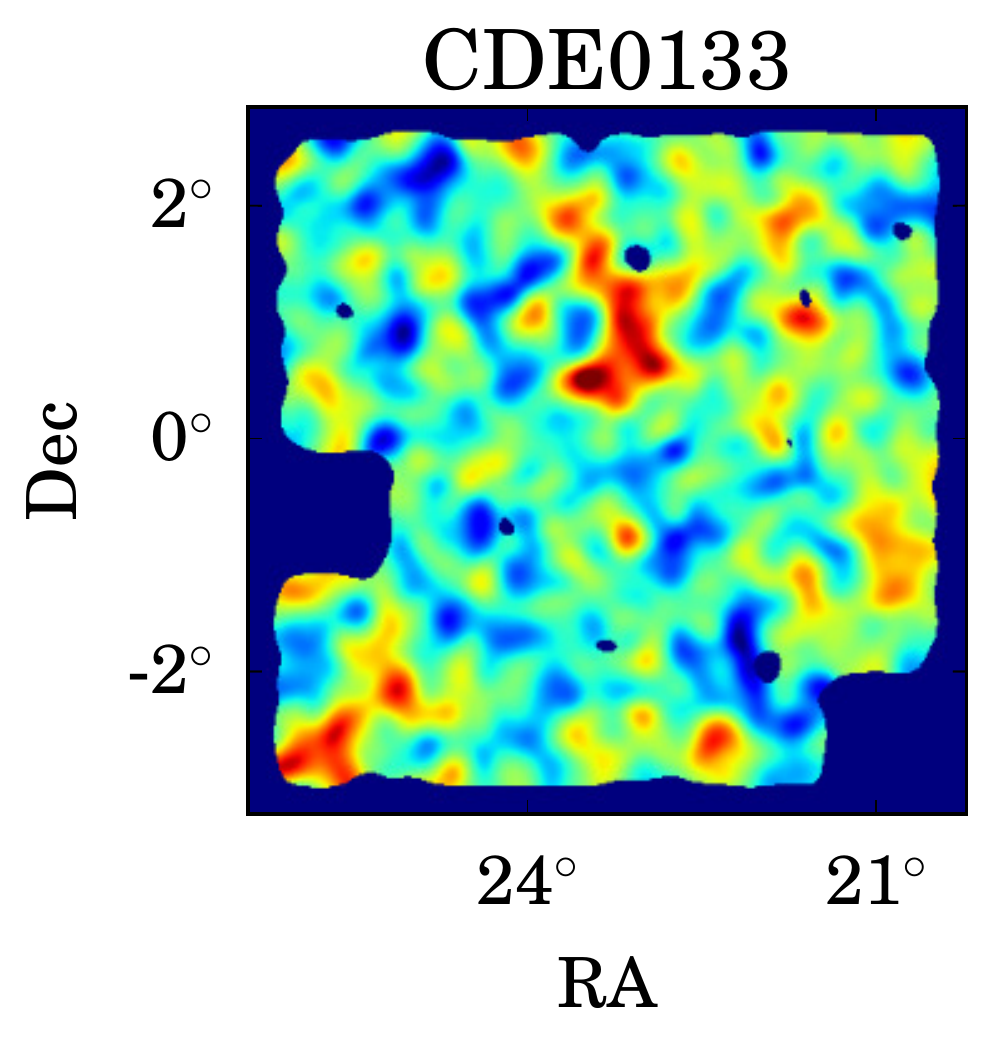}
\includegraphics[valign=t,scale=0.33]{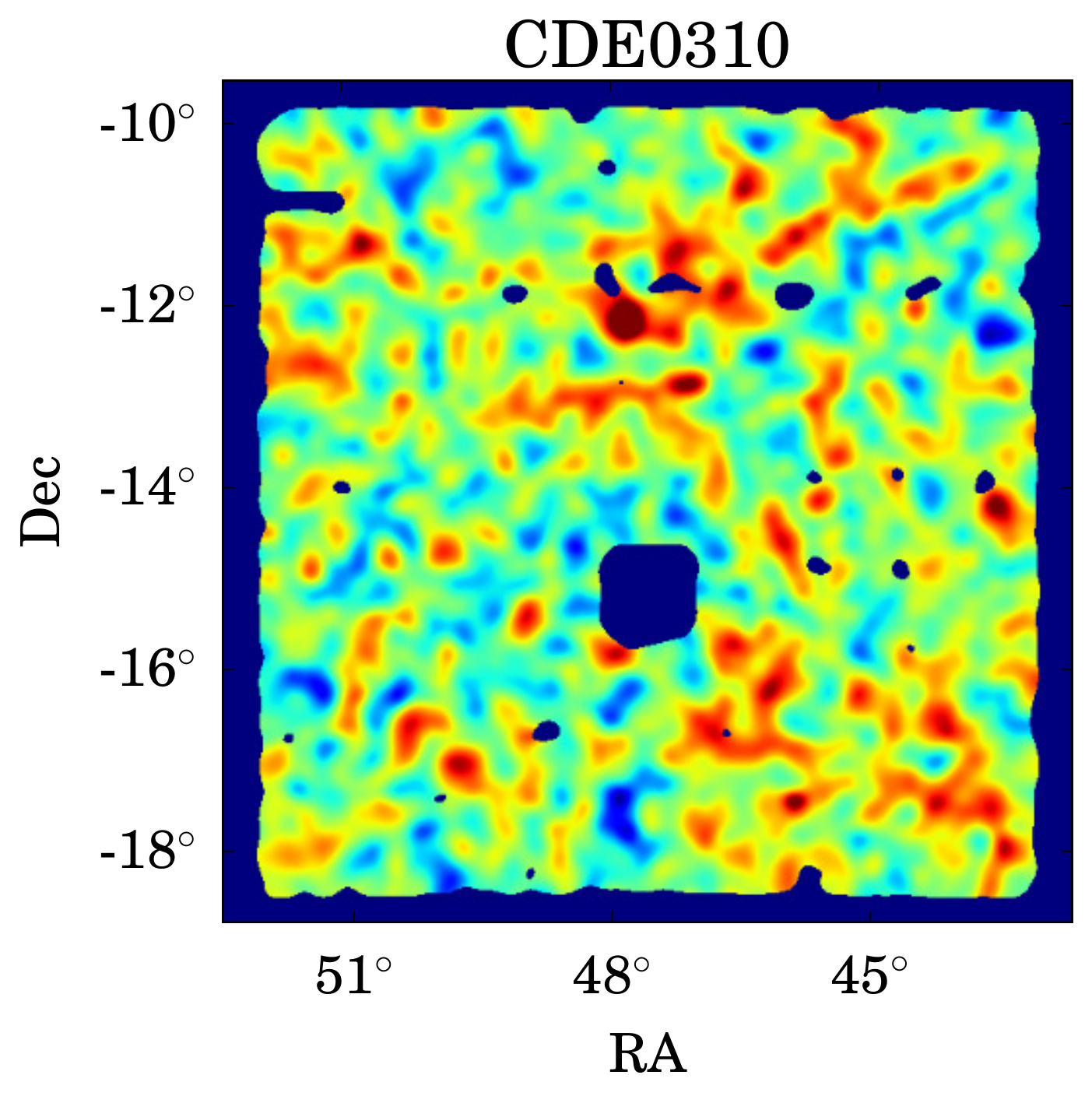}
\includegraphics[valign=t,scale=0.33]{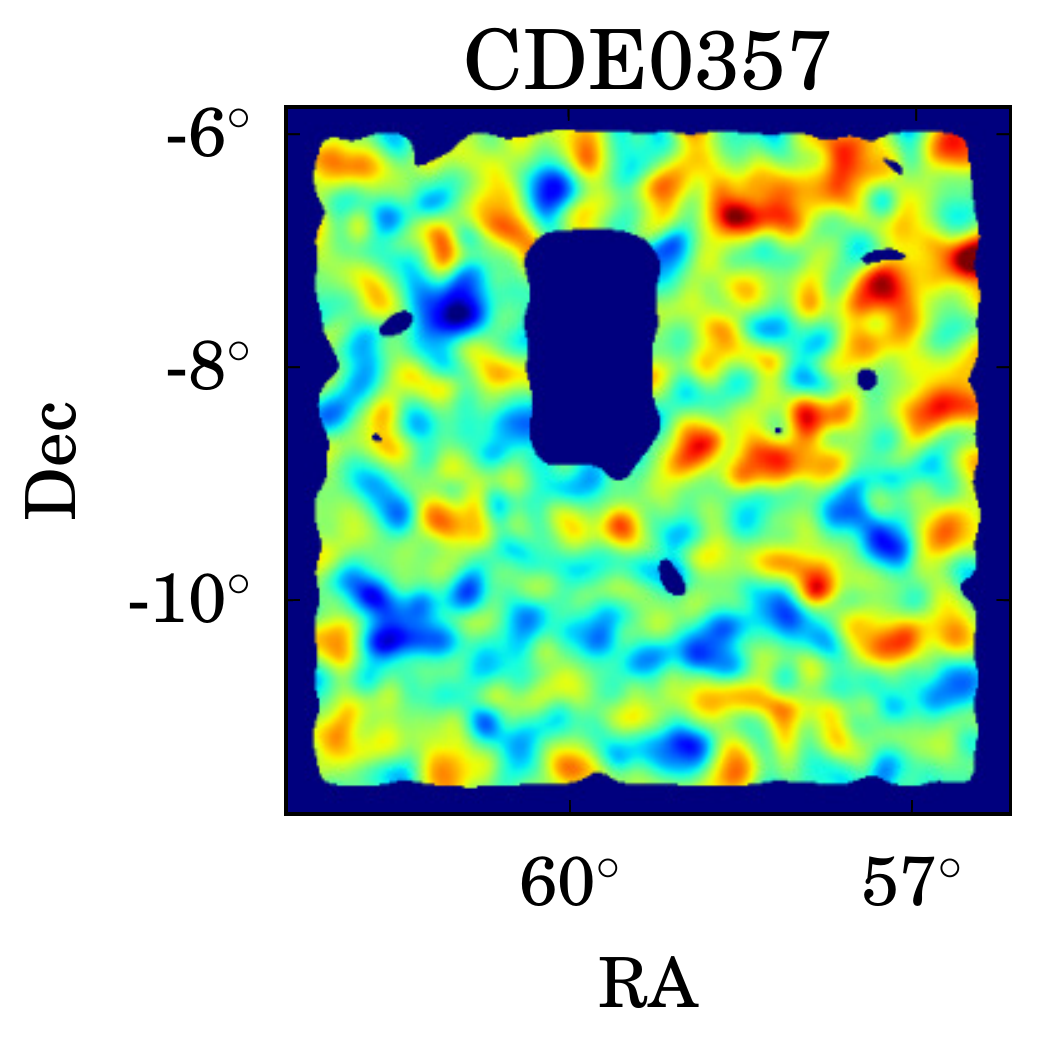}
\includegraphics[valign=t,scale=0.33]{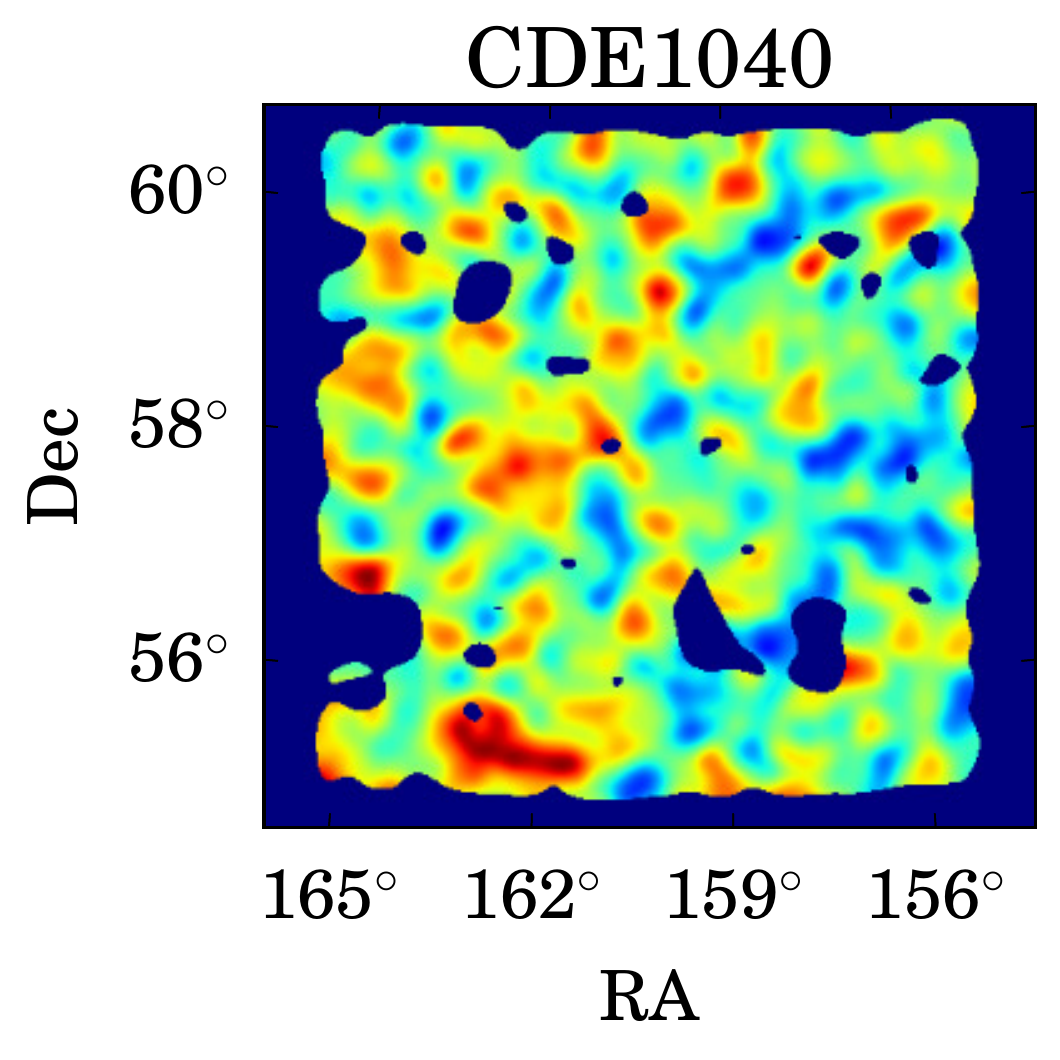}
\includegraphics[valign=t,scale=0.33]{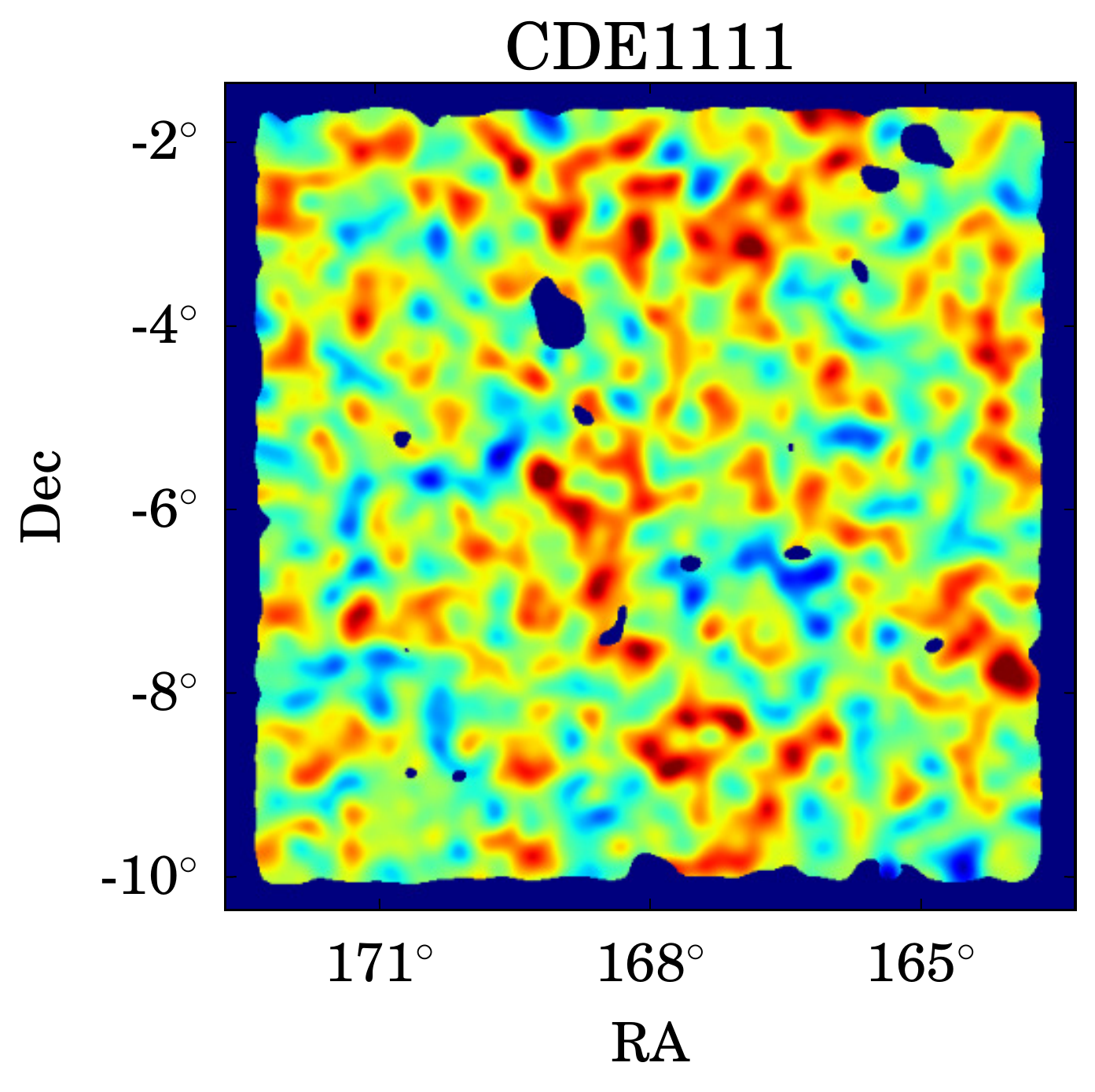}
\includegraphics[valign=t,scale=0.33]{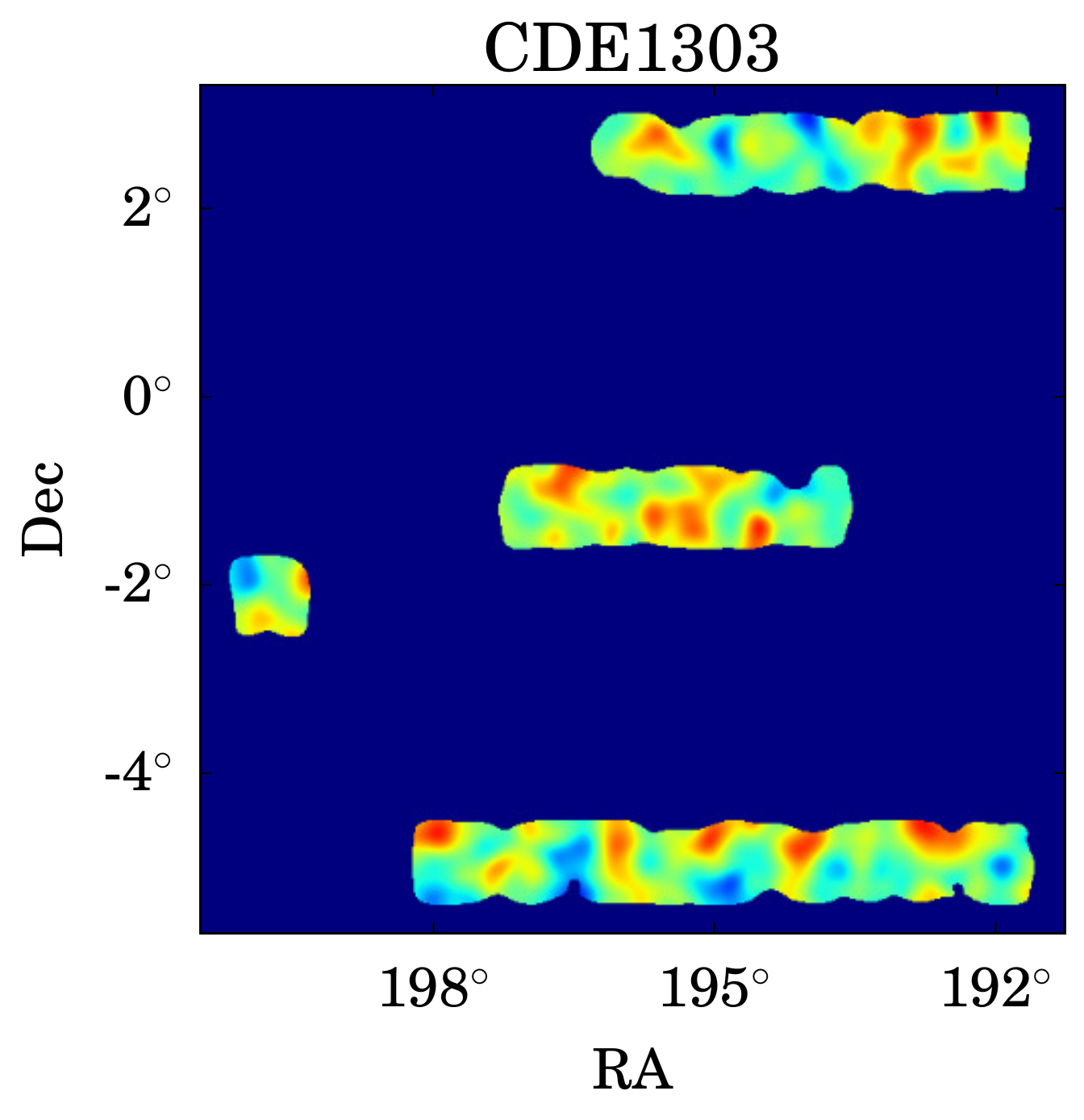}
\includegraphics[valign=t,scale=0.33]{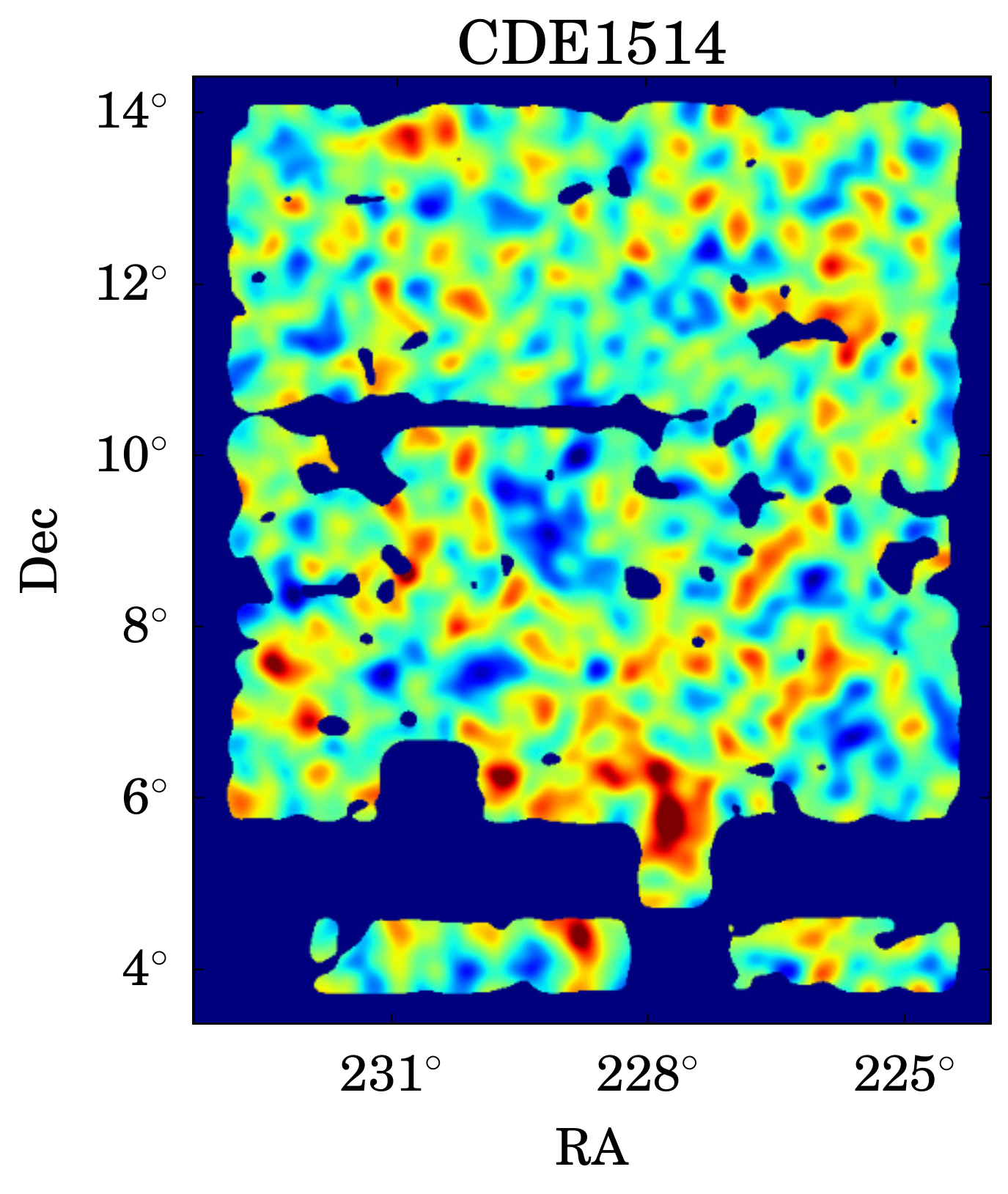}
\includegraphics[valign=t,scale=0.33]{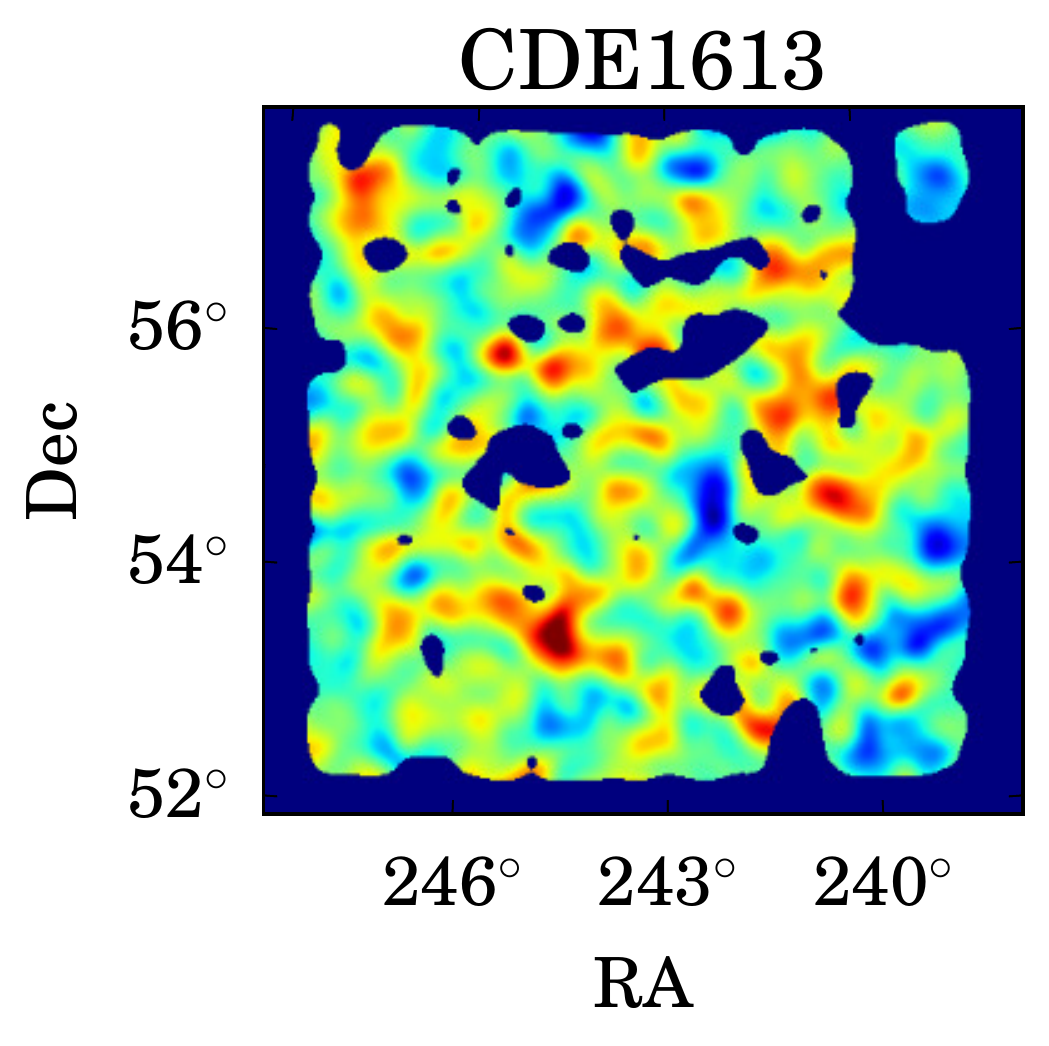}
\includegraphics[valign=t,scale=0.33]{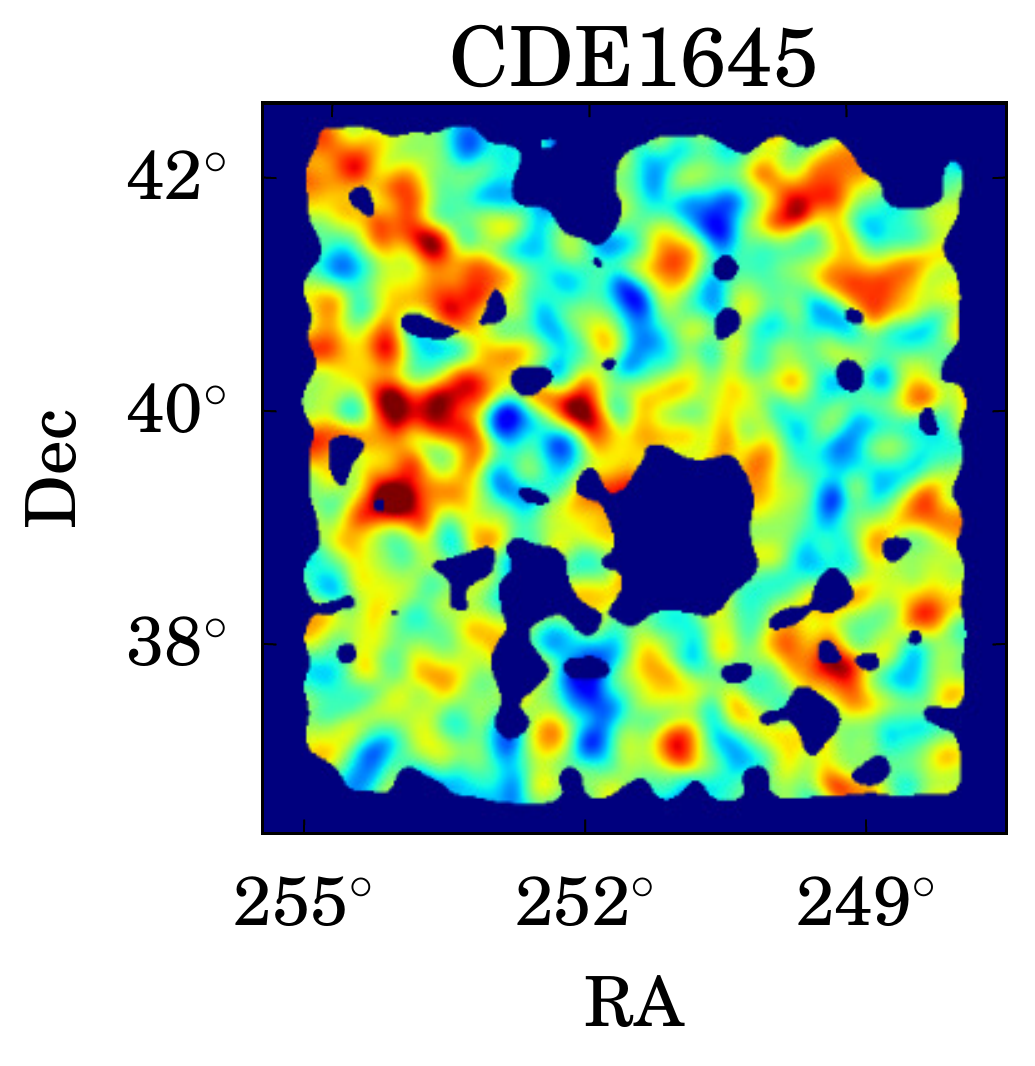}
\includegraphics[valign=t,scale=0.33]{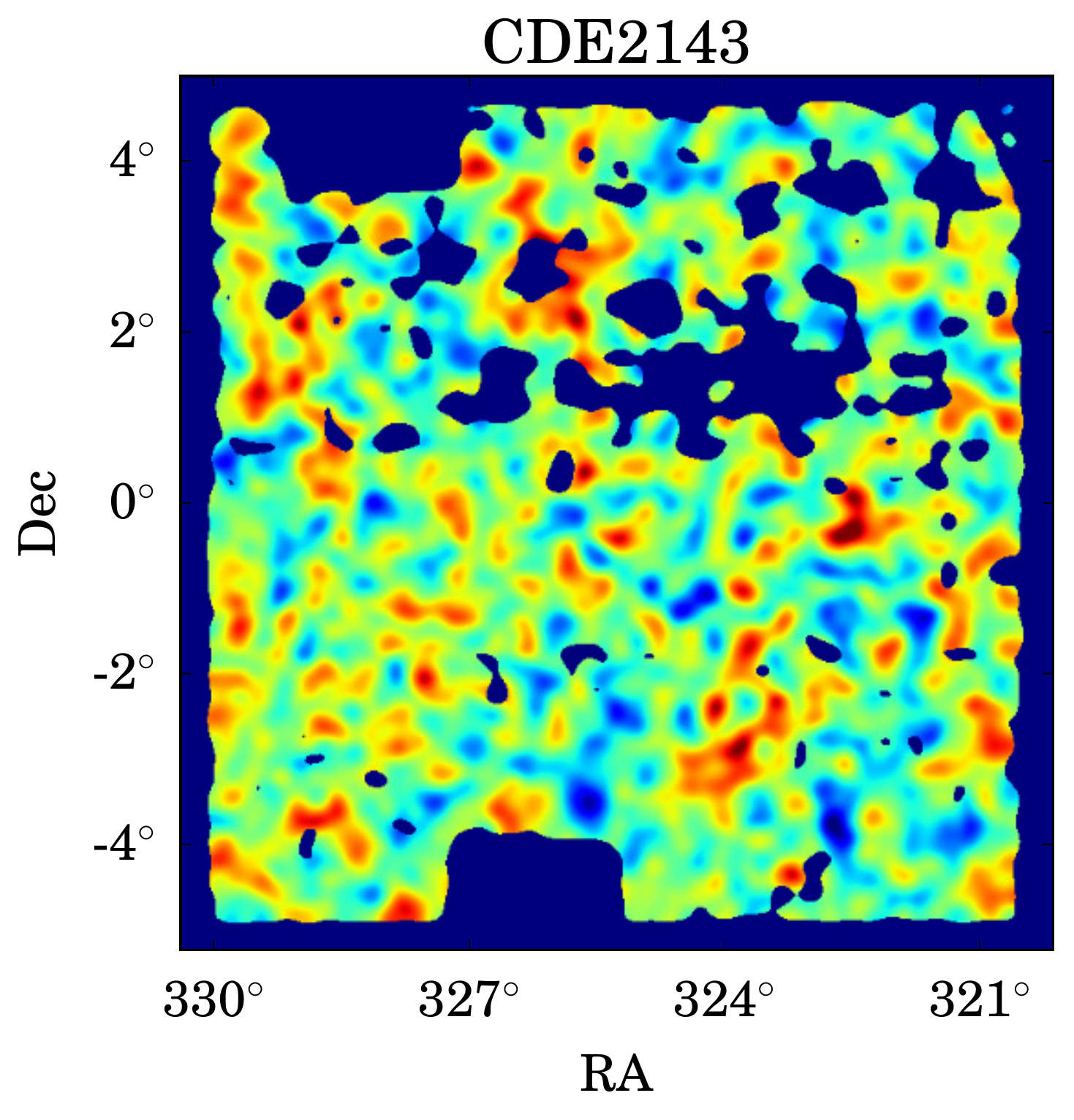}\\
\includegraphics[valign=t,scale=0.33]{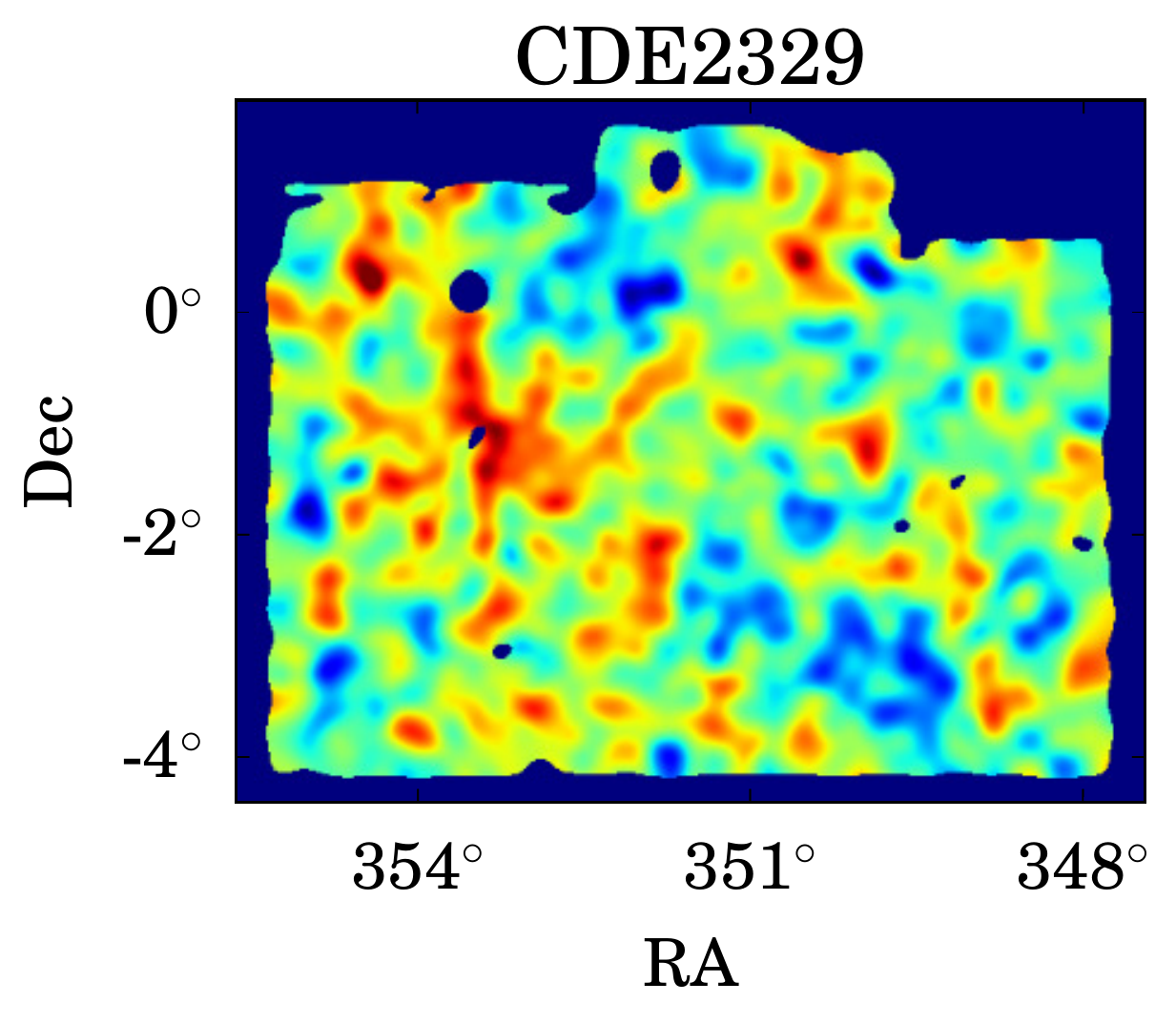}
\includegraphics[valign=t,scale=0.33]{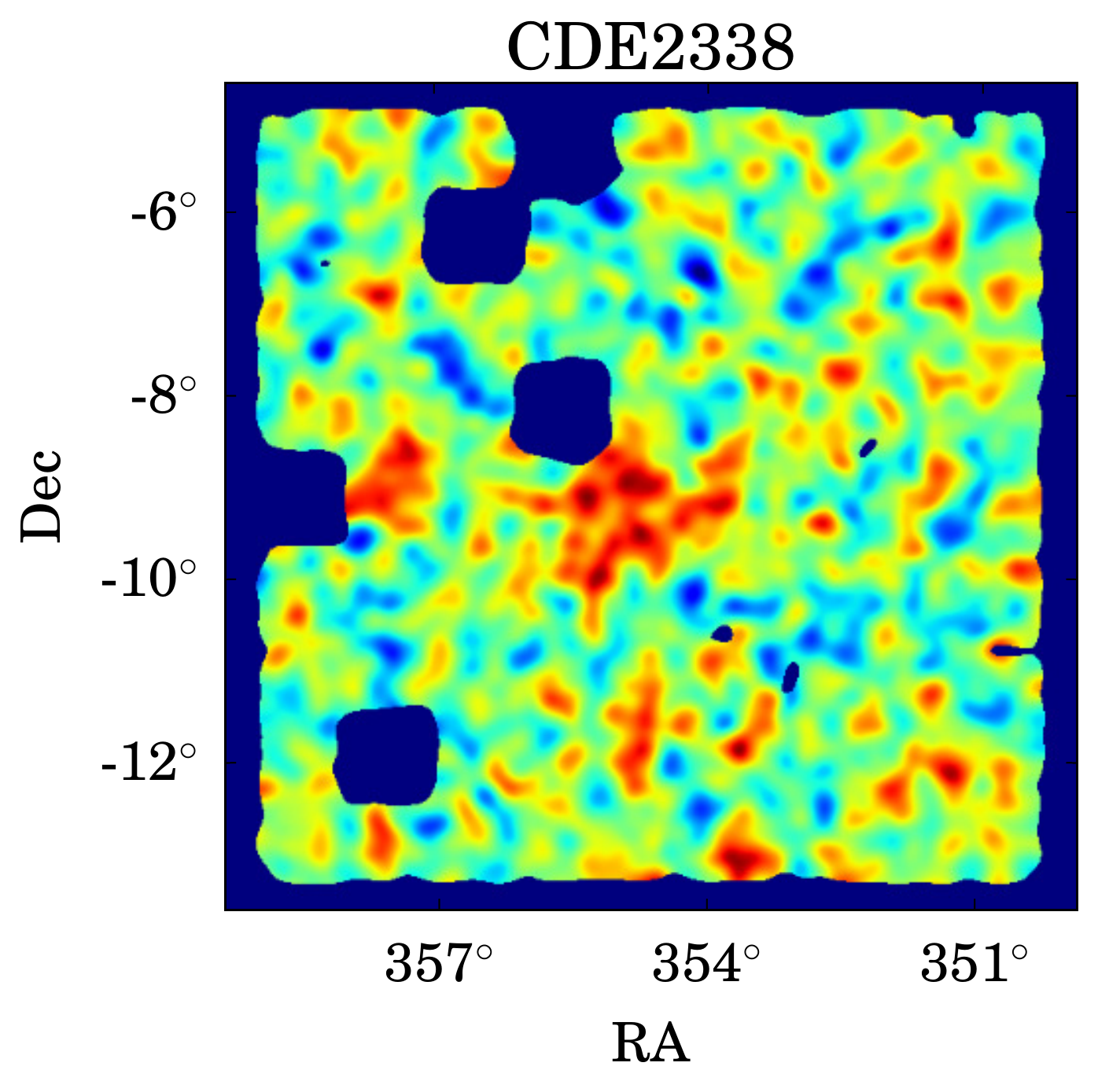}
\includegraphics[valign=t,scale=0.33]{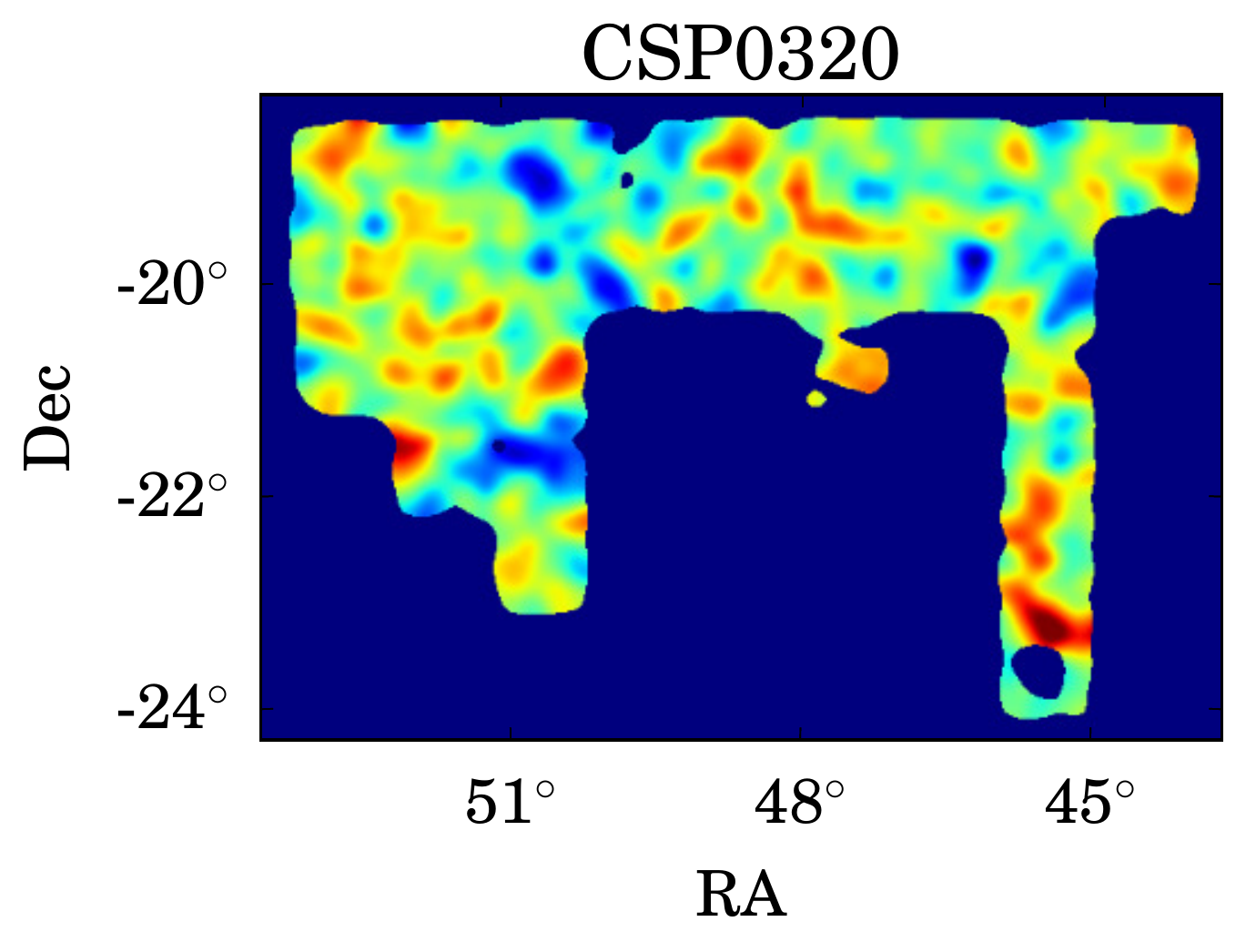}\\
\vspace{-10cm}
\hfill
\includegraphics[valign=t,scale=0.4]{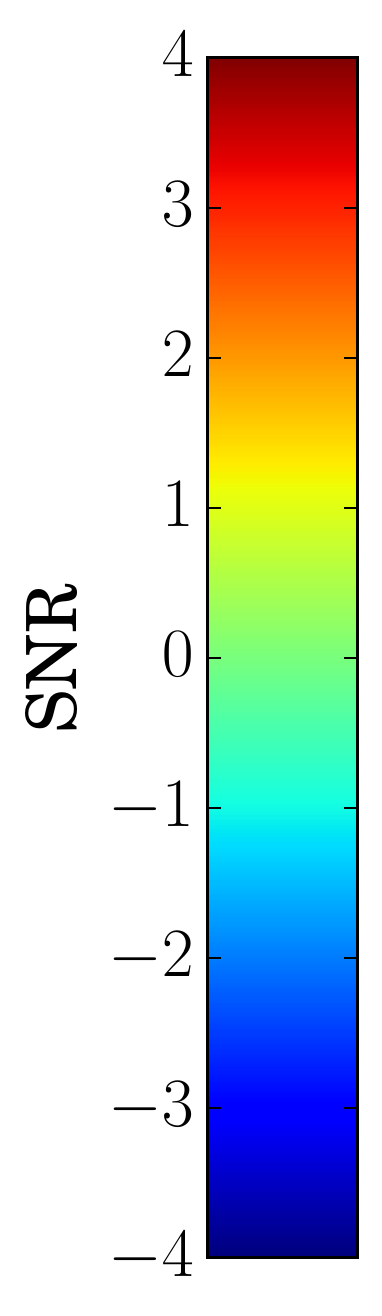}\\\vspace{4.7cm}
\caption{Weak lensing mass reconstructions of the 14 RCSLenS patches. Shown are signal-to-noise maps of the E-mode component of the shear field smoothed on a scale of $\sim$16$'$. The SNR scale of $-4<{\rm SNR}<4$ roughly corresponds to $-2.5\%<\kappa<2.5\%$ depending slightly on position and patch. The total reconstructed area is 571.7\,deg$^2$}
\label{fig:DM_maps}
\end{figure*}

As an application of the RCSLenS shear catalogue we present weak lensing mass reconstructions of the full 761 fields. The creation of these maps follows closely the approach described in \cite{2013MNRAS.433.3373V}. In Fig.~\ref{fig:DM_maps} signal-to-noise maps for the 14 RCSLenS patches smoothed at a scale of $\sim$16$'$ are shown. These maps correspond to the E-modes in the shear field. The noise level is estimated from randomising the galaxy ellipticities in many realisations.
 
\begin{figure}
\centering
\includegraphics[width=0.8\hsize]{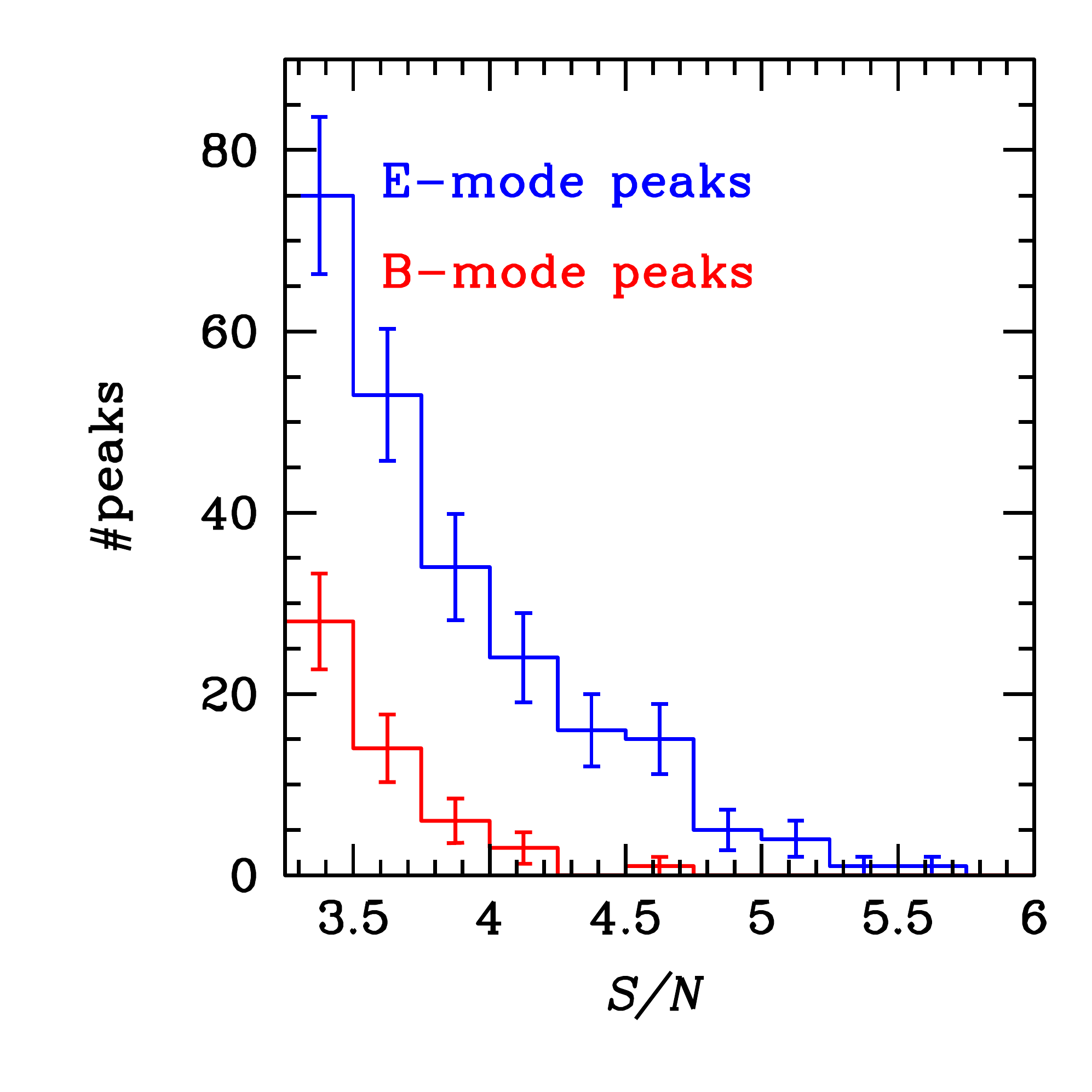}
\caption{Distribution of peaks as a function of signal-to-noise ratio in the E-mode (red) and the B-mode maps (blue). The smoothing scale is the same as in Fig.~\ref{fig:DM_maps}, i.e. $\sim16'$.}
\label{fig:peaks}
\end{figure}

Similar maps are created from a catalogue where all galaxies have been rotated by 45 degrees. Those maps correspond to B-modes in the shear field and should ideally be consistent with noise. We show a quick sanity check based on the number of peaks found in the E- and B-mode maps. In Fig.~\ref{fig:peaks} the distribution of peaks as a function of their significance is shown. At high significance (SNR$>3$) the number of E-mode peaks greatly exceeds the number of B-mode peaks. This qualitative result is further analysed in a more quantitative way in the next section.

The mass maps are used in several forthcoming cross-correlation studies. \cite{2016arXivHarnois} use the unprecedented statistical power of RCSLenS in combination with the lensing re-constructions from the Planck cosmic microwave background maps \citep{2015arXiv150201591P} to study large-scale-structure and cosmology. Hojjati et al. (in preparation) look at a cross-correlation between the RCSLenS mass maps and the tSZ signal from the Planck data \citep{2015arXiv150201596P} revealing new insights about the missing baryons in the warm hot intergalactic medium.

\begin{figure*}
\centering
\includegraphics[width=0.45\hsize]{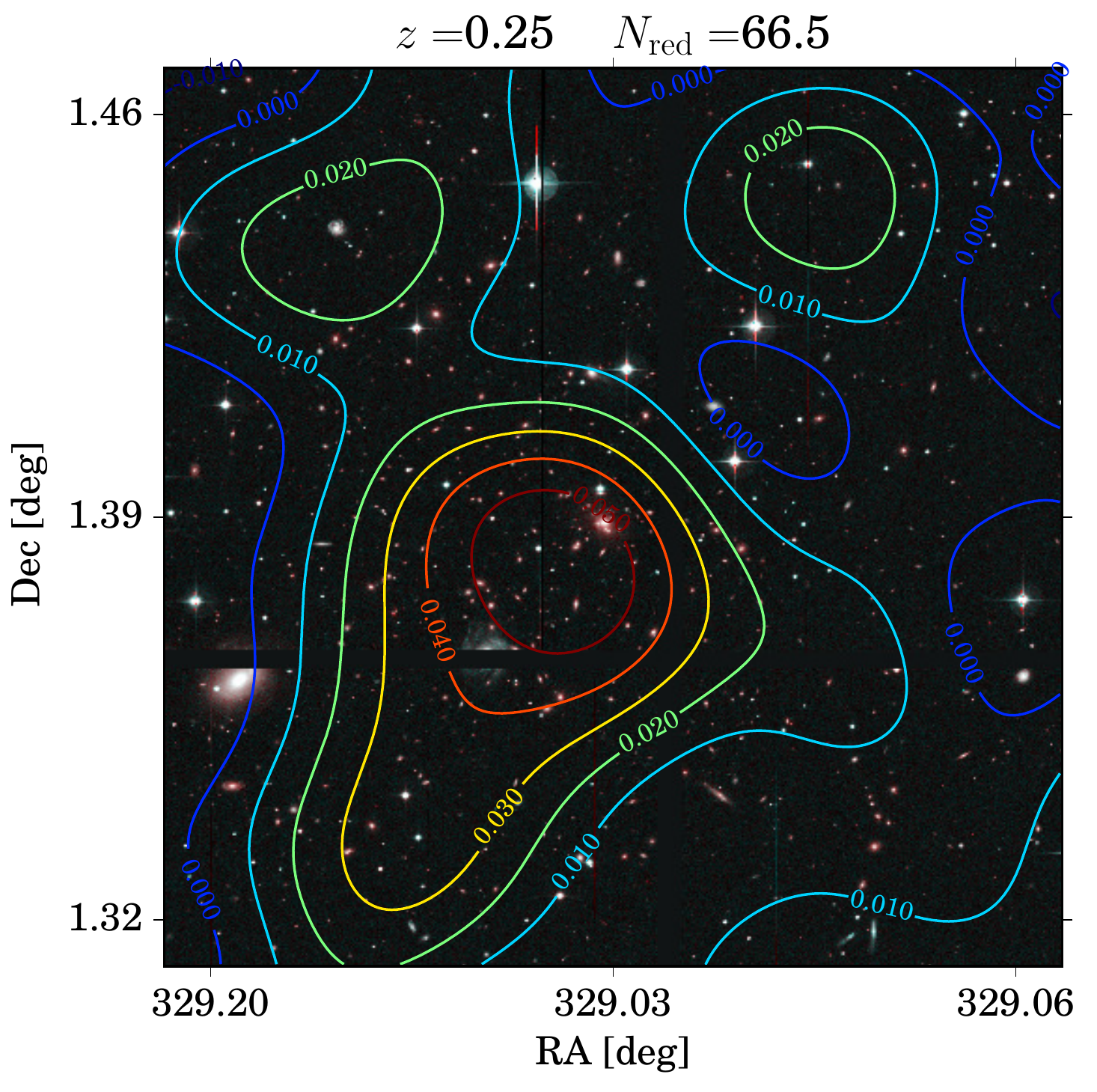}
\includegraphics[width=0.45\hsize]{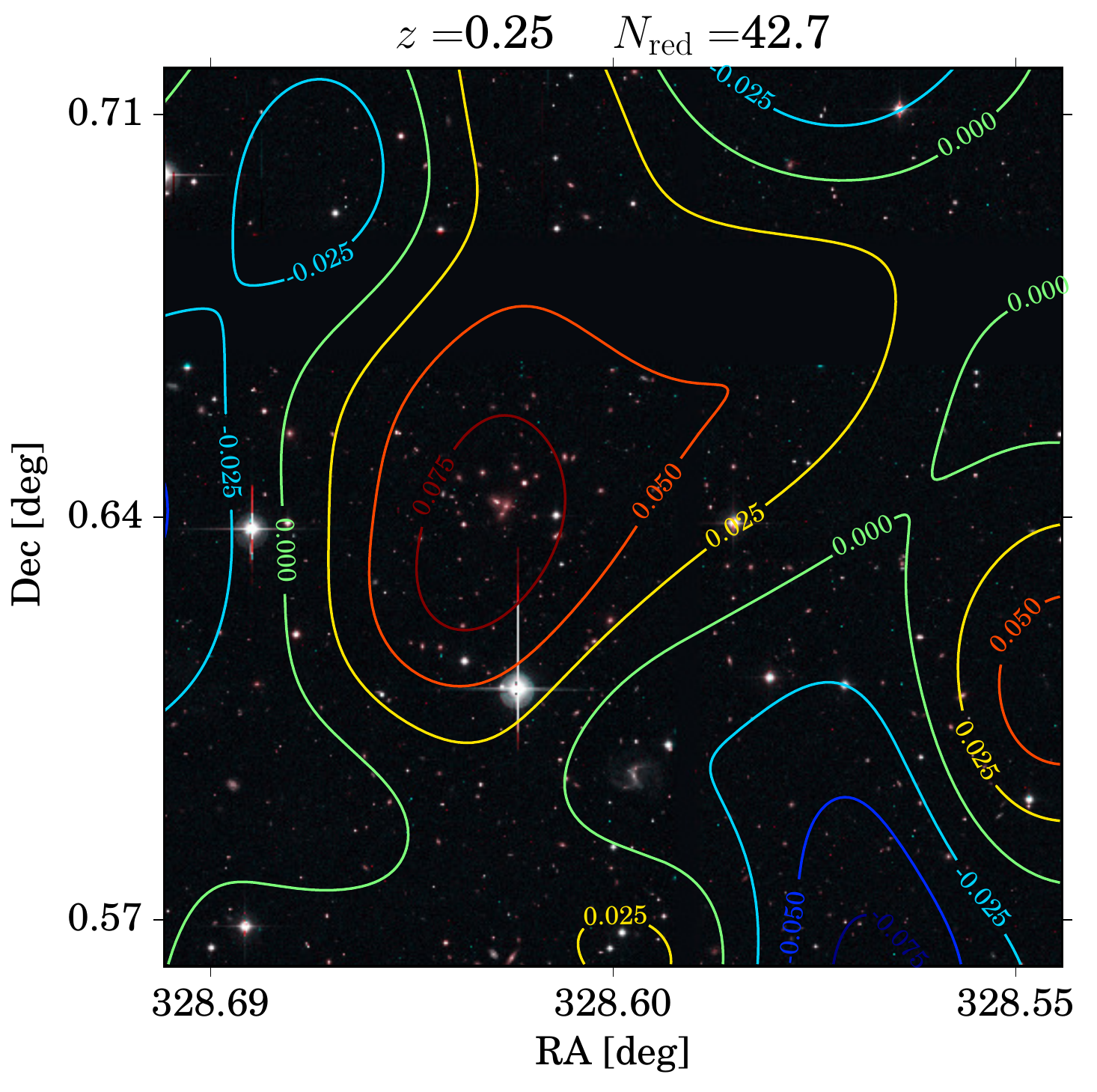}
\includegraphics[width=0.45\hsize]{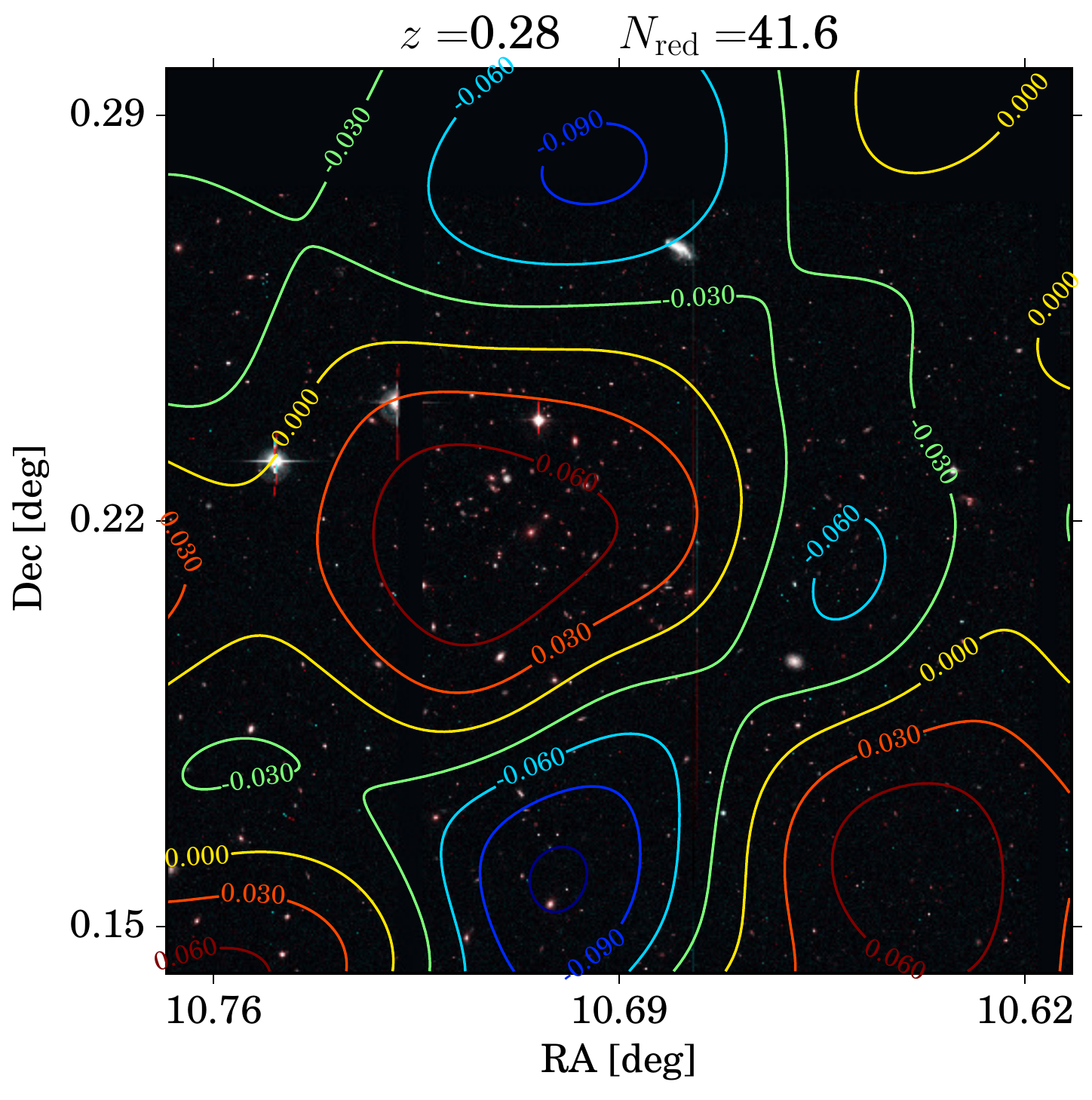}
\includegraphics[width=0.45\hsize]{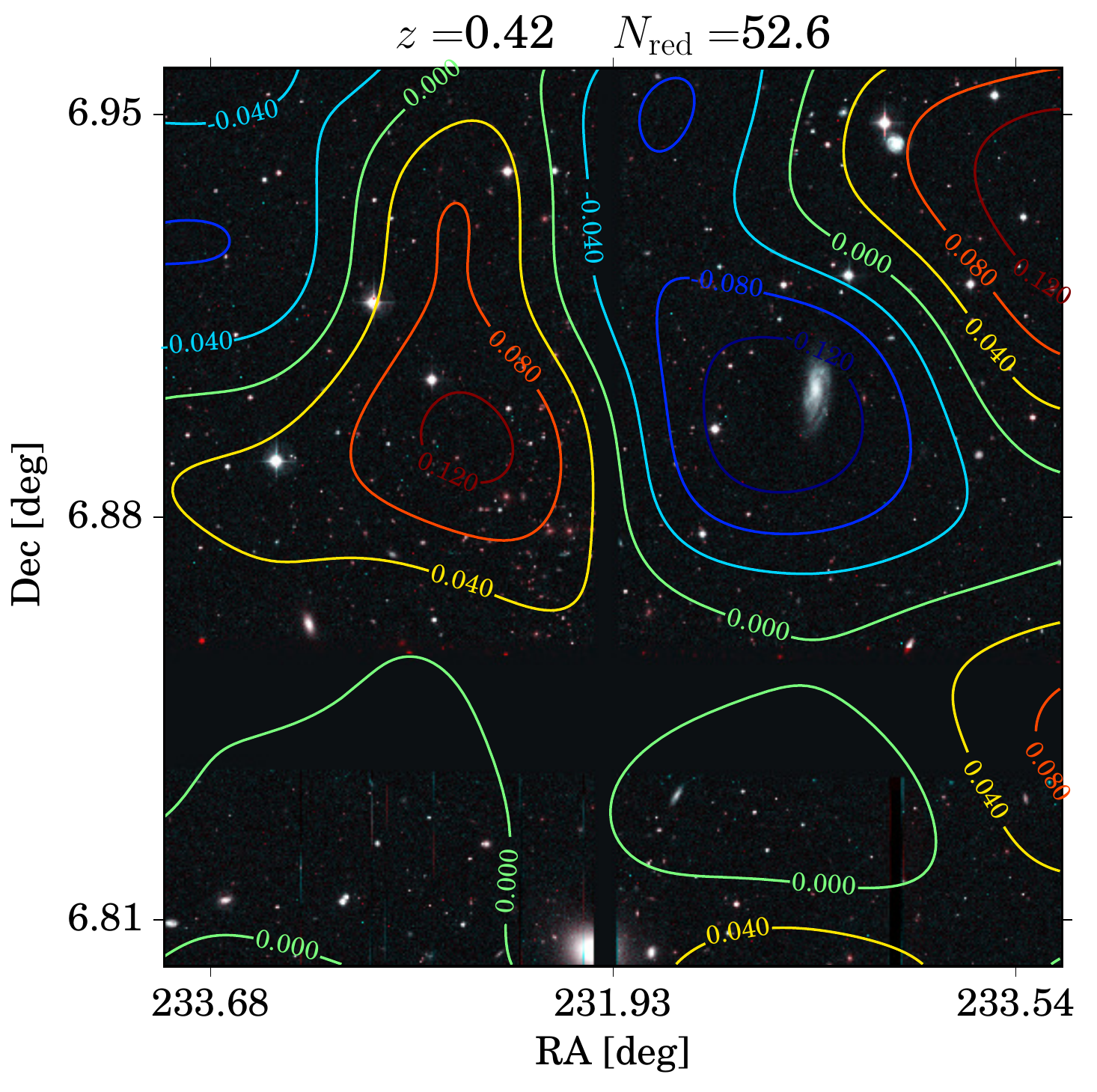}
\caption{Examples of rich, low-$z$ clusters from the RCS2 cluster catalogue. Shown are three-colour composite images based on $gri$-data. Contours show a $\kappa$ reconstruction with the actual $\kappa$ values indicated by the numbers. The smoothing scale is $\sim1'$ here. Redshift and richness estimates for each cluster are reported above the panels.}
\label{fig:clusters}
\end{figure*}

Some smaller-scale examples of cluster mass reconstructions with the \cite{1993ApJ...404..441K} method are shown in Fig.~\ref{fig:clusters}. Cluster positions are based on the RCS2 cluster catalogue \citep[see][]{2011AJ....141...94G,2015arXiv150603817V} from which we chose a few rich low-$z$ clusters for illustration. Unlike the large-scale mass maps, where all galaxies were used as sources, here we only use objects whose photo-$z$ estimate is greater than the redshift estimate for the cluster (based on its red-sequence).

\section{Residual B-modes}
\label{sec:Bmode}
\begin{figure}
\includegraphics[width=\hsize]{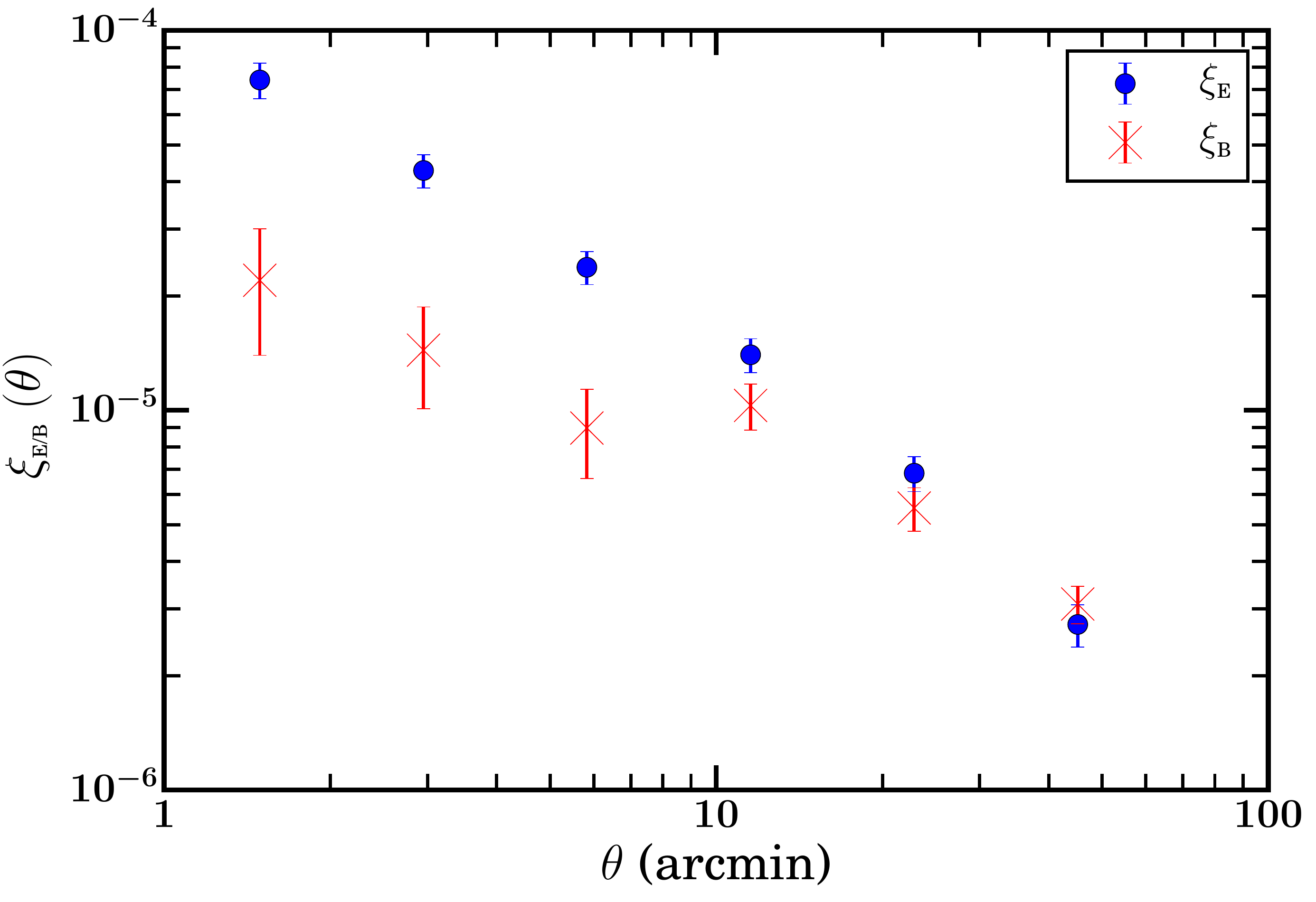}
\caption{\label{fig:xi_EB}Two-point correlation functions $\xi_{\rm E}$ (blue circles) and $\xi_{\rm B}$ (red crosses) for the full RCSLenS shear catalogue.}
\end{figure}

In the absence of residual systematics, the scalar nature of the gravitational potential leads to a vanishing B-mode pattern in the shear field. As one of the systematic checks we investigate the level of residual B-modes in the data by estimating the two-point correlation functions $\xi_{\rm E}$ and $\xi_{\rm B}$ \citep[see e.g. Eq.~27 of][]{kuijken/etal:2015}.\footnote{The integrals over $\xi_-$required to calculate $\xi_{\rm E}$ and $\xi_{\rm B}$ extend to infinite angular scales. Here we use theoretical estimates for scales $>2$deg to calculate those integrals.} Those are shown in Fig.~\ref{fig:xi_EB} for a sample of galaxies selected from the pass fields. A significant B-mode signal is detected at all angular scales that becomes comparable to the E-mode at a scale of $\theta\ga1$deg.

Additionally we also check the convergence B-mode signal from the B-mode convergence maps described above. Consistency of $\kappa_{\rm B}$ with zero after subtraction of noise (estimated from 100 noise realisations) is a way to check that the residual systematics are negligible. The results of the $\kappa_{\rm E/B}$ two-point correlation functions are very similar to the findings with $\xi_{\rm E/B}$ mentioned above. There are residual B-modes in the RCSLenS data and their strength differs from patch to patch. Independent analysis of projected 3D shear power spectra \citep[for an application of this technique to CFHTLenS see][]{2014MNRAS.442.1326K} also confirms the presence of an excess residual B-mode signal consistent with that found from $\xi_{\rm E/B}$ and $\kappa_{\rm E/B}$. Considering the significance of the additive systematic shear measurement bias found in Sect.~\ref{sec:noise_bias} this is perhaps unsurprising. Until the origin of this systematic is understood and resolved we have to conclude that the RCSLenS data in the current form are not suitable for accurate cosmic shear measurements or more generally for measurements that correlate shear with shear.

We find that the cross-correlation between the E- and B-modes for the different probes ($\xi_{\rm E/B}$, $\kappa_{E/B}$, power spectra) is always consistent with zero, but this is not a sufficient condition to conclude that the E-mode maps (or the shear catalogues) are free from systematics. Fortunately, the RCSLenS lensing data set can be used for cross-correlation studies: the requirement for cross-correlation studies is far less restrictive than for auto-correlation studies. The only requirement is that the residual systematics do not cross-correlate with the external data set. This is the case for any residual systematics originating from the PSF anisotropy. The residual systematics that cross-correlate with the shear and hence an external data set is the shear calibration "m". The latter is shown to be small \citep{2013MNRAS.429.2858M}. Galaxy-galaxy lensing \citep{2016MNRAS.456.2806B} and CMB-galaxy lensing \citep{2016arXivHarnois} studies show, that the cross-correlation signal, i.e. the E-mode with an external data set, is consistent with the predictions, while the B-modes cross-correlated with the external data sets are consistent with zero. We are therefore confident that the RCSLenS data can be used for cross-correlation studies, even though they cannot be used for auto-correlation studies.

\section{Summary}
\label{sec:summary}
The RCSLenS project applied the methods from CFHTLenS to the RCS2 data set. A data reduction of all $griz$-band images with the THELI pipeline is followed up by shape measurements with the \emph{lens}fit code and photo-$z$ estimates with BPZ. 

Multiplicative and additive biases in the shear measurements are analysed and calibrated. In this process we find a more complicated behaviour of the data compared to CFHTLenS. In particular, the additive bias depends strongly on the Strehl ratio of the PSF, and we need a two-stage calibration scheme to remove the bias to tolerable levels. We attribute these problems to the fact that RCS2 is a single-exposure survey which results in stronger systematics. While a sophisticated analysis of the ellipticity cross-correlation of stars and galaxies yields very encouraging results (only 8\% of the data have to be rejected), a subsequent analysis of B-mode patterns in the shear field reveals significant residual systematics. At this point we have to conclude that this data set cannot be analysed with the techniques presented here in a way that completely removes the B-modes. However, as the E-modes are uncorrelated with the B-modes the data set is still extremely valuable for cross-correlation studies with other data sets that are not based on the same shear catalogue.

The photometry is calibrated against SDSS where available and with stellar locus regression otherwise. This results in a homogeneous data set that shows a residual RMS scatter of 3-5\% with respect to SDSS after calibration. We attribute this to a varying zeropoint over the mosaic that is not completely removed in the ELIXIR pre-reduction (note that this is an old version of ELIXIR that is superseded now). Photometric redshifts are well-behaved in the range of $0.4<z_{\rm phot}<1.1$ with large degeneracies outside this range.

We show dark matter maps for the full data set reconstructed from the measured ellipticities of background galaxies. These maps represent the largest area that has been mapped hitherto in this way by a weak lensing mass re-construction on scales of up to ten degrees.

The amount of large-scale B-modes in the RCSLenS data - while comparable to many previous surveys \citep[e.g.][]{2007MNRAS.381..702B} - prevents the use of this data set for precision cosmic shear science or an auto-correlation analysis of the E-mode maps at the moment. The shear-shear (or $\kappa$-$\kappa$) cross-correlation amplifies these systematics and we are not confident in the results from such analyses. However, different cross-correlation analyses of these shear measurements with other non-shear probes have been carried out and shown to be free of systematics:
\begin{itemize}
\item \cite{2016MNRAS.456.2806B} used the RCSLenS galaxy-galaxy-lensing signal around BOSS and WiggleZ lenses in combination with their redshift-space distortion signal to constrain the gravitational slip $E_G$ finding consistency with general relativity.
\item \cite{2015arXiv151203627K} applied the shear ratio method around RCS2 clusters to come up with a cosmology independent measurement of the geometry of the Universe out to $z\sim1$.
\item \cite{2016MNRAS.456.3886B} used the same galaxy-galaxy-lensing and clustering signals of BOSS galaxies that were already used in \cite{2016MNRAS.456.2806B} to measure their galaxy bias with a new data compression scheme.
\item \cite{2016arXivHarnois} cross-correlated the RCSLenS $\kappa$ maps with $\kappa$ from Planck CMB lensing to test structure formation through these combined probes with unprecedented precision.
\end{itemize}
The data presented here are ideal for such cross-correlation studies between independent data sets given their large volume and otherwise high quality. Therefore, we make the data public in a similar way as the CFHTLenS data to further enable studies with this data set that represents the largest public weak-lensing survey to date.

In order to trigger further use of RCSLenS the data are being released to the public via CADC in a very similar fashion to the CFHTLenS data \citep{2013MNRAS.433.2545E}. Details of the released data are covered in appendix~\ref{sec:release}.

\section*{Acknowledgements}
Matthias Bartelmann acted as the external blind setter for RCSLenS. We would like to thank him for his great help in this crucial area.

We are grateful to the RCS2 team for planning the survey, applying for observing time, and conducting the observations. We acknowledge use of the Canadian Astronomy Data Centre operated by the Dominion Astrophysical Observatory for the National Research Council of Canada's Herzberg Institute of Astrophysics. 

We would like to thank Mischa Schirmer for his work on the THELI pipeline.

HH is supported by an Emmy Noether grant (No. HI 1495/2-1) of the Deutsche Forschungsgemeinschaft. TE is supported by the BMBF through project "GAVO III" and by the DFG through project ER 327/3-1 and the TR 33.  LVW is supported by NSERC and CIfAR. AC and CH acknowledges support from the European Research Council under the EC FP7 grant numbers 240185 and 647112. AH is supported by an NSERC postdoctoral fellowship. BJ is supported by an STFC Ernest Rutherford Fellowship, grant reference ST/J004421/1. CB acknowledges the support of the Australian Research Council through the award of a Future Fellowship. JHD received support from NSERC and from a Marie Sklodoska-Curie Fellowship (Grant 656869). TK is supports by a Royal Society URF. RN acknowledges support from the German Federal Ministry for Economic Affairs and Energy (BMWi) provided via DLR under project no.50QE1103. MV acknowledges support from the European Research Council under FP7 grant number 279396 and the Netherlands Organisation for Scientific Research (NWO) through grants 614.001.103. NJ was funded by EPCC.

Computations for the N-body simulations were performed on the GPC supercomputer at the SciNet HPC Consortium. SciNet is funded by: the Canada Foundation for Innovation under the auspices of Compute Canada; the Government of Ontario; Ontario Research Fund - Research Excellence; and the University of Toronto.

For data processing we used EDIM1, a machine built for data-intensive research as a joint collaboration between EPCC \& Informatics at the University of Edinburgh.

\small

{\bf Author contributions:}
H. Hildebrandt led this paper and manages the RCSLenS collaboration. A. Choi oversaw the data analysis and catalogue production for RCSLenS. C. Heymans led the analysis of shear systematics.

The remaining co-authors are split into three alphabetical groups. The first group represents the RCSLenS core team who spent a significant amount of time on technical development work for the project. C. Blake ran many tests for systematic errors on the data. T. Erben reduced the complete RCS2 data set with the THELI pipeline. L. Miller developed and adapted the \emph{lens}fit algorithm. R. Nakajima implemented the sanity checks, the masking, and the stellar locus regression. L. van Waerbeke produced the mass maps. M. Viola implemented and tested different algorithms for the star galaxy separation.

The second group represents the extended RCSLenS team who participated in the scientific exploitation of the data set and/or the technical development. The third group are external collaborateurs who organised the data release at CADC (S. Gwyn), produced colour images of the RCS2 clusters (A. Tudorica), provided a key piece of sotware (K. Kuijken, Z. Sheikhbahaee), designed and conducted the RCS2 observations (H.~K.~C. Yee), and helped with the implementation of the data processing on HPC facilities (N.~Johnson).

\normalsize

\bibliographystyle{mn2e_mod}

\bibliography{RCSLenS_data_paper}

\appendix

\section{Layout of the RCSLenS patches}
\label{sec:app_layout}
\begin{figure*}
\includegraphics[valign=t,width=0.29\textwidth]{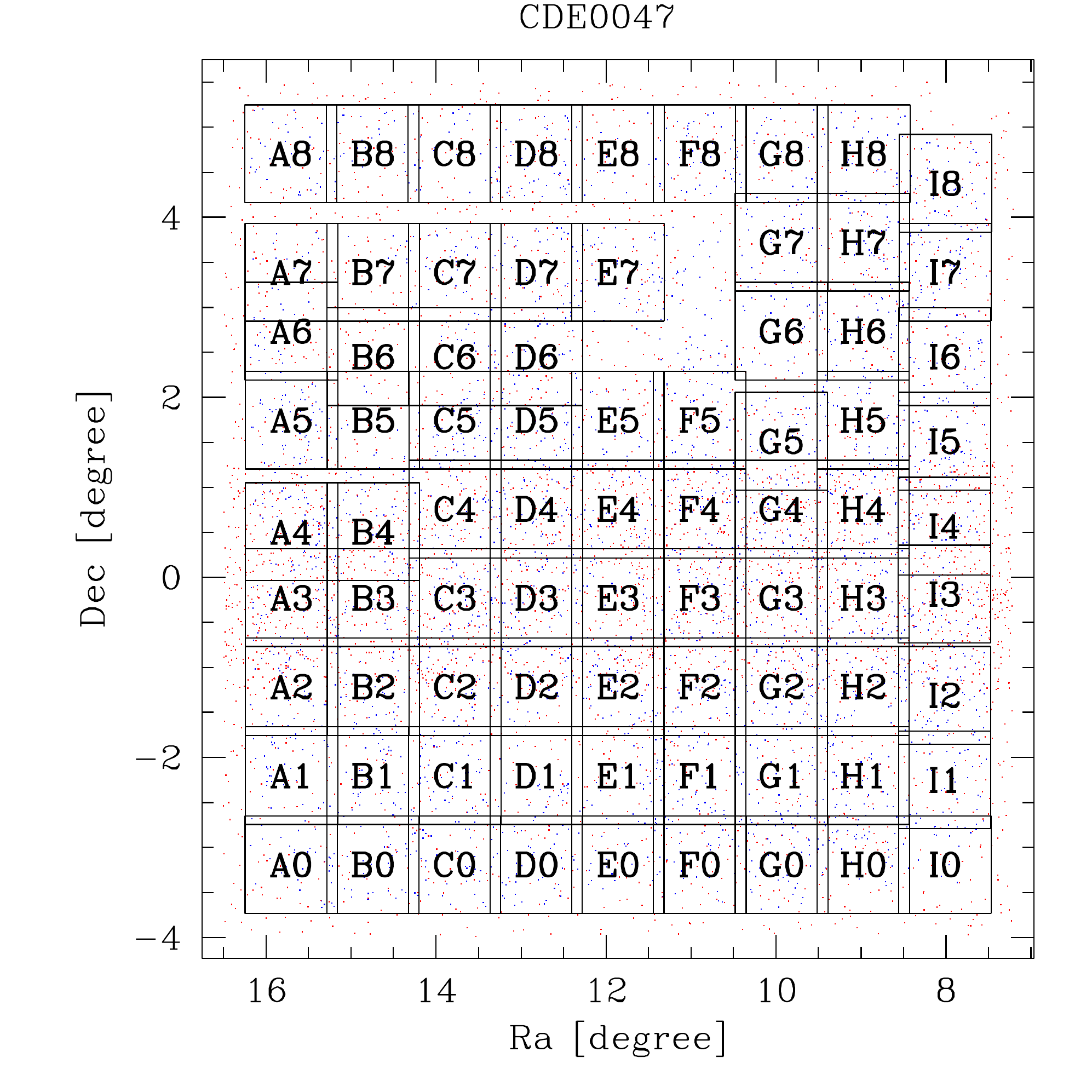}\,
\includegraphics[valign=t,width=0.20\textwidth]{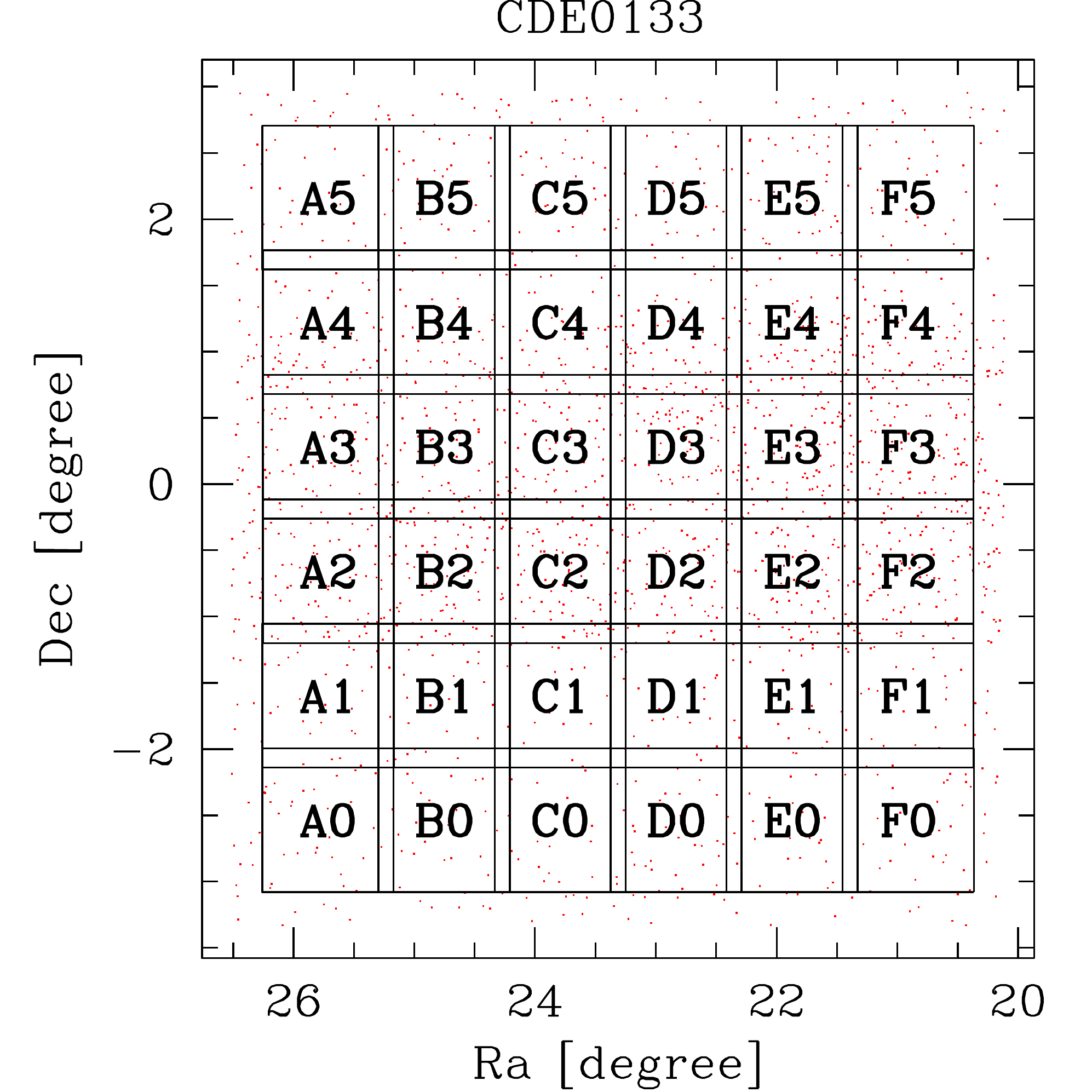}\,
\includegraphics[valign=t,width=0.29\textwidth]{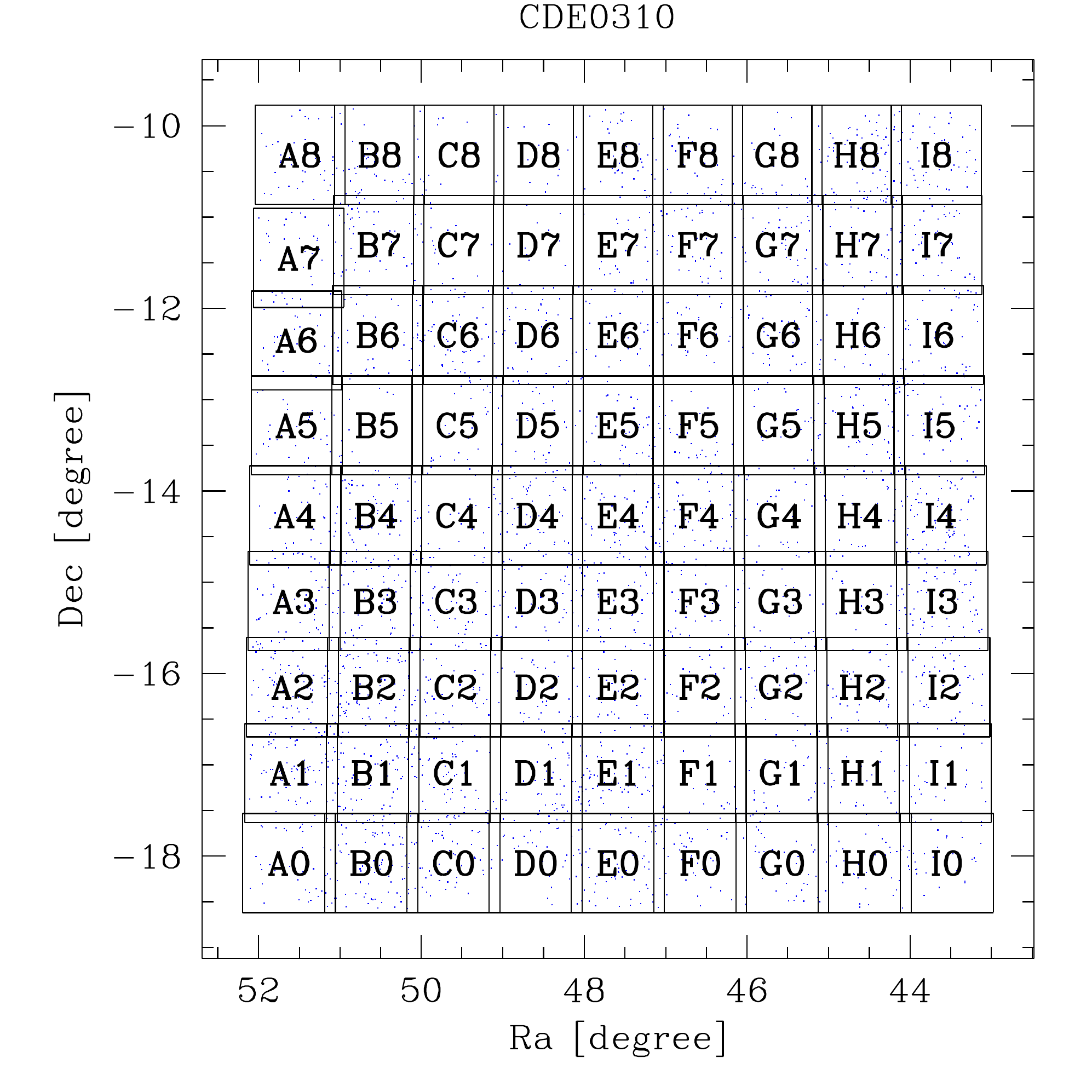}\,
\includegraphics[valign=t,width=0.20\textwidth]{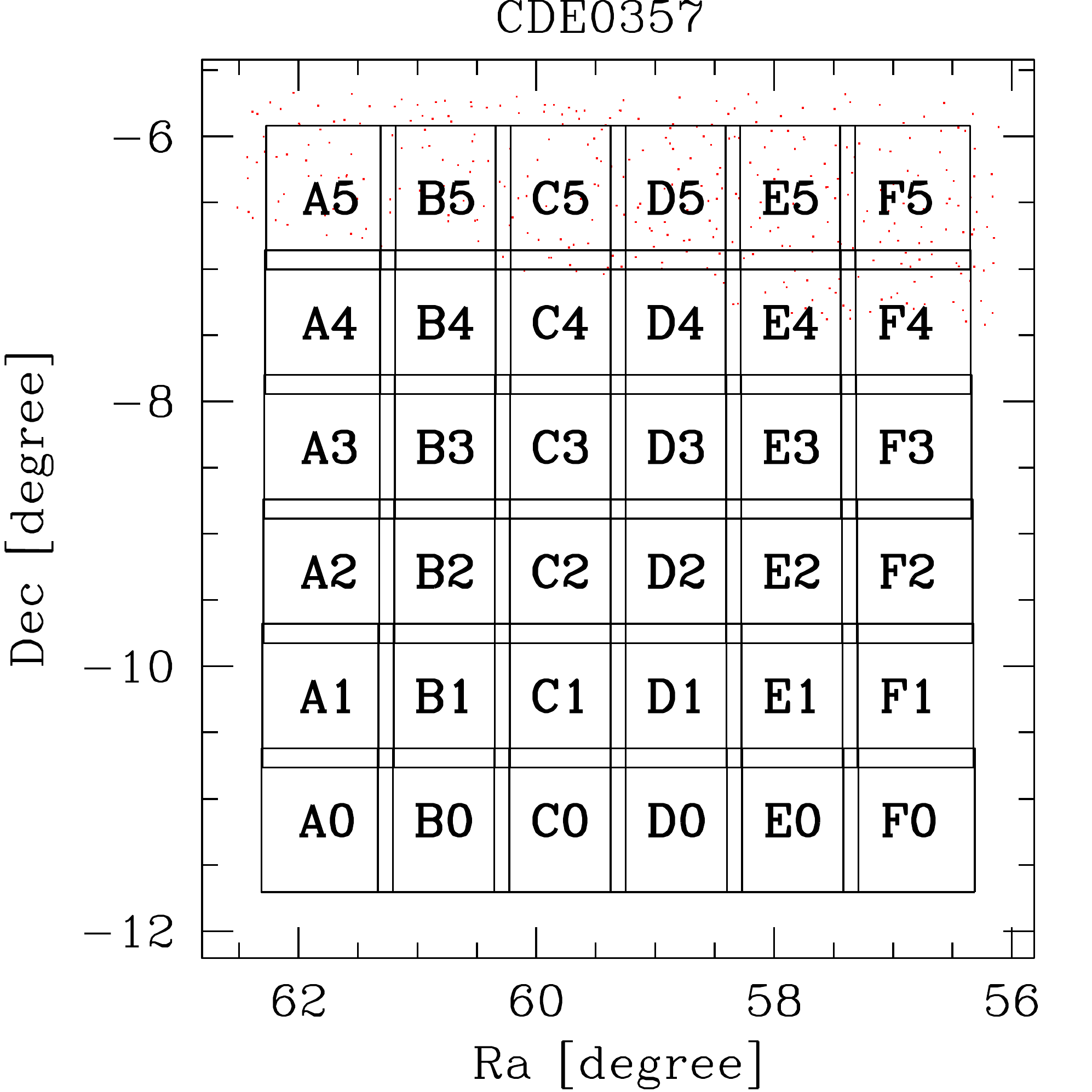}\\\vspace{0.5cm}
\includegraphics[valign=t,width=0.20\textwidth]{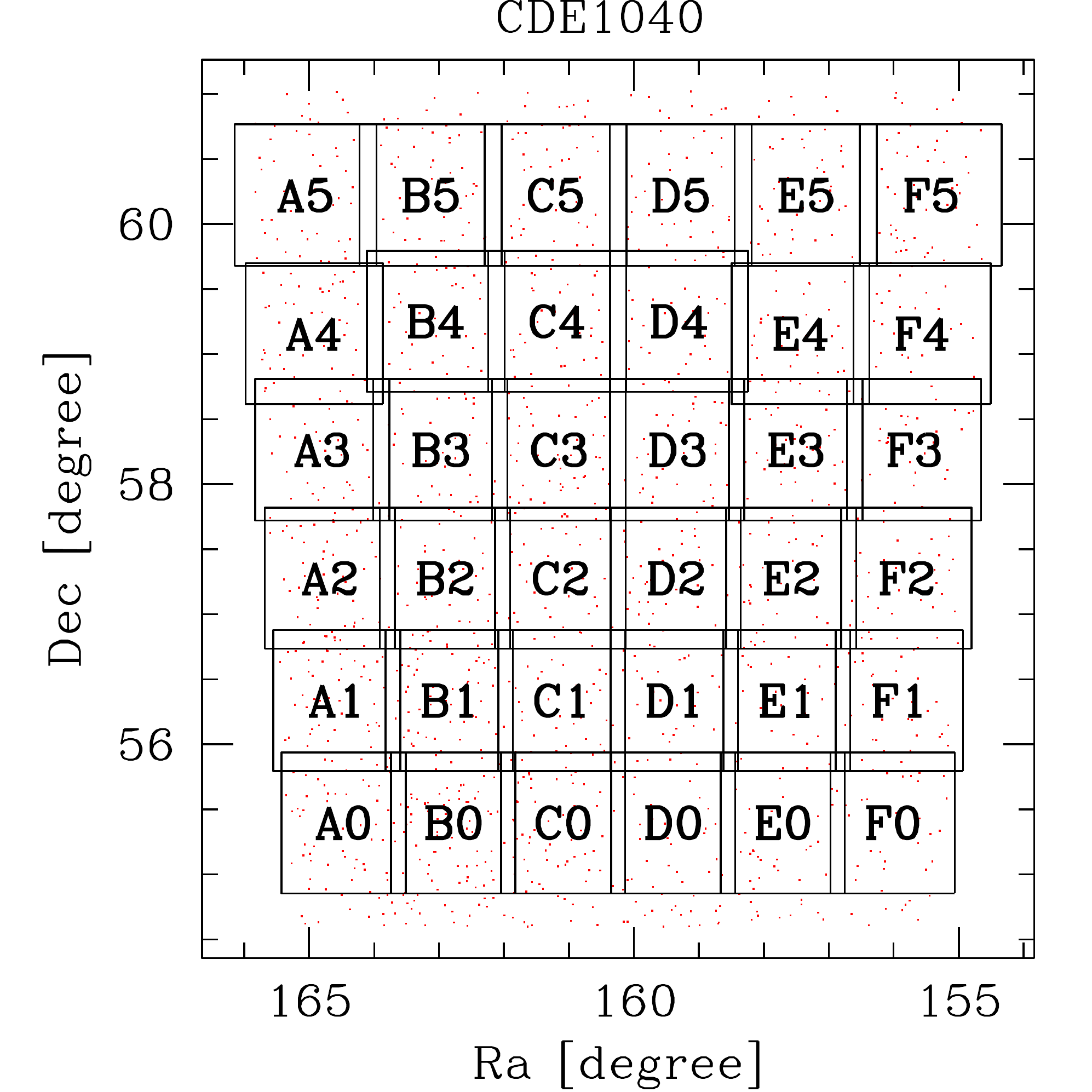}\,
\includegraphics[valign=t,width=0.29\textwidth]{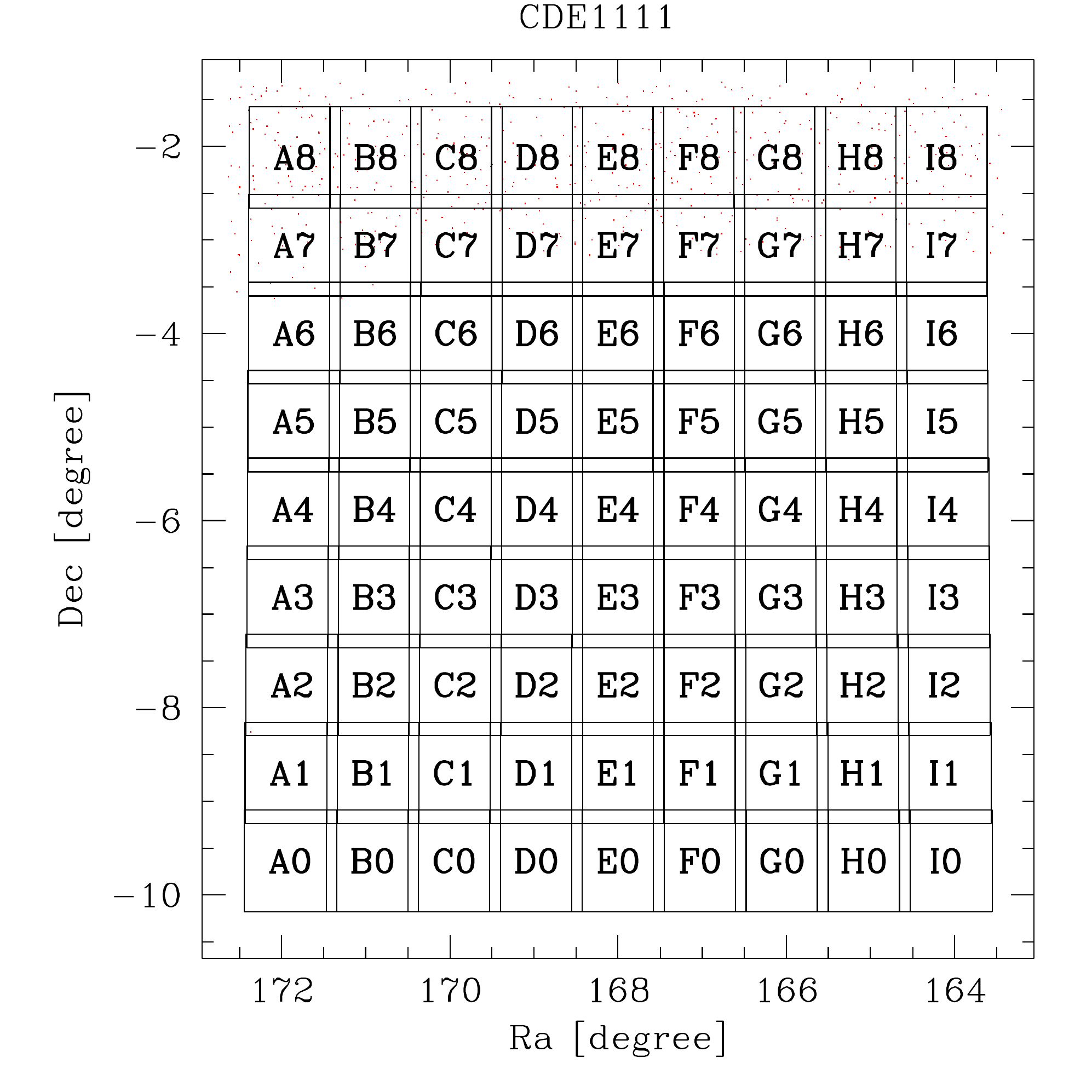}\,
\includegraphics[valign=t,width=0.20\textwidth]{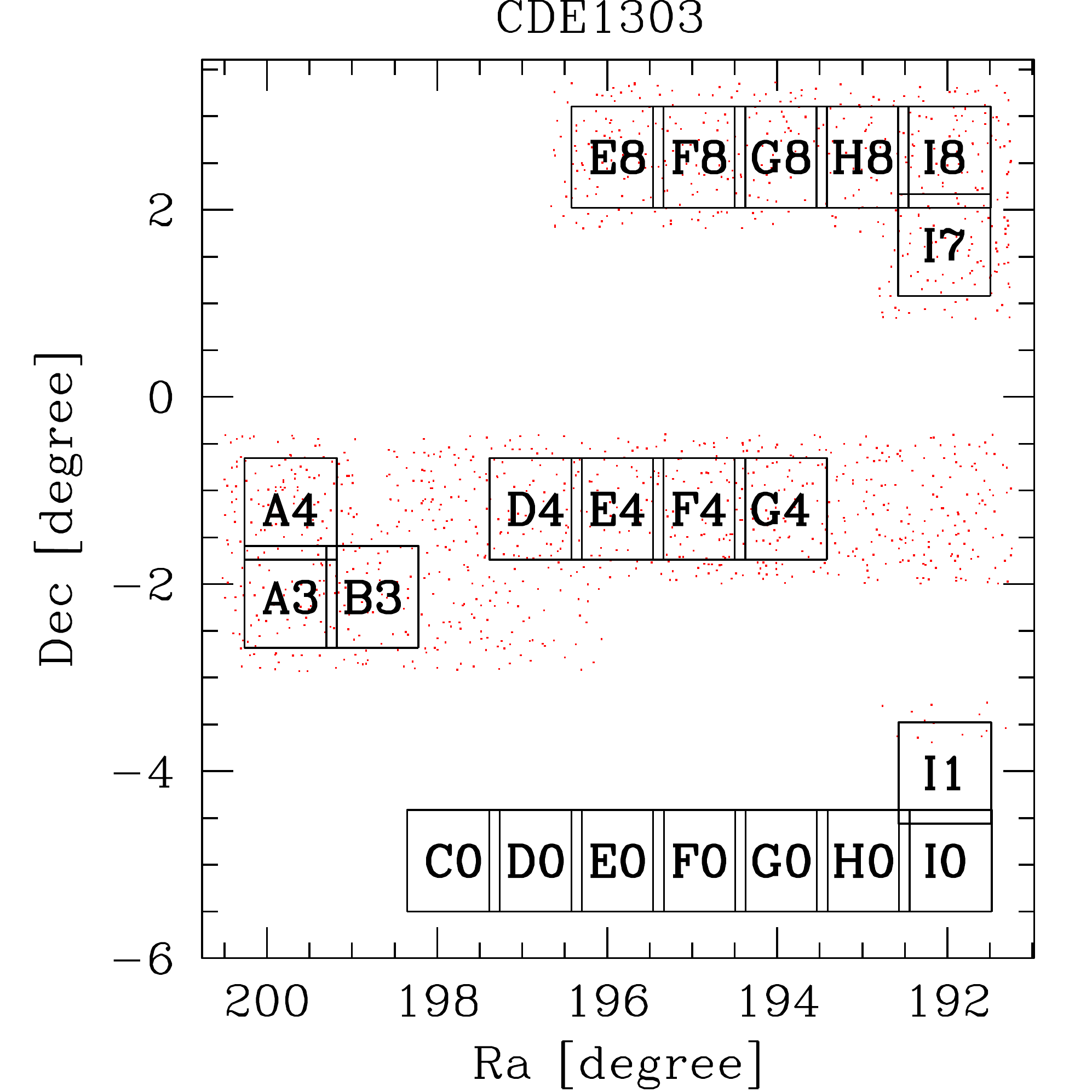}\,
\includegraphics[valign=t,width=0.29\textwidth]{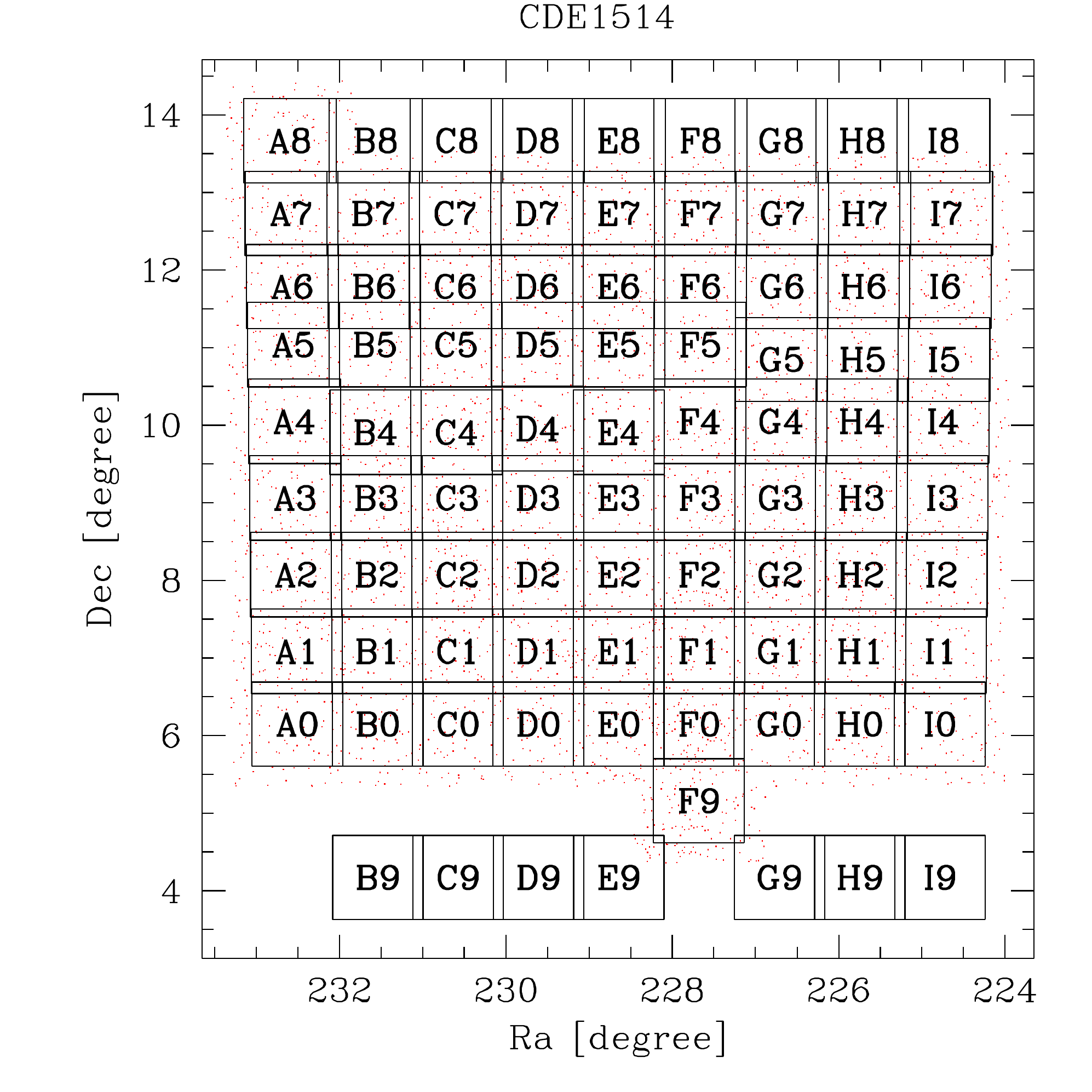}\\\vspace{0.5cm}
\includegraphics[valign=t,width=0.20\textwidth]{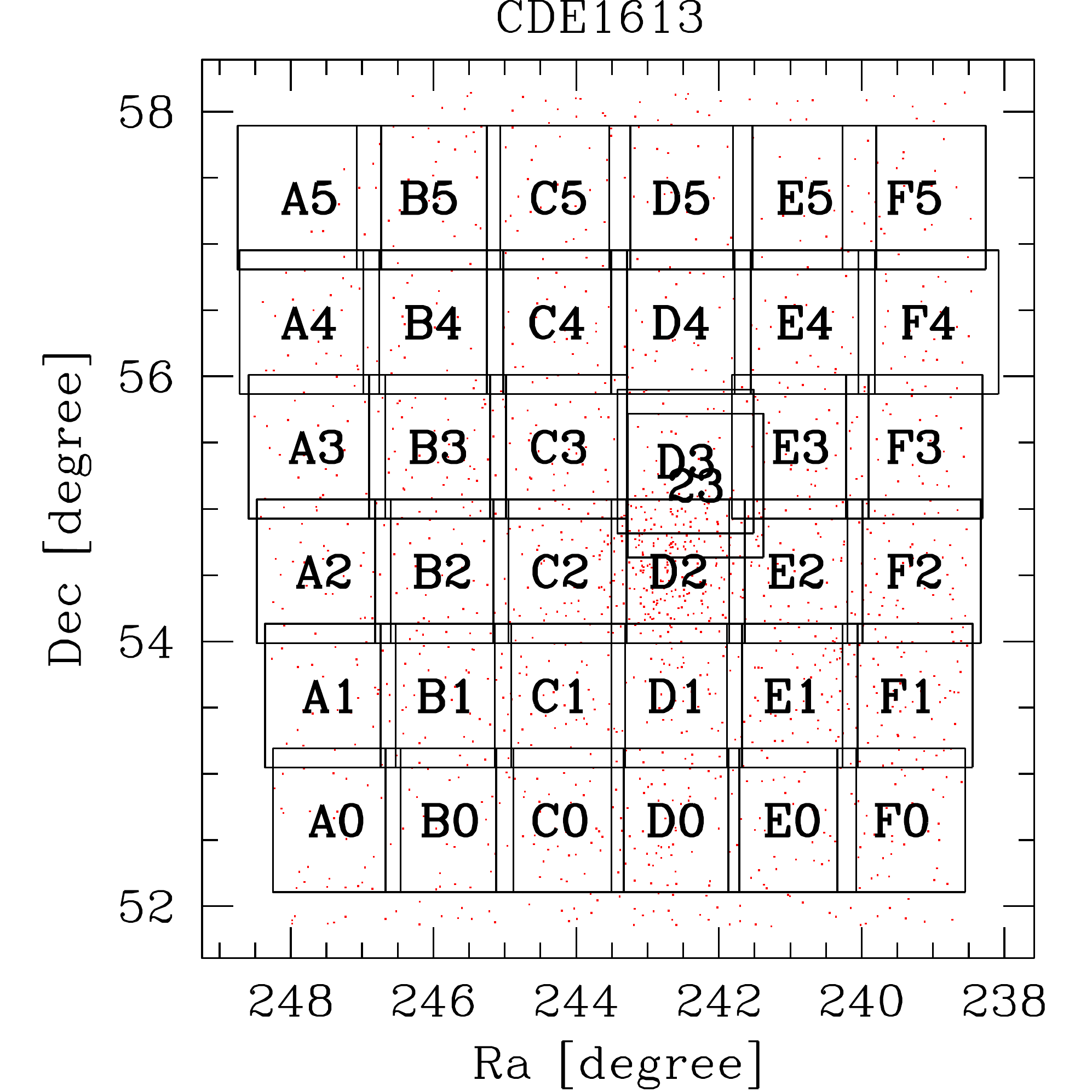}\,
\includegraphics[valign=t,width=0.20\textwidth]{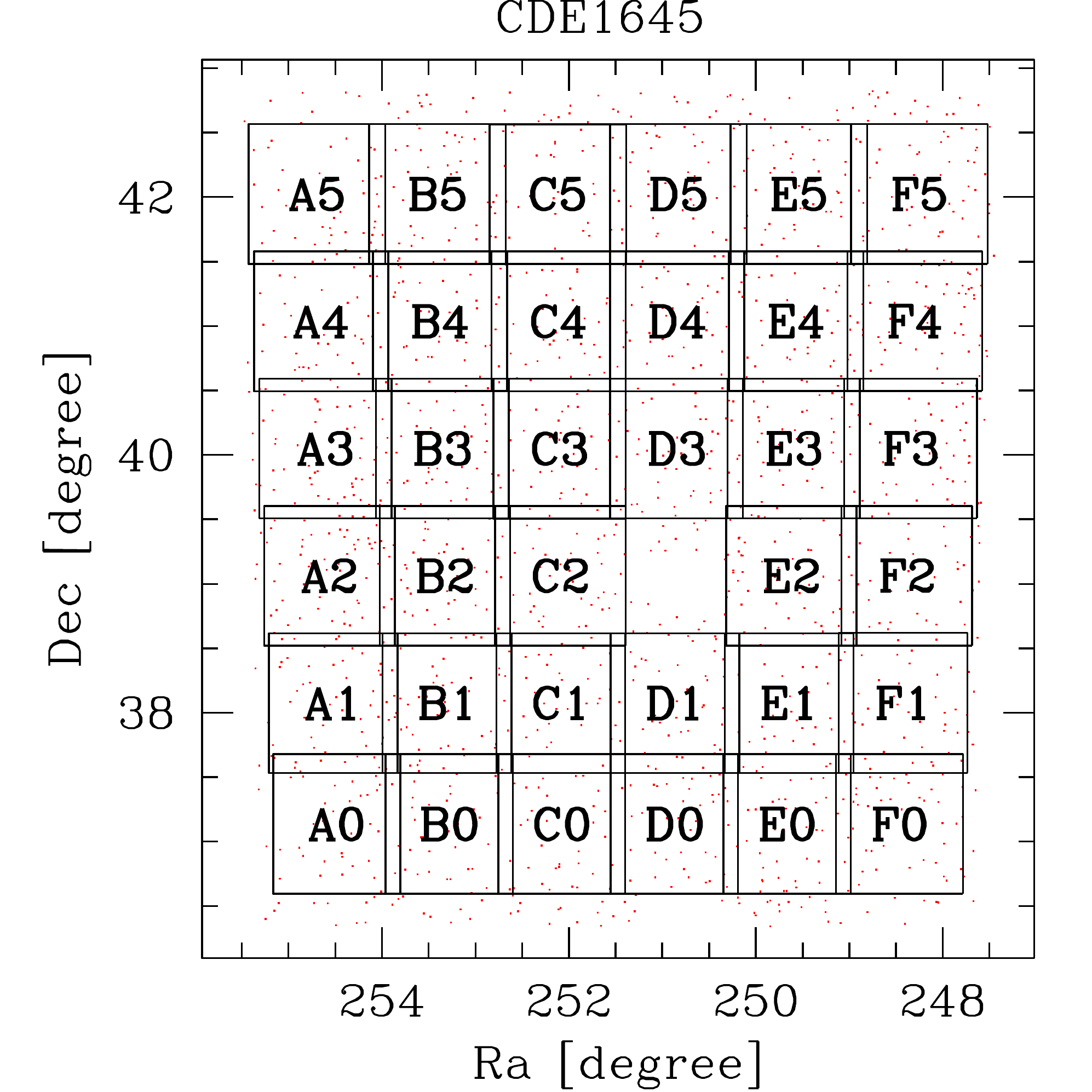}\,
\includegraphics[valign=t,width=0.29\textwidth]{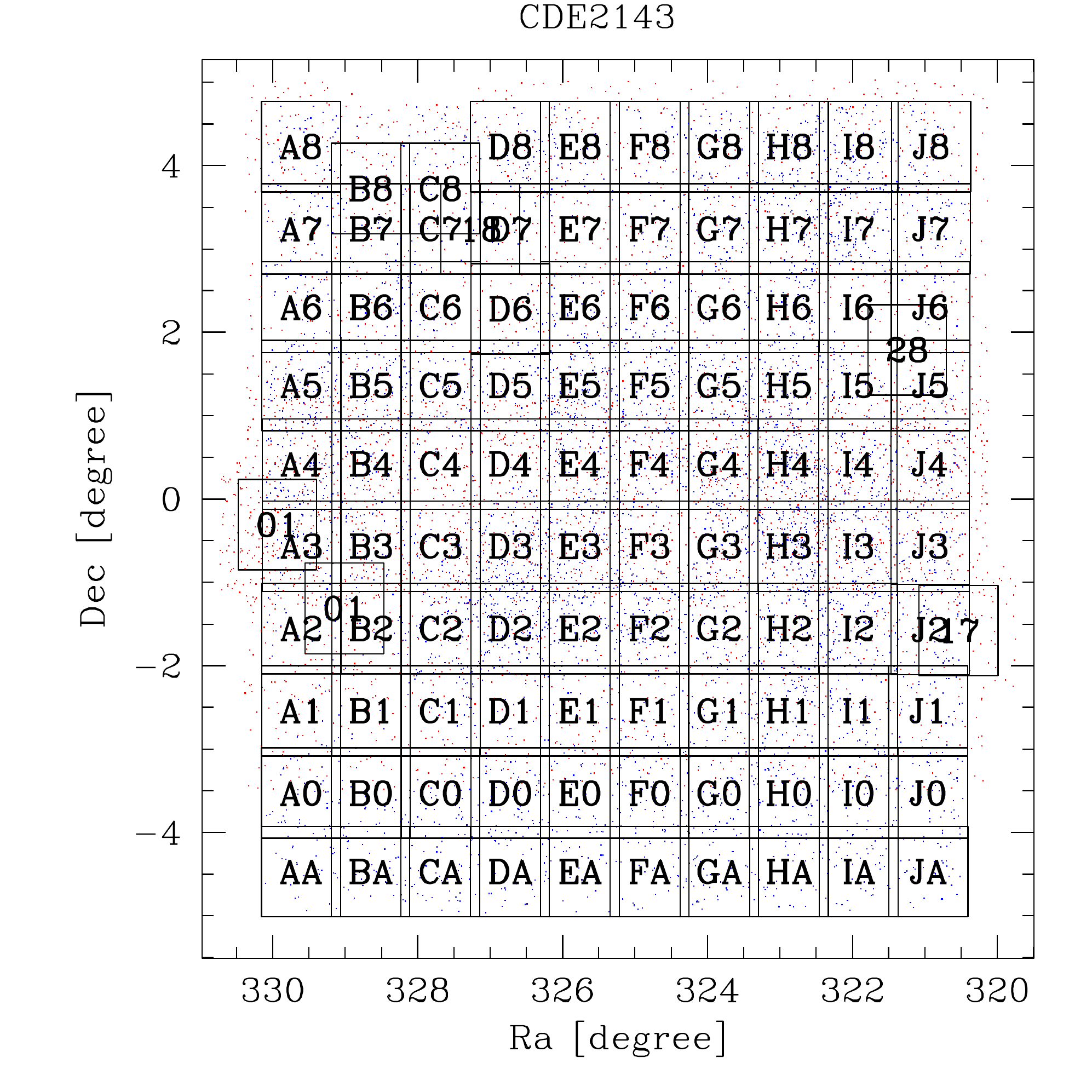}\\\vspace{0.5cm}
\includegraphics[valign=t,width=0.20\textwidth]{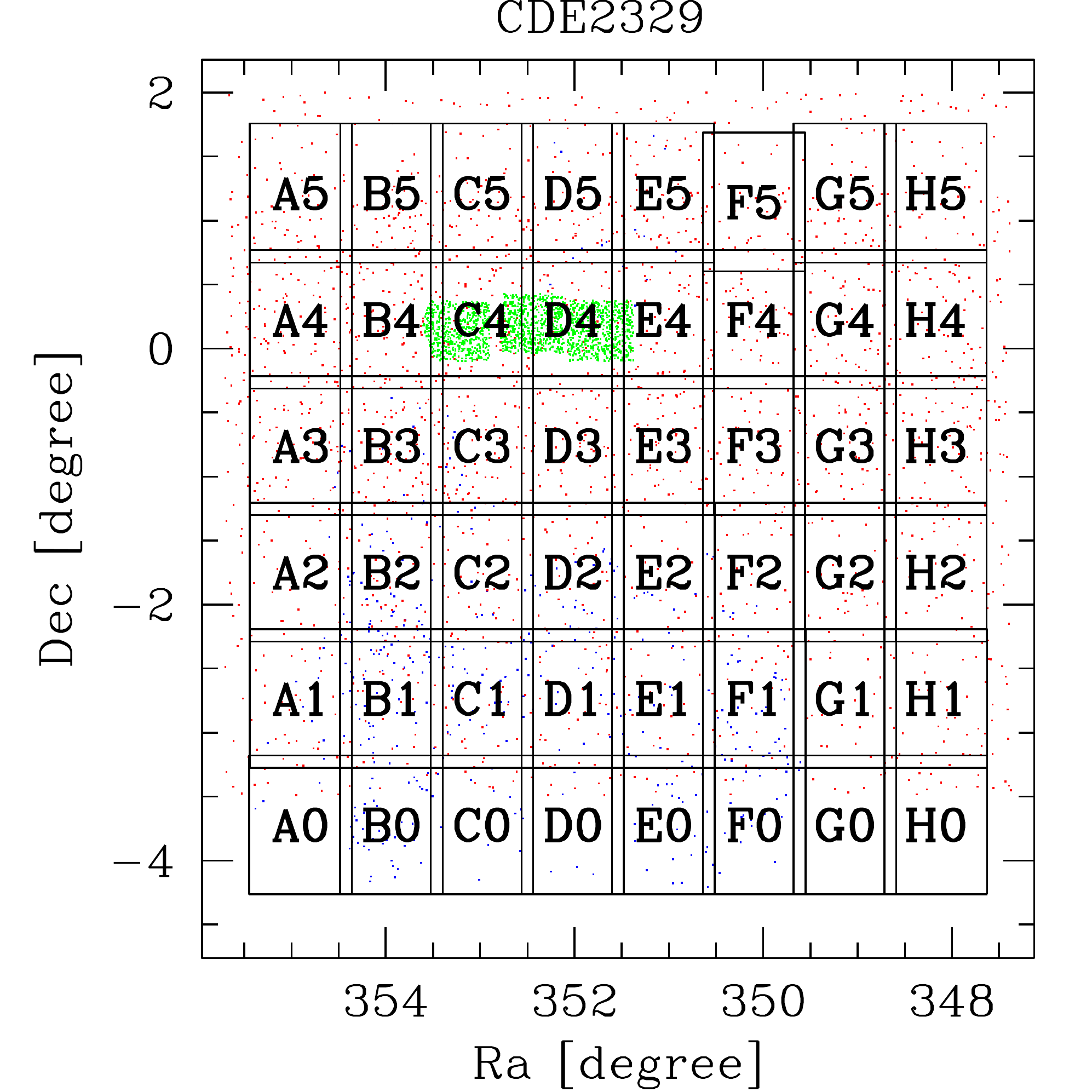}\,
\includegraphics[valign=t,width=0.29\textwidth]{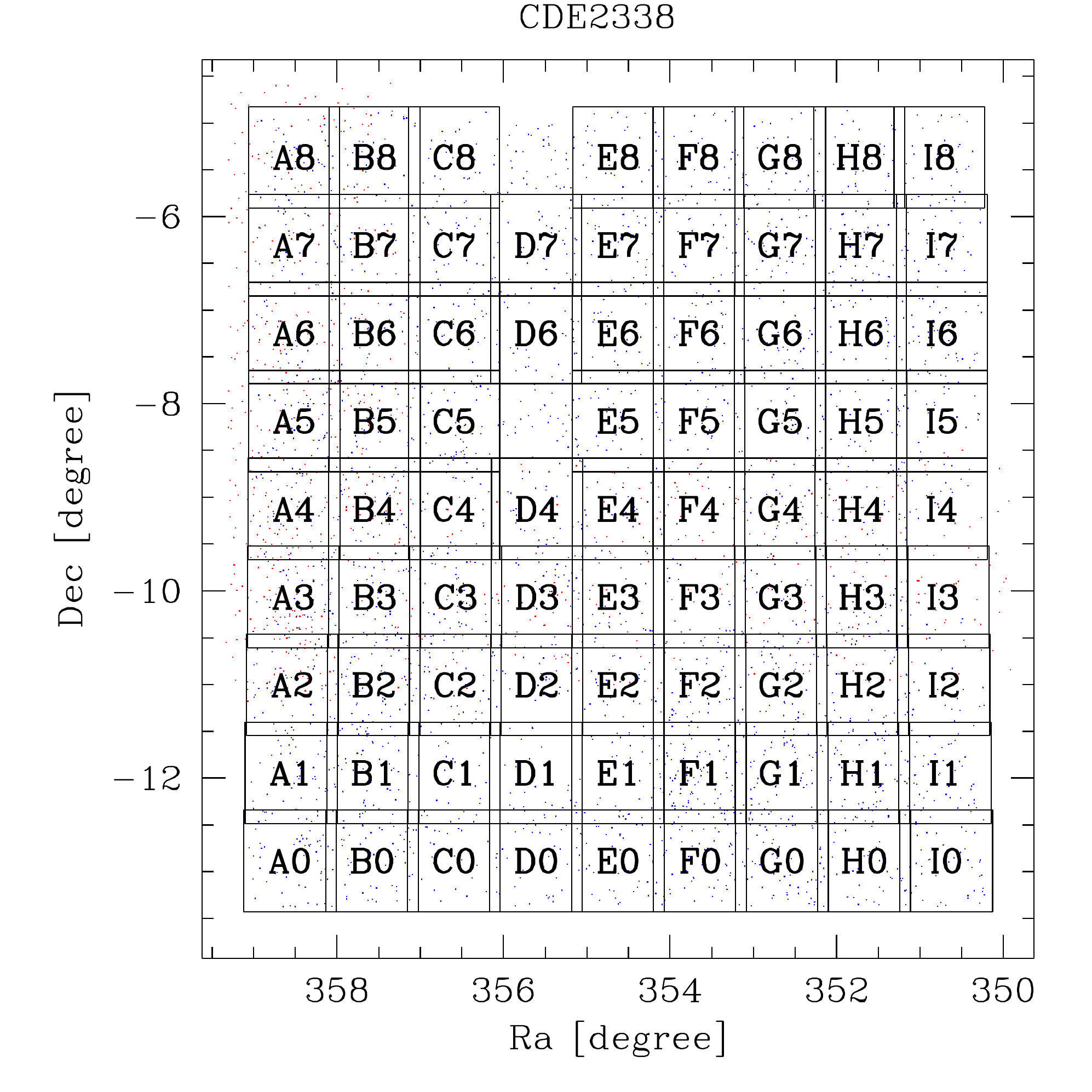}\,
\includegraphics[valign=t,width=0.20\textwidth]{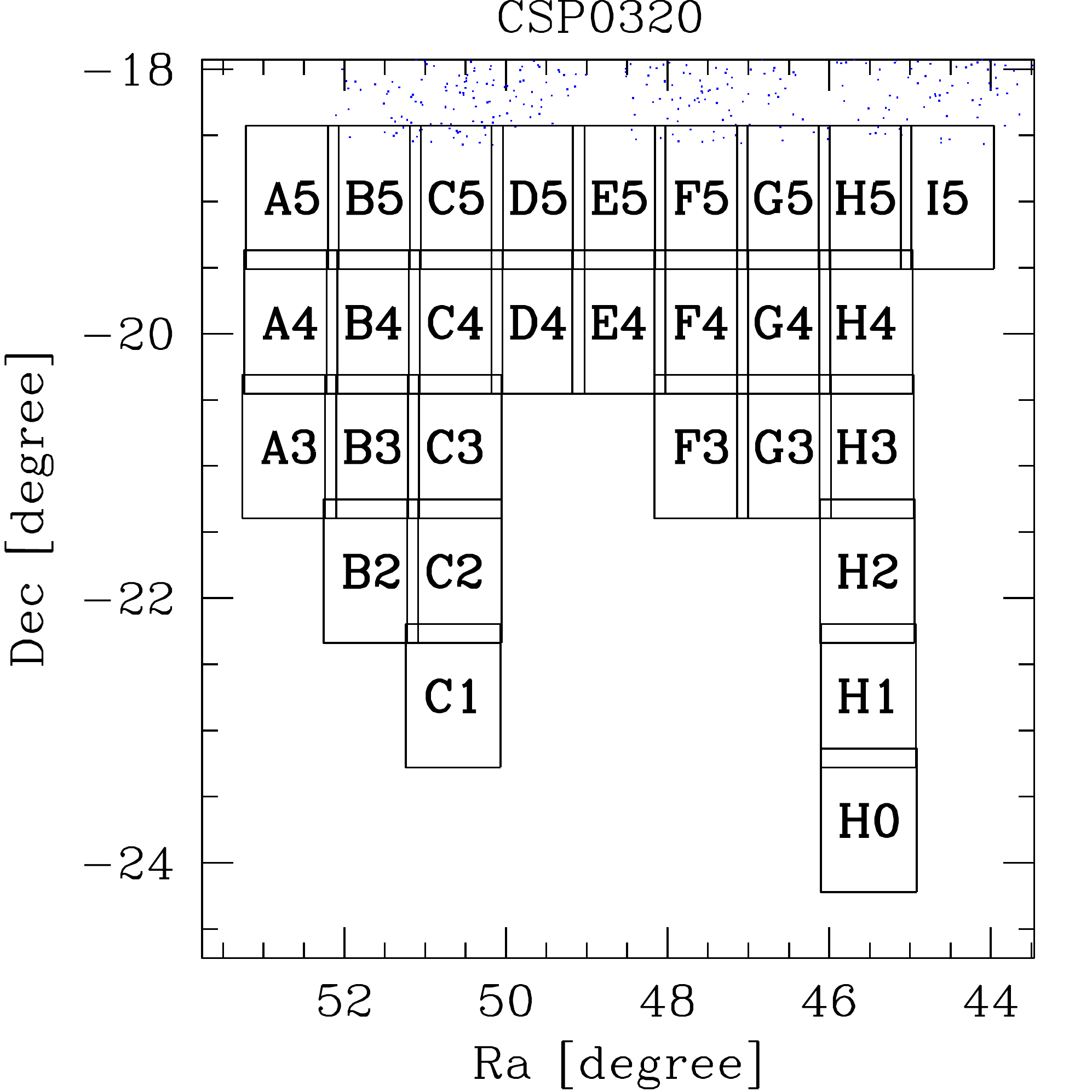}
\caption{Layout plots of the 14 RCS2 patches. Note that this plot contains more pointings than are included in RCSLenS since we only reduce the pointings where $r$-band data is available. Over-plotted are the positions of objects with spectroscopic redshifts (red: SDSS; blue: WiggleZ; green: DEEP2).}
\label{fig:layout}
\end{figure*}

Figure~\ref{fig:layout} shows the layout of the 14 RCSLenS patches and the naming scheme for the individual fields/pointings.

\section{Quality control}
\label{sec:app_sanity}
An example of a quality control plot for the field CDE0047B0 is shown in Fig.~\ref{fig:CDE0047B0_sanity}. A full description of each panel can be found in appendix B of \cite{kuijken/etal:2015}.

\begin{figure*}
\includegraphics[width=0.9\hsize]{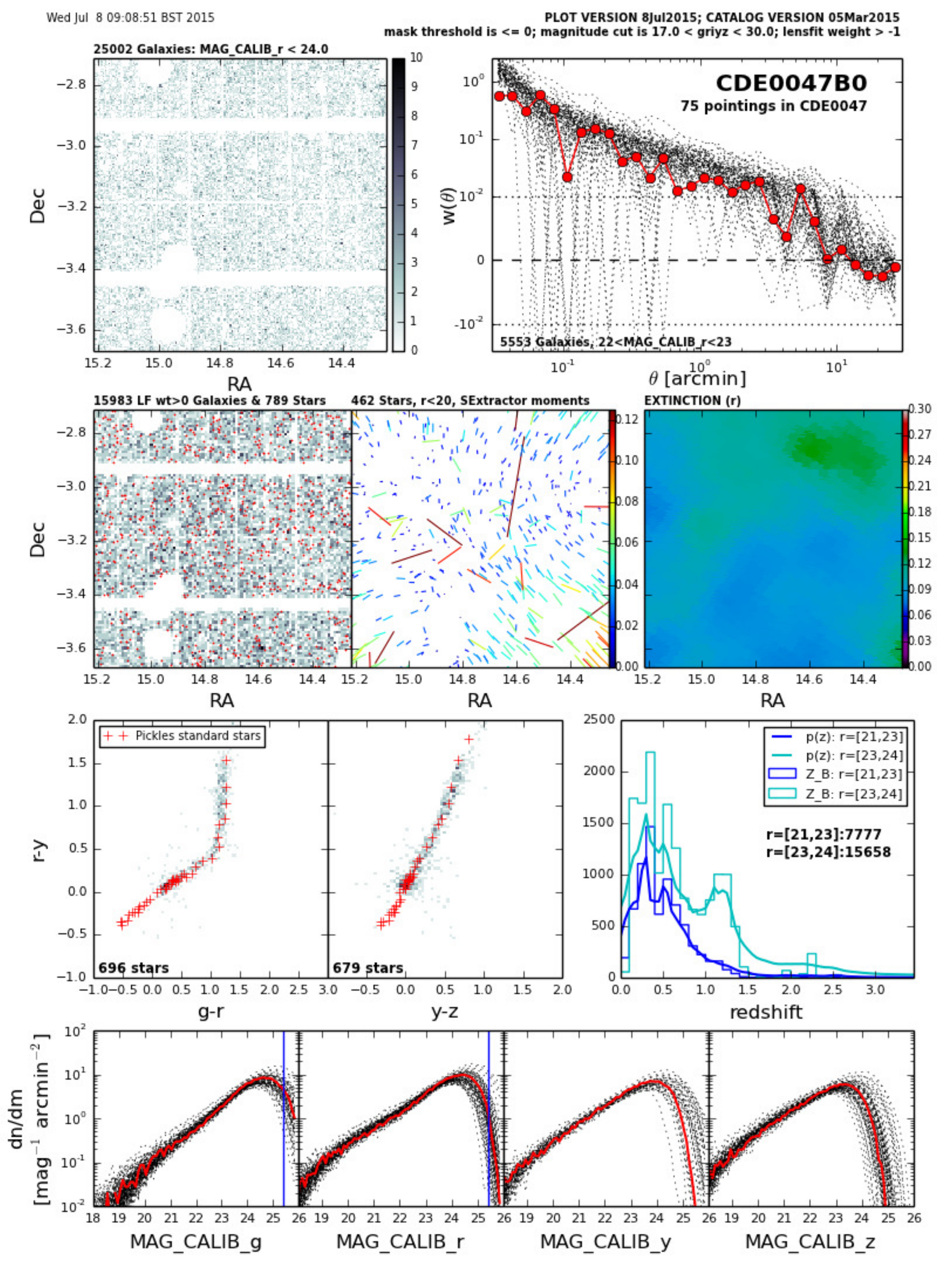}
\caption{Quality control plots for the field CDE0047B0 (see appendix~\ref{sec:app_layout} for the field naming scheme). Shown are the sky distribution of objects with $r<24$, the angular auto-correlation function of galaxies, the sky distribution of galaxies and stars, a whisker plot of the PSF ellipticity measured on stars, an extinction map, colour-colour diagrams of stars, photometric redshifts distributions along with the stacked PDFs, and magnitude number counts in the different bands. For the angular correlation function and the number counts we additionally plot all other fields (grey dashed lines) from the same patch for comparison.}
\label{fig:CDE0047B0_sanity}
\end{figure*}

\section{Data release}
\label{sec:release}
The RCSLenS data release contains the following data products:
\begin{enumerate}
  \item Calibrated science images in all bands.
  \item Weight maps describing the noise properties of the science images.
  \item Flag maps which represent a binary version of the weight maps.
  \item Sum images containing information about how many exposures went into a pixel (mostly 1 for RCSLenS).
  \item Masks images that contain information about image defects.
  \item Multi-colour catalogues containing information on photometry and shapes.
\end{enumerate}
The data are accessible through the Canadian Astronomy Data Centre at \url{http://www.cadc-ccda.hia-iha.nrc-cnrc.gc.ca/en/community/rcslens/query.html}. Since RCSLenS was processed with essentially the same pipeline as CFHTLenS the data structure is very similar to that survey. We refer the reader to the CFHTLenS data paper \citep{2013MNRAS.433.2545E} for many of the details about the data that also apply to RCSLenS. Here we summarise the differences in the two data releases.

\subsection{Images}
The structure of the RCSLenS images (i.e. science images, weight maps, flag maps, sum images, and mask images) follows CFHTLenS. The main difference is the photometric re-calibration which is based on SDSS and stellar locus regression (SLR) in RCSLenS. Hence besides the original AB magnitude zeropoint (FITS header keyword {\tt MAGZP}) the images contain an additional header keyword {\tt MAGZPCOR} that includes the re-calibration and was used for the extraction of the catalogues. Attached to that is another keyword, {\tt PHOREF}, to indicate whether the photometry in this particular image was re-calibrated with SDSS or SLR.

Owing to the different filter set of RCSLenS compared to CFHTLenS ($griz$ vs. $ugriz$) the mask bits have been distributed slightly differently. See Table~\ref{tab:mask_bits} for a summary of the meaning of the bit value in the mask images.

\begin{table}
\caption{Bit coding of the RCSLenS FITS masks. The $R$-band magnitude for bits 1, 2, and 4 corresponds to the Guide Star Catalogue 1/2.}
\label{tab:mask_bits}
\begin{tabular}{ll}
\hline
\hline
Bit  & reason for mask\\
\hline
1    & star halos with $10.35<R<11.2$ \\
2    & star halos with $R<10.35$      \\
4    & stars with      $R<17.5$       \\
8    & saturated pixels\\
16   & asteroid tracks\\
32   & areas of significant underdensity in $g$-, $i/y$-, $z$-bands\\
64   & areas of significant underdensity in $r$-band\\
128  & manual masks\\
256  & $g$-band flag map\\
512  & $r$-band flag map\\
1024 & $i/y$-band flag map\\
2048 & $z$-band flag map\\
8192 & pixels outside the RA/Dec cuts of this pointing\\
\end{tabular}
\end{table}

Table~\ref{tab:filter_names} summarises the RCSLenS filter naming convention with respect to the official CFHT filter names. For photo-$z$ estimation we use the filter curves available under \url{http://www.cfht.hawaii.edu/Instruments/Imaging/Megacam/data.MegaPrime/MegaCam_Filters_data.txt} properly convolved with the transmission curves of the other optical elements and the CCD.

\begin{table}
\caption{\label{tab:filter_names}RCSLenS filter names and their official CFHT identifiers.}
\begin{tabular}{ll}
\hline
\hline
RCSLenS name & CFHT identifier \\
\hline
$g$ & g.MP9401\\
$r$ & r.MP9601\\
$i$ & i.MP9701\\
$y$ & i.MP9702\\
$z$ & z.MP9801\\
\end{tabular}
\end{table}

\subsection{Catalogues}
Most quantities in the RCSLenS catalogues are also available in the CFHTLenS catalogues. A detailed description can be found in \cite{2013MNRAS.433.2545E} in appendix C. Additional quantities in RCSLenS are (an \_$x$ is a placeholder for different filter names like $griz$):
\begin{itemize}
\item {\tt SG\_FLAG}: Star-galaxy classifier as described in Sect.~\ref{sec:SG_sep}.
\item {\tt c1\_DP, c2\_DP}: 1st pass c-correction as described in Sect.~\ref{sec:noise_bias}.
\item {\tt c1\_NB, c2\_DP}: 2nd pass c-correction as described in Sect.~\ref{sec:noise_bias}.
\item {\tt LP\_kcor\_}$x$: k-corrections in the RCSLenS bands estimated with the LePhare code (see Sect.~\ref{sec:abs_mag}).
\item {\tt LP\_log10\_SFR\_MED, LP\_log10\_SFR\_SUP, LP\_log10\_SFR\_INF}: Star formation rates (median, upper 95\% confidence bound, lower 95\% confidence bound) estimated with the LePhare code (see Sect.~\ref{sec:abs_mag}).
\end{itemize}


\label{lastpage}
\end{document}